\PassOptionsToPackage{sort&compress}{natbib}   % elsarticle loads natbib (numbered)
\documentclass[3p,times]{elsarticle}   % single-column journal layout; use [preprint,12pt] for a double-spaced review copy

\journal{Journal of Computational Physics}

\usepackage{mathtools} % Includes amsmath functionality
\usepackage{amssymb}
\usepackage{amsthm}
\usepackage{bm}

\usepackage{booktabs}
\usepackage{array}
\usepackage{makecell}
\usepackage{subcaption}

\usepackage{pgf}
\usepackage{tikz}

\usepackage{caption}

\usepackage{xcolor}

\usepackage{hyperref}
\hypersetup{
    colorlinks=true,
    linkcolor=blue,
    filecolor=magenta,
    urlcolor=cyan,
    citecolor=blue,
    }
\DeclareMathOperator{\dvg}{div}
\DeclareMathOperator{\grad}{grad}
\DeclareMathOperator{\curl}{curl}
\DeclareMathOperator{\Id}{Id}

\newcommand{\bR}{\mathbb{R}}
\newcommand{\rd}{\mathrm{d}}

\newcommand{\ttimes}{\!\times\!}

\newcommand{\Hdiv}{H^{\dvg}}
\newcommand{\Hcurl}{H^{\curl}}
\newcommand{\Hone}{H^1}

\newcommand{\Energy}{\mathcal E}
\newcommand{\Helicity}{\mathcal H}
\newcommand{\parameter}{\alpha}
\newcommand{\Leray}{\Pi^{\dvg = 0}}

\newcommand{\Domain}{\Omega}
\newcommand{\ip}[2]{\langle #1, #2 \rangle_{\Domain}} 

\newcommand{\norm}[1]{\| #1 \|_{\Domain}}
\newcommand{\normrel}[1]{\| #1 \|_{\mathrm{rel}}}     % relative error against a reference (2026-09-22)
\newcommand{\normnorm}[1]{\| #1 \|_{\mathrm{norm}}}   % the force residual, normalised by the magnetic pressure gradient (2026-09-22)
\newcommand{\vavg}[1]{\langle #1 \rangle_{\Domain}}   % the volume average
\newcommand{\Mapping}{\Phi}

\newcommand{\nfp}{n_{\mathrm{fp}}}                          % field periods
\newcommand{\betavol}{\beta_{\mathrm{vol}}}                 % volume-averaged beta
\newcommand{\mesh}[3]{#1 \ttimes #2 \ttimes #3}             % a spline resolution (n_r, n_theta, n_zeta)
\newcommand{\dB}{\delta B}                                  % a field perturbation ...
\newcommand{\dA}{\delta A}                                  % ... and the potential it comes from
\newcommand{\aopt}{a^{\star}}                               % the energy-optimal seed amplitude
\newcommand{\res}[2]{r_{#1#2}}                              % the radius of the (m, n) resonance
\newcommand{\iotabar}{\mathord{\mbox{\ooalign{$\iota$\cr\hidewidth\rule[0.47ex]{0.32em}{0.045em}\hidewidth\cr}}}}

\newtheorem{theorem}{Theorem}
\newtheorem{corollary}[theorem]{Corollary}
\newtheorem{lemma}[theorem]{Lemma}
\newtheorem{definition}[theorem]{Definition}
\newtheorem{example}[theorem]{Example}
\theoremstyle{remark}
\newtheorem*{remark}{Remark}
\theoremstyle{plain}
\newtheorem{proposition}[theorem]{Proposition}

\DeclareMathOperator*{\range}{range}

\usepackage{import}
\usepackage{placeins}   % \FloatBarrier: every float before the references

\begin{document}

\begin{frontmatter}

\title{MRX: A differentiable 3D MHD equilibrium solver without nested flux surfaces}

\author[nyu]{Tobias Blickhan\corref{cor1}}
\ead{tobias.blickhan@nyu.edu}
\cortext[cor1]{Corresponding author}

\author[nyu]{Julianne Stratton}

\author[nyu,uw]{Alan A. Kaptanoglu}

\affiliation[nyu]{organization={Courant Institute of Mathematical Sciences, New York University},
            city={New York},
            postcode={10012},
            state={NY},
            country={USA}}

\affiliation[uw]{organization={William E. Boeing Department of Aeronautics and Astronautics, University of Washington},
            city={Seattle},
            postcode={98195},
            state={WA},
            country={USA}}

\begin{abstract}
    This article introduces the 3D magnetohydrodynamic (MHD) equilibrium solver MRX, based on relaxation of a magnetic field to a lower energy equilibrium state via admissible variations.
    We describe the mathematical theory behind this method and discuss what can be transferred to the fully discrete setting of a relaxation code.
    Our code is designed to address a number of traditional challenges to 3D MHD equilibrium solvers: enforcing physical constraints such as divergence-free magnetic field, fast convergence, handling strongly shaped polar geometries, and controlling topology changes of the magnetic field.
    Building on the JAX framework, we address questions of computational efficiency on modern computing architectures, user accessibility, and differentiability at each step.
    The solver is verified on manufactured vacuum solutions and against the vacuum field of a VMEC equilibrium of a quasi-axisymmetric stellarator.
    Automatic differentiation with respect to the boundary shape is demonstrated by optimizing stellarator shape for quasi-axisymmetry in vacuum.
    Relaxation is demonstrated on a finite-$\beta$ configuration, where we study $h$ refinement and the targeted formation of magnetic island chains at rational surfaces following an energy criterion. 
    An inexact Newton method enables the computation of high-resolution finite-$\beta$ equilibria with islands and chaos in minutes on a single GPU.
\end{abstract}

% Graphical abstract: 13 x 5 cm, text-light, built from figs/pgf/figure1_mesh3d/mesh_sections_3d.pgf by graphical_abstract.tex
% \begin{graphicalabstract}
% \centering
% \includegraphics[width=13cm]{figs/graphical_abstract.pdf}
% \end{graphicalabstract}

% Research highlights: 3 to 5 bullets of at most 85 characters each, also in highlights.txt
% \begin{highlights}
% \item A differentiable 3D MHD equilibrium solver in JAX that runs on CPU, GPU, and TPU
% \item Structure-preserving discretization retains the de~Rham structure of electrodynamics
% \item Formation of magnetic islands and chaos through targeted reconnection
% \item An inexact Newton method computes finite-$\beta$ equilibria in minutes on a single GPU
% \end{highlights}


\end{frontmatter}

% Opening figure, placed directly below the front matter (not floated)
\begin{center}
    \begin{minipage}[c]{0.07\linewidth}
        \centering
        \includegraphics{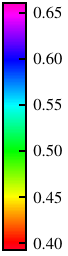}\\[0.3em]
        $\iota$
    \end{minipage}\hfill
    \begin{minipage}[c]{0.78\linewidth}
        \centering
        \resizebox{\linewidth}{!}{\includegraphics{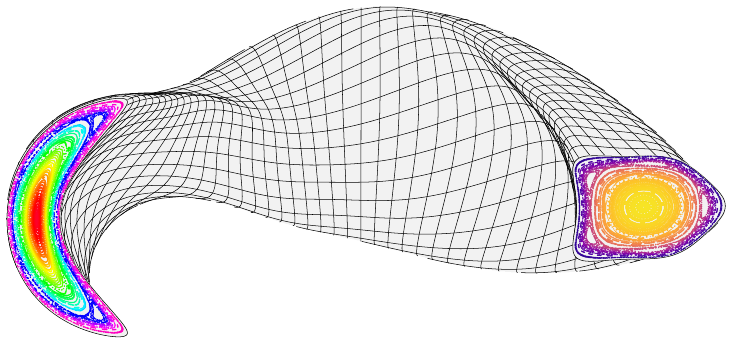}}
    \end{minipage}\hfill
    \begin{minipage}[c]{0.13\linewidth}
        \centering
        \includegraphics{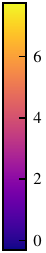}\\[0.3em]
        $p_{\mathrm{norm}} \times 100$
    \end{minipage}
    \captionof{figure}{Visualization of a non-nested equilibrium with MRX, the seeded state of Section.~\ref{subsec:seeding} at $\zeta = 0$ (left) and $\zeta = 0.5$ (right), colored by rotational transform $\iota$ (Definition~\ref{def:rotational_transform}, left) and normalized pressure (Equation \ref{eq:pressure_problem}, right). 
    The geometry is that of the National Compact Stellarator Experiment~\cite{ncsx2001physics}.
    Starting from a nested configuration, a large $\tfrac35$ island chain has formed after island seeding and relaxation.}
    \label{fig:ncsx_sections}
\end{center}

\section{Introduction}
\label{sec:introduction}

The magnetohydrodynamic (MHD) equations are partial differential equations (PDEs) that treat the plasma as a single fluid, used to model its dynamics.
This work is concerned with finding special solutions to this set of PDEs: equilibrium states.

In this introduction section, after presenting the problem in Section~\ref{subsec:problem}, we discuss solution strategies under additional assumptions in Section~\ref{subsec:special_cases}. 
In Section~\ref{subsec:goals}, we discuss the overarching aim of this work, Section~\ref{subsec:existing_codes} presents existing codes, and Section~\ref{subsec:contributions} lists our contributions.

\subsection{Problem statement}
\label{subsec:problem}

Many approximately steady-state space and laboratory plasmas are well-described by the time-independent limit of the MHD equations with zero equilibrium flow.

\paragraph{Magnetohydrostatics}
A three-dimensional, static, ideal magnetohydrodynamic equilibrium state is defined by the force balance equation coupled with Maxwell's equations. 
This constitutes a search for $B: \Domain \to \bR^3$, with a bounded Lipschitz domain $\Domain \subset \bR^3$ such that
\begin{align}
    \label{eq:mhs}
    J \times B = \grad p, \quad \curl B = J, \quad \dvg B = 0,
\end{align}
together with suitable boundary conditions. 
We will refer to~\eqref{eq:mhs} as the magnetohydrostatic (MHS) problem. 
Throughout this work, we take units so that vacuum permeability $\mu_0 = 1$. 
Plasma pressure is denoted by $p$ and points in $\Domain$ are denoted $x$.

Figure~\ref{fig:ncsx_sections} illustrates an equilibrium magnetic field configuration for a stellarator geometry, with a mix of magnetic surfaces and island chains computed with MRX (Section~\ref{subsec:seeding}).

The boundary conditions we use in this work are $B \cdot n = J \times n = 0$ on $\partial \Domain$, where $n$ is the unit vector normal to the boundary.
In equilibrium, it follows that $\grad p \cdot n = 0$ on $\partial \Domain$.
Physically, $\partial \Domain$ is a magnetic flux surface in vacuum. 
Our discretization also allows boundary conditions of the form $(B - B_{\mathrm{vac}}) \cdot n = 0$ for vacuum fields (Section~\ref{subsec:manufactured_vacuum}). Their use in relaxation will be investigated in future work.

\paragraph{Applications}
It is hard to overstate the importance of computing realistic magnetic equilibria. 
MHD equilibria form the foundation of the design of magnetic-confinement fusion devices by facilitating: (i) plasma state reconstructions from experimental data~\cite{lao_reconstruction_1985}, (ii) MHD and other stability calculations~\cite{sanchez_cobra_2000}, (iii) neoclassical~\cite{van_rij_variational_1989} and gyrokinetic~\cite{mandell_gx_2024} transport calculations, and (iv) the initialization of extended MHD time-dependent codes such as JOREK3D~\cite{nikulsin_jorek3d_2022}, {M3D-C1}~\cite{jardin_multiple_2012} and {NIMSTELL}~\cite{sovinec_modeling_2024}.

Plasma state reconstruction in particular is important for the performance of plasma control systems. 
These systems can allow for real-time instability and heat flux control and are crucial to fusion device design and operation~\cite{goncalves_real-time_2010}. 

\subsection{Special cases}
\label{subsec:special_cases}

Finding a solution to~\eqref{eq:mhs} is challenging analytically and there are open questions about the existence of MHS solutions in toroidal geometry without the assumption of axisymmetry. 
The Grad conjecture~\citep{grad_toroidal_1967} states that smooth MHS solutions with nested toroidal pressure surfaces that come in continuous families can only exist in the presence of symmetry (axial, helical, or by reflection)~\citep[Conjecture 1]{constantin_flexibility_2021}. 
Explicit smooth stellarator-symmetric solutions without axial, helical or reflection symmetry, with nested surfaces and finite pressure, were recently constructed in~\citep{landreman_analytic_2026} (\ref{sec:landreman_verification}). 
These solutions exhibit very low shear, i.e., their rotational transform $\iota$ (Definition~\ref{def:rotational_transform}) is nearly constant.

\paragraph{Grad--Shafranov}
An important class of fusion experiments, tokamaks, are based on magnetic confinement in axisymmetric toroidal geometry.
In this case, flux surfaces are guaranteed and can be labeled by a magnetic flux function $\psi$. The MHS condition~\eqref{eq:mhs} reduces to the Grad--Shafranov equation, a two-dimensional, elliptic, nonlinear partial differential equation.
Various reliable solvers for the Grad--Shafranov equation are in use, see~\cite{hansen_tokamaker_2024} and references therein.

\paragraph{Nested flux surfaces}
Toroidal devices without axisymmetry are called stellarators. 
Nested flux surfaces are not guaranteed in a 3D toroidal volume $\Domain$, and therefore the full 3D equations must be solved.
Nonetheless, the most commonly used codes to compute stellarator equilibria, such as {VMEC}~\citep{hirshman_steepest-descent_1983}, {GVEC}~\citep{hindenlang_computing_2025, Hindenlang2026}, and {DESC}~\cite{dudt_desc_2020} assume that nested magnetic flux surfaces exist and therefore the surfaces can be labeled by a toroidal magnetic flux $\psi: \Domain \to \bR$. MRX takes these solutions as initial conditions (Section~\ref{par:maps_initial_fields}) and reproduces them in the vacuum limit (Section~\ref{subsec:vmec_convergence}).

The existence of nested flux surfaces implies the foliation of the plasma volume into nested tori.
Magnetic field lines (i.e., integral curves of the vector field $B$) lie on these tori and $\grad p$ is orthogonal to them: the dot product of the force balance with $B$ gives $B \cdot \grad p = 0$.
This implies that the pressure $p(x) = p_{\mathrm{profile}}(\psi(x))$ and the rotational transform $\iota(x) = \iota_{\mathrm{profile}}(\psi(x))$ are flux functions $[0,1] \to \bR$.
The rotational transform describes the number of poloidal rotations per single toroidal rotation of a magnetic field line. 

Nested flux surfaces are often a useful assumption because they simplify the problem, establish a convenient flux coordinate system, and reflect the expectation that stellarators with good confinement should exhibit large volumes of nested flux surfaces.
Force balance generally can only hold in a very weak sense, as singular currents will appear at rational values of $\iota_{\mathrm{profile}}(\psi)$~\cite{loizu_magnetic_2015, lazerson_verification_2016}.
This will be discussed in detail in Section~\ref{subsec:rational_surfaces}.

\subsection{Goals}
\label{subsec:goals}

The goal of this project is to develop a robust, fast, and conceptually sound solution method for the MHS problem. This allows modeling more realistic plasma equilibria featuring magnetic islands and chaos, out of scope for nested flux surface solvers. Furthermore, we want to be able to explore the optimization landscape of the MHS problem and make statements about the characteristics of equilibrium states, their dependence on initial conditions, and sensitivity with respect to design parameters.

\paragraph{Magnetic islands}
As stellarator optimization and experiments increasingly scale to more realistic geometries representing fusion-scale devices, there is an increasing need for finding equilibria with islands and chaos.
The former are closed tubes (non-nested with nearby flux surfaces) formed by magnetic field lines, the latter are volume-filling field lines.
A five-fold island chain can be seen near the plasma edge in Figure~\ref{fig:ncsx_sections}.

Modeling magnetic islands and chaos is important for capturing the dynamics of real experiments, where these features often play a large part in transport and divertor operation~\cite{coelho_global_2022}.
The presence of uncontrolled magnetic islands and chaotic regions primarily leads to a decrease in the quality of confinement of the plasma.
However, the experimental device Wendelstein 7-X is engineered to take advantage of a particular magnetic island chain at its edge, paired with an island divertor to allow for a controlled release of heat from the plasma~\cite{pedersen_confirmation_2016}.
Sophisticated design of such island divertors~\cite{feng_understanding_2021} or non-resonant divertors~\cite{garcia_exploration_2023} is crucial as the divertor nominally controls critical device properties such as large heat fluxes to the material surfaces~\cite{hegna_infinity_2025}.
Resolving island regions can also help to initialize extended time-dependent MHD codes, since the formation of islands is very slow at fusion-relevant resistivity. 

\paragraph{MHS landscape}
Under which conditions the three-dimensional MHS problem admits a unique solution for a given boundary, profiles, and toroidal flux is unknown. 
The empirical evidence suggests that it does not in general.

For the same fixed boundary, pressure and rotational transform, an axisymmetric and a helical-core equilibrium of comparable energy coexist~\cite{Cooper2010HelicalCore}, pointing towards a bifurcated state.
Under the assumption of nested surfaces, continua of near-equilibria have been found in~\cite{Thun2026Narrow}. 
For fixed boundary and rotational transform, continuous families of states with force residuals below the discretization error exist.
Most convincingly, the Large Helical Device (LHD) experiment shows hysteresis between magnetic-island states over overlapping parameter ranges. 
These are long-lived states that differ in topology and remember how they were reached~\cite{Narushima2015Hysteresis}.

The consequence for an iterative solver like MRX is that its result is determined by the initial condition.
Ideal relaxation preserves the magnetic topology of its initial condition exactly (Section~\ref{sec:magnetic_relaxation}) in the continuous setting. Discretely, the topology changes only through grid-scale reconnection (Section~\ref{subsec:rational_surfaces}). Hence the topology has to be supplied (Section~\ref{subsec:seeding}), and only resistivity lets differently seeded states approach one another. We demonstrate this explicitly in Section~\ref{subsec:reconnection_demo}.

\paragraph{Parametric dependence}
Consider an equilibrium state $B_{\mathrm{eq}, \parameter}$ computed for a certain set of parameters such as domain geometry, boundary condition, or physical parameters, collected in a variable $\parameter$. 
Questions of optimization and stability require the derivative $\partial_\parameter B_{\mathrm{eq}, \parameter}$.
Making this quantity computationally available is hugely beneficial for these applications. 
MRX is built to provide it through automatic differentiation (Section \ref{subsec:shape_optimization}).

\subsection{Existing 3D MHD equilibrium codes}
\label{subsec:existing_codes}
We have already mentioned a class of 3D MHD equilibrium codes that assume nested flux surfaces.
The solution is then determined by the flux coordinate system together with a handful of scalar functions, see~\eqref{eq:vmec_vector_potential}.

\paragraph{Breaking nested flux surfaces}
Beyond this class of methods, there are a few other codes that can produce 3D toroidal MHD equilibria with islands, which fall broadly into the classes of codes relying on multi-region relaxed MHD (MRxMHD) with stepped pressure profiles (SPEC, SPECTRE~\cite{loizu_verification_2016, balkovic2026spectre}), Picard-like fixed point iterations (e.g., PIES~\cite{drevlak_pies_2005}), resistive dynamics approaching a steady state (HINT~\cite{suzuki_development_2006, suzuki_hint_2017}), and magnetic relaxation codes (e.g., SIESTA~\cite{hirshman_siesta_2011}, and the MRX code presented here).
We give a summary of current and historical 3D MHS codes we are aware of in Table~\ref{tab:codes}.
Some of these codes can perform free-boundary calculations, but the table summary and our focus in this work are on fixed-boundary equilibria.

\begin{table}[t]
\caption{Three-dimensional MHD equilibrium codes for stellarator geometry, ordered by year of first publication. FD stands for finite differences, FR for Fourier representation, FE for finite elements.
Codes restricted to Cartesian geometry used to study the Parker problem are omitted, such as~\cite{pontin_parker_2020, smiet_ideal_2017, he_topology-preserving_2025, pontin_lagrangian_2009, candelaresi_mimetic_2014, mao2026structure}.
Also omitted are codes for time-dependent MHD, e.g., JOREK3D~\cite{nikulsin_jorek3d_2022}, {M3D-C1}~\cite{jardin_multiple_2012} and {NIMSTELL}~\cite{sovinec_modeling_2024}.
MRX is the only code in this list that is differentiable and does not assume nested flux surfaces.
}
\centering
\scriptsize
\setlength{\tabcolsep}{4pt}
\providecommand{\child}{\hspace{1em}$\vdash$\ }
\begin{tabular*}{\textwidth}{@{\extracolsep{\fill}} l l c l l l c @{}}
\toprule
\textbf{Code} & \textbf{Repository} & \textbf{Year} & \textbf{Method} & \textbf{Discretization} & \textbf{Back-end} & \textbf{Islands?} \\
\midrule
BETA / BETAS~\cite{bauer_computational_1978, bauer_magnetohydrodynamic_1984, betancourt_betas_1988, betancourt_comparison_1983} & -- & 1978 & relaxation (momentum) & FD $\times$ FR$^2$ & Fortran & \\
\child NSTAB~\citep{taylor_high_1992} & -- & 1992 & energy minimization & FD $\times$ FR$^2$ & Fortran & \\
\addlinespace
Chodura \& Schl\"uter~\cite{chodura_3d_1981} & -- & 1981 & relaxation (CG) & FD (not field-aligned) & Fortran & \checkmark \\
\child NEAR~\cite{hender_calculation_1985} & -- & 1985 & relaxation (CG) & FD $\times$ FR$^2$ & Fortran & \\
\addlinespace
VMEC~\citep{hirshman_steepest-descent_1983, hirshman_preconditioned_1991} & \texttt{gh:hiddenSymmetries/VMEC2000} & 1983 & energy minimization & FD $\times$ FR$^2$ & Fortran & \\
\child VMEC++~\cite{schilling2025numericsvmecpp} & \texttt{gh:proximafusion/vmecpp} & 2025 & energy minimization & FD $\times$ FR$^2$ & C++ & \\
\child VMEX & \texttt{gh:uwplasma/vmex} & 2026 & energy minimization & FD $\times$ FR$^2$ & JAX & \\
\addlinespace
PIES~\cite{reiman_calculation_1986} & -- & 1986 & field-line Picard & FD $\times$ FR$^2$ & Fortran & \checkmark \\
\addlinespace
HINT~\cite{park_three-dimensional_1986, harafuji_computational_1989} & -- & 1989 & resistive time-stepping & FD (not field-aligned) & Fortran & \checkmark \\
\child HINT2~\cite{suzuki_development_2006, suzuki_hint_2017} & -- & 2006 & resistive time-stepping & FD (not field-aligned) & Fortran & \checkmark \\
\addlinespace
SIESTA~\cite{hirshman_siesta_2011} & \texttt{gh:ORNL-Fusion/SIESTA} & 2011 & relaxation (Newton) & FD $\times$ FR$^2$ & Fortran & \checkmark \\
\addlinespace
SPEC~\cite{hudson_computation_2012} & \texttt{gh:PrincetonUniversity/SPEC} & 2012 & stepped pressure & Chebyshev $\times$ FR$^2$ & Fortran & \checkmark \\
\child SPECTRE~\cite{balkovic2026spectre} & \texttt{gl:spectre-eq/spectre} & 2026 & stepped pressure & Chebyshev $\times$ FR$^2$ & Numba & \checkmark \\
\addlinespace
% PSI-TET~\citep{hansen_mhd_2014} & -- & 2014 & Taylor eigenproblem & mixed FE (1st order) & Fortran, PETSc & \checkmark \\
% \addlinespace
DESC~\cite{dudt_desc_2020, panici_desc_2023} & \texttt{gh:PlasmaControl/DESC} & 2020 & nested collocation & Zernike $\times$ FR & JAX & \\
\addlinespace
GVEC~\cite{hindenlang_computing_2025} & \texttt{gh:gvec-group/gvec} & 2025 & energy minimization & B-splines $\times$ FR$^2$ & Fortran & \\
\midrule
This work & \texttt{gh:ToBlick/mrx} & 2026 & relaxation (Newton) & B-spline FE & JAX & \checkmark \\
\bottomrule
\end{tabular*}
\label{tab:codes}
\end{table}

The MRxMHD stepped pressure approach relies on solving Taylor relaxation problems in nested domains with pressure jumps at the interfaces.
It has the advantage of reducing the problem to coupled eigensolves of $\curl B = \lambda B$, $\lambda \in \bR$. In contrast to the ideal, topology-preserving relaxation of this work, the field relaxes to a Taylor state in each subdomain.
Hence, the result depends on the partition of $\Domain$ into nested subdomains, and convergence with respect to the number of subdomains is generally unclear in 3D toroidal geometry.

The PIES and HINT approaches are general approaches for non-nested equilibria, but each exhibits potential numerical disadvantages.
PIES is a Picard iteration on the magnetic field, and such fixed-point schemes converge at a linear rate.
In practice this rate degrades and under-relaxation of the current-profile updates is required for stable iteration~\cite{reiman_computation_1990}.
A Newton variant of PIES~\cite{oliver_solving_2006} computes the Newton derivative analytically in magnetic coordinates and therefore assumes nested flux surfaces. It shows a significant advantage over Picard iteration in many cases, but has not been adapted to magnetic islands and stochastic regions.

HINT instead relaxes artificial dissipative MHD equations toward a steady state, alternating a pressure update at fixed $B$ with a field update at fixed $p$.
Since the field is advanced explicitly on an Eulerian grid, a CFL condition limits the step size while the relaxation must run over the much longer dissipative timescale, which is costly.

Out of the relaxation codes, SIESTA is the most similar code to that proposed in this work. 
Similarities and differences will be worked out throughout the presentation, in particular in Sections~\ref{subsec:newton} and~\ref{subsec:seeding}.

\subsection{Our contributions}
\label{subsec:contributions}
In this work, we present a new code to solve magnetic relaxation problems in realistic geometries using a mixed finite element method built on top of the JAX computational framework.
Its contributions, in the order of the sections that demonstrate them, are:
\begin{enumerate}
    \item A structure-preserving B-spline discretization of the de~Rham complex on strongly shaped toroidal domains with a polar map following~\citep{buffa_isogeometric_2010, toshniwal_multi-degree_2017}, in which $\dvg B = 0$ holds to machine precision. Its vacuum solutions converge at order $h^p$ and agree with the vacuum field of a VMEC equilibrium (Section~\ref{subsec:vmec_convergence}).
    \item Derivatives of the vacuum field with respect to the map by automatic differentiation, used to optimize a stellarator for quasi-axisymmetry at fixed mean rotational transform, volume, and minor radius (Section~\ref{subsec:shape_optimization}).
    \item Incompressible relaxation of finite-$\beta$ fields by admissible variations, with the pressure as a Lagrange multiplier, accelerated by an inexact Newton method on the second variation of the energy (Section~\ref{sec:magnetic_relaxation}). This computes finite-$\beta$ NCSX equilibria to a normalized force residual of $10^{-8}$ in minutes on a single GPU (Section~\ref{sec:relaxation_numerics}), and its discretization is verified against an exact finite-$\beta$ equilibrium (\ref{sec:landreman_verification}).
    \item Targeted island seeding at rational surfaces following an energy criterion, which yields relaxed finite-$\beta$ states with island chains and stochastic regions (Section~\ref{sec:reconnection}).
    \item Resistive relaxation guided by a prescribed current, which carries relaxed states of different topology to a common equilibrium (Section~\ref{sec:resistive}).
\end{enumerate}
The two paragraphs below describe the foundations these rest on, the finite element setting and the code framework.

\paragraph{Structure preserving finite elements}
The finite element framework we employ guarantees the preservation of crucial features of the continuous problem after discretization.
Computational electrodynamics in general and magnetohydrodynamics in particular are rich with geometric structure, as we will discuss in Section~\ref{sec:framework}.
It is by now well-understood how this structure can be retained in the corresponding discrete problems thanks to advances in the theory of mixed finite element methods~\citep{alain_bossavit_computational_1998, arnold_finite_2006, buffa_isogeometric_2010}.
Even seemingly benign problems such as the computation of a vector potential $A$ such that $\curl A = B$ given $B$ can pose difficulties after discretization~\citep[Chapter 5]{boffi_mixed_2013}.
The same is true for handling the harmonic fields present in non-contractible domains that span the kernel of the vector Laplace operator.
These fields are extremely important as the magnetic field in stellarators is often harmonic to $> 90 \%$: for the relaxed finite-$\beta$ NCSX field of Section~\ref{sec:relaxation_numerics}, is $96.4\%$ harmonic.

Preserving this structure requires a suitable choice of discrete vector spaces.
While it is impossible to preserve all features of the continuous problem in a finite-dimensional approximation, it is possible to use approximation spaces where some identities (e.g., $\dvg B = 0$) hold to machine precision (Section~\ref{subsec:discretization}).

Structure-preserving FE methods have been applied to model MHD phenomena in a number of works~\cite{hu_stable_2017, gawlik_structure-preserving_2021, hu_helicity-conservative_2021, gawlik_finite_2022, holderied_magneto-hydrodynamic_2022, carlier_variational_2025} and also in magnetic relaxation codes~\cite{bressan_2023_relaxation, he_topology-preserving_2025, mao2026structure}.
However, to our knowledge, there are no works that tackle the relaxation problem in the practically relevant toroidal geometry using structure-preserving finite elements. 
While we focus on the case of solid toroidal geometry, there are no structural obstacles to transfer this work to arbitrarily shaped domains, including ones with multiple cavities and tunnels. 

All operations are performed matrix-free.

\paragraph{Code framework}
We built the codebase for this work on the JAX framework~\cite{frostig_compiling_2019, noauthor_jax_nodate}, a tracing just-in-time compiler for generating high-performance accelerator code from pure Python and NumPy programs.
The benefit of this is threefold.

(1) The compiled code is highly performant on CPUs, GPUs, and TPUs.
It allows us to run MRX on the latest computational hardware available by leveraging the abstractions of the compilers in the OpenXLA framework.
We are able to generate magnetic equilibria on relevant, highly shaped geometries in minutes, see Section~\ref{subsec:newtons_method}.

(2) JAX supports automatic differentiation, i.e., the computation of gradients of functions with respect to their input arguments by tracing the primitive operations encountered throughout function evaluation.
This allows, for example, the computation of derivatives of equilibrium fields with respect to geometry inputs without resorting to costly finite difference approximations.
Section~\ref{subsec:shape_optimization} uses the derivative of the vacuum field with respect to the map to optimize a stellarator for quasi-axisymmetry.
This is a very attractive feature for PDE-constrained optimization and other applications, e.g., incorporation of MRX into stellarator optimization routines or uncertainty quantification tasks.

(3) The MRX code is open-source and highly accessible as all dependencies can be installed via the \texttt{pip} package manager after cloning the code repository.\footnote{\texttt{github.com/ToBlick/mrx}}
Interfaces for initializing equilibria from GVEC and VMEC are provided.

\subsection{Running MRX}
\label{subsec:running}

A standard MRX run follows a relax--seed--relax schedule with three steps (Section~\ref{sec:reconnection}).
First, a nested-surface equilibrium, for example a VMEC output file, which also provides the geometry, is relaxed to the floor of the force residual.
Second, the energy criterion of Section~\ref{subsec:seeding} seeds every rational surface in the range of the rotational transform at its optimal amplitude~\eqref{eq:seed_amplitude}.
Third, the seeded state is relaxed to the floor again.
Every run writes a record and a checkpoint per chunk, from which it can be restarted or analyzed.
The required user input is an equilibrium file, other parameters that can be set and their default values are shown in Table~\ref{tab:hyperparameters}.

\subsection{Outline}
\label{sec:outline}

In this document, we will introduce the theoretical concepts one-by-one, followed by numerical experiments to demonstrate and validate their impact on the code.
The following list covers the remaining sections of this work.

\begin{itemize}
    \item[\ref{sec:framework}] Notation and formalization of the problem statement.
    \item[\ref{sec:vacuum_numerics}] Numerical verification on vacuum equilibria and shape optimization.
    \item[\ref{sec:magnetic_relaxation}] Magnetic relaxation and its structure-preserving discretization.
    \item[\ref{sec:relaxation_numerics}] Numerical relaxation examples.
    \item[\ref{sec:reconnection}] Reconnection and targeted topology changes.
    \item[\ref{sec:resistive}] Resistive relaxation.
    \item[\ref{sec:discussion}] Summary and pointers to future work.
\end{itemize}
The notation is summarized in~\ref{sec:notation} and all run parameters and code settings are given in~\ref{sec:hyperparameters}.

\section{Mathematical and computational framework}
\label{sec:framework}

At the center of MRX stands the de~Rham complex of function spaces and its faithful discrete approximation.
This approximation yields important tools such as the discrete Hodge--Helmholtz decomposition, the crucial identities $\curl \grad = \dvg \curl = 0$, discretely exact, and the identification of harmonic fields.
The latter constitute MHD equilibria with $J = p = 0$, called vacuum solutions.
A thorough treatment of the mathematical theory is given in~\citep[Section 2.2]{arnold_finite_2006} or~\citep[Chapter 7.5]{abraham_manifolds_1988}.

We will discuss the properties of the de~Rham complex in Section~\ref{subsec:derham_complex}, the encoding of geometry in Section~\ref{subsec:domain_mapping}, and the crucial Hodge--Helmholtz decomposition in Section~\ref{subsec:hodge_decomposition}.
In Sections~\ref{subsec:discretization} and~\ref{subsec:implementation}, we present the discretization strategy and some practical implementation choices.

\subsection{The de~Rham complex}
\label{subsec:derham_complex}

Denote the boundary of $\Domain$ by $\partial \Domain$ and the outward unit normal at $x \in \partial \Domain$ by $n(x)$. 
The Hilbert space of square-integrable, vector-valued functions on this domain is denoted $L^2(\Domain; \mathbb{R}^3)$.
We equip all of the following function spaces with the $L^2$ inner product and its induced norm. 
We also introduce notation for a volume average over $\Domain$, with $|\Domain|$ denoting the volume of the domain.
For $u, v \in L^2(\Domain; \mathbb{R}^3)$, $f \in L^1(\Domain; \bR)$,
\begin{align}
    \label{eq:inner_product}
    \ip{u}{v} = \int_\Domain u \cdot v \, \rd x, \qquad
    \norm{u} = \sqrt{ \ip{u}{u} }, \qquad
    \vavg{f} = \frac{1}{|\Domain|} \int_\Domain f \, \rd x . 
\end{align}
The point-wise norm $|u|^2(x)$ is denoted with single vertical lines. 
We will also use the notation $\langle f \rangle_{\theta, \zeta}(r)$ to denote an average over surfaces of constant $r$.
We will, from now on, write $L^2(\Domain)$ for short for both vectorial and scalar spaces and omit $\Domain$ if the domain is clear from context.

\paragraph{Function spaces}
Computational electrodynamics is set in scalar- and vector-valued Sobolev spaces.

\begin{definition}[Function spaces]
    The spaces of vector fields on $\Domain$ with weak grad, curl and divergence are defined as:
    \begin{align}
        \Hone(\Domain)  = \{ f \in L^2 : \grad f \in L^2 \},  \;
        \Hcurl(\Domain) = \{ u \in L^2 : \curl u \in L^2 \}, \;
        \Hdiv(\Domain)  = \{ v \in L^2 : \dvg v \in L^2 \}.
    \end{align}
    The corresponding spaces of vector fields with homogeneous Dirichlet boundary conditions are defined as:
    \begin{align}
        \Hone_0(\Domain)  = \{ f \in \Hone : f|_{\partial \Domain} = 0 \},  \;\;
        \Hcurl_0(\Domain) = \{ u \in \Hcurl : u \times n|_{\partial \Domain} = 0 \},  \;\;
        \Hdiv_0(\Domain)  = \{ v \in \Hdiv : v \cdot n|_{\partial \Domain} = 0 \}.
    \end{align}
    The expressions $|_{\partial \Domain} = 0$ should be understood in the sense of a trace operator, see~\citep[Lemma 2.1.1]{boffi_mixed_2013}.
\end{definition}

The 3D de~Rham complex
\begin{align}
    \label{eq:derham_complex}
    0 \xrightarrow[]{} \Hone \xrightarrow[]{\grad} \Hcurl \xrightarrow[]{\curl}\Hdiv \xrightarrow[]{\dvg} L^2 \xrightarrow[]{} 0
\end{align}
is a special case of a closed Hilbert complex
\begin{align}
    \label{eq:derham_complex_general}
    0 \xrightarrow[]{} V^0 \xrightarrow[]{\rd^0} V^1 \xrightarrow[]{\rd^1} V^2 \xrightarrow[]{\rd^2} V^3 \xrightarrow[]{} 0.
\end{align}
It satisfies the following properties for all $k$: (i) $\range \rd^k \subset V^{k+1}$, (ii) $\rd^{k+1} \circ \rd^{k} = 0$, and (iii) the range of $\rd^k$ is closed in $V^{k+1}$~\citep[Section 3.1.3]{arnold_finite_2010}.

For the de~Rham complex, (ii) describes the central vector calculus identities $\curl \grad = \dvg \curl = 0$.
We anticipate already that (iii) is crucial to retain during discretization in order to arrive at well-posed discrete saddle point problems~\citep[Theorem 4.15]{boffi_mixed_2013}. Elements of the space $V^k$ are called $k$-forms.

\paragraph{Dual complex}
Each $\rd^k$ has a Hilbert adjoint, the \emph{co-differential} $\rd^*_{k+1} \colon V^{k+1}_* \to L^2(\Domain)$, defined by
\begin{align}
    \label{eq:codifferential}
    \langle \rd^k u, v \rangle_{V^{k+1}} = \langle u, \rd^*_{k+1} v \rangle_{V^{k}} \quad \text{for all} \quad u \in V^k,\ v \in V^{k+1}_*,
\end{align}
where the domain $V^{k+1}_*$ of $\rd^*_{k+1}$ is a dense subspace of $L^2(\Domain)$ that carries the boundary conditions. For the de~Rham complex~\eqref{eq:derham_complex}, $\rd^*_1 = -\dvg$ on $\Hdiv_0$, $\rd^*_2 = \curl$ on $\Hcurl_0$, and $\rd^*_3 = -\grad$ on $H^1_0$.
The co-differentials form the dual complex
\begin{align}
    \label{eq:dual_complex}
    0 \xleftarrow[]{} L^2 \xleftarrow[]{\rd^*_1} V^1_* \xleftarrow[]{\rd^*_2} V^2_* \xleftarrow[]{\rd^*_3} V^3_* \xleftarrow[]{} 0,
\end{align}
which is again a closed Hilbert complex: $\rd^*_{k} \circ \rd^*_{k+1} = 0$ and $\range \rd^*_{k+1} = (\ker \rd^k)^\perp$ is closed.

\subsection{Domain and mapping}
\label{subsec:domain_mapping}
The domain of interest in this work is $\Domain \subset \bR^3$, the volume of space that encloses the plasma.
This domain is typically shaped in a moderately complicated manner as in Figure~\ref{fig:ncsx_sections}.
We will define approximation spaces in the logical domain, and therefore need to define appropriate operations to translate between the logical and physical domains. 

\paragraph{Coordinate systems}
The approach we take in this work is to describe $\Domain$ as the image of the logical domain $\hat \Domain = [0, 1]^3$ after application of the mapping $\Mapping: \hat \Domain \to \bR^3$.
We will denote the coordinates on the logical domain by $(r, \theta, \zeta) = \hat x \in \hat \Domain$.
The map $\Mapping$ is assumed to be a $C^1$ diffeomorphism everywhere except at $r = 0$.
The Jacobian matrix of the mapping $\Mapping$ is defined as $D\Mapping$ with entries $(D\Mapping(\hat x))_{ij} = {\partial \Mapping_i(\hat x)} / {\partial {\hat x_j}}$, $1 \leq i,j \leq 3$.
The metric is defined as $g = D\Mapping^T D\Mapping$, with $\sqrt g = \det D\Mapping$, which is positive for all maps in this work.

We denote $\Mapping(\hat x) = x = (X, Y, Z)$, these are points in physical space.
We also introduce a cylindrical coordinate system $(R, \phi, Z) = (\sqrt{X^2 + Y^2}, \, \arctan_2(Y, X), \, Z)$. 

\paragraph{Pull-back and push-forward}
Points in the logical domain $\hat x \in \hat \Domain$ and functions defined in logical coordinates such as $\hat f: \hat \Domain \to \bR$ carry a hat.
The hat decorator is omitted on coordinate names $r, \theta, \zeta$, the map $\Mapping$ with its Jacobian $D\Mapping$, and the basis functions $\Lambda^k_i$ introduced in Section~\ref{subsec:discretization}. 
Through the following push-forward and pull-back operations, we can associate to every function $\hat f$ on the logical domain a function $f: \Domain \to \bR$ on the physical domain.

\begin{definition}[Push-forward]
\label{def:pushforward}
    For functions $\hat f: \hat \Domain \to \bR$ we define two push-forward operations to the physical domain as follows: For all $x \in \Domain$,
    \begin{align}
    \label{eq:pushforward}
        f(x) = (\Mapping^0_* \hat f)(x) = \hat f(\Mapping^{-1}(x)), \quad  
        (\Mapping^3_* \hat f)(x) = \frac{\hat f(\Mapping^{-1}(x))}{ \det D\Mapping(\Mapping^{-1}(x))}.   
    \end{align}
    We furthermore define two push-forward operators on vector fields $\hat u, \hat v: \hat \Domain \to \bR^3$ as:
    \begin{align}
    \label{eq:pushforward_eb}
        u(x) = (\Mapping^1_* \hat u)(x) = \left( D\Mapping(\Mapping^{-1}(x)) \right)^{-T} \hat u(\Mapping^{-1}(x)),   \quad
        v(x) = (\Mapping^2_* \hat v)(x) = \frac{ D\Mapping(\Mapping^{-1}(x)) \hat v(\Mapping^{-1}(x))}{ \det D\Mapping(\Mapping^{-1}(x)) }.   
    \end{align}
    The corresponding pull-back operations are obtained when replacing the mapping $\Mapping$ with its inverse $\Mapping^{-1}$.
\end{definition}
\begin{lemma}[Properties of push-forward and pull-back] 
    \label{lem:pushforward_properties}
    The operations grad, div and curl are \emph{natural} with respect to the push-forward under $C^1$-diffeomorphisms, that is, the push-forward of the gradient/curl/divergence is the gradient/curl/divergence of the push-forward.

    Furthermore, when $\hat f \in \Hone_0(\hat \Domain)$, $\hat u \in \Hcurl_0(\hat \Domain)$, and $\hat v \in \Hdiv_0(\hat \Domain)$, 
    then $\Mapping^0_* \hat f \in \Hone_0(\Domain)$, $\Mapping^1_* \hat u \in \Hcurl_0(\Domain)$, and $\Mapping^2_* \hat v \in \Hdiv_0(\Domain)$.
\end{lemma}
The first statement is~\cite[Theorem 6.4.4]{abraham_manifolds_1988}. The proof for the second is given in~\ref{sec:proofs}. Nowhere do we assume that $\Mapping$ is an orthogonal mapping.

\paragraph{Covariant and contravariant components}
The columns $\partial_r \Mapping$, $\partial_\theta \Mapping$, $\partial_\zeta \Mapping$ of $D\Mapping$ are tangent to the coordinate lines. The rows of $D\Mapping^{-1}$ are the gradients $\grad \, \{r, \theta, \zeta\}$ and $\grad \hat x_i \cdot \partial_j \Mapping = \delta_{ij}$.
In components, $g_{ij} = \partial_i \Mapping \cdot \partial_j \Mapping$, and its inverse $g_{ij}^{-1} = g^{ij} = \grad \hat x_i \cdot \grad \hat x_j$.
A vector field $u$ has the contravariant components $u^i = u \cdot \grad \hat x_i$ and the covariant components $u_i = u \cdot \partial_i \Mapping$.
The vector pull-backs of Definition~\ref{def:pushforward} hence select one or the other for $u = \Mapping^1_* \hat u$ and $v = \Mapping^2_* \hat v$:
\begin{align}
    \label{eq:covariant_contravariant}
    \hat u = D\Mapping^T \, u \circ \Mapping = (u_r, u_\theta, u_\zeta), \qquad
    \hat v = \sqrt g \, D\Mapping^{-1} \, v \circ \Mapping = \sqrt g \, (v^r, v^\theta, v^\zeta).
\end{align}
As a result, $B \cdot \grad r = B^r = \hat B^r / \sqrt g$ vanishes exactly where $B$ is tangent to the logical surfaces $r = \mathrm{const}$.
Their unit normal is $\grad r / |\grad r|$, with $|\grad r|^2 = g^{rr}$. It is parallel to the tangent $\partial_r \Mapping$ only where $g_{r\theta} = g_{r\zeta} = 0$, which the maps of this work do not satisfy.

\paragraph{Stellarator geometry}
All calculations in this work take place in solid toroidal geometry.
All mappings share a polar singularity, i.e., $\Mapping (r, \theta, \zeta) {\to} x_\text{axis}(\zeta)$ for all $\theta$ as $r \to 0$. As the limit does not depend on $\theta$, the Jacobian of the map is not invertible at $r=0$. The handling of this singularity is discussed in Section~\ref{subsec:discretization}.

Given functions $R: \hat \Domain \to \bR_{ > 0}$ and $Z: \hat \Domain \to \bR$ (the cylindrical coordinates of the image point), we obtain a basic stellarator map for a device with $\nfp{}$ field periods as follows:
\begin{align}
    \label{eq:stellarator_map}
    \Mapping:
    \begin{bmatrix} r \\ \theta \\ \zeta \end{bmatrix} \mapsto
    \begin{bmatrix}
        R(r, \theta, \zeta) \cos(2\pi \zeta / \nfp{}) \\
        - R(r, \theta, \zeta) \sin(2\pi \zeta / \nfp{}) \\
        Z(r, \theta, \zeta)
    \end{bmatrix}.
\end{align}
The image of $\hat x$ under this map covers one period. Additional stellarator symmetry, i.e., invariance of the domain under $(r, \theta, \zeta) \mapsto (r, -\theta, -\zeta)$ or equivalently $(R, \phi, Z) \mapsto (R, -\phi, -Z)$, is also exploited in MRX and discussed in~\ref{sec:stellarator_symmetry}.

\paragraph{Interface with nested flux surface solvers}
The functions $R, Z : \hat \Domain \to \bR$ are typically obtained from nested flux surface solvers such as VMEC, GVEC, or DESC.
In MRX, the map is represented as a $C^1$ spline map, explained in more detail in Section~\ref{subsec:discretization}.

We emphasize here that we do not assume that $\Mapping$ describes a flux coordinate system.
All that is required is that $\Mapping(\hat \Domain) = \Domain$.
In general, the magnetic axis and the axis of the coordinate system, $\Mapping(\hat x) |_{r = 0}$, need not coincide. 

For every equilibrium solve, the mapping $\Mapping$ is held constant.
In contrast, for codes that operate under a nested flux surface assumption, the mapping is itself the unknown of the optimization and changes during solving for equilibrium.
Anticipating future use of this code for stellarator optimization (in which the boundary shape and therefore the mapping changes each iteration), MRX is built with the ability to modify the geometry mid-run without the need for re-compilation.

\begin{figure}[thb]
\centering
\includegraphics[width=0.66\linewidth]{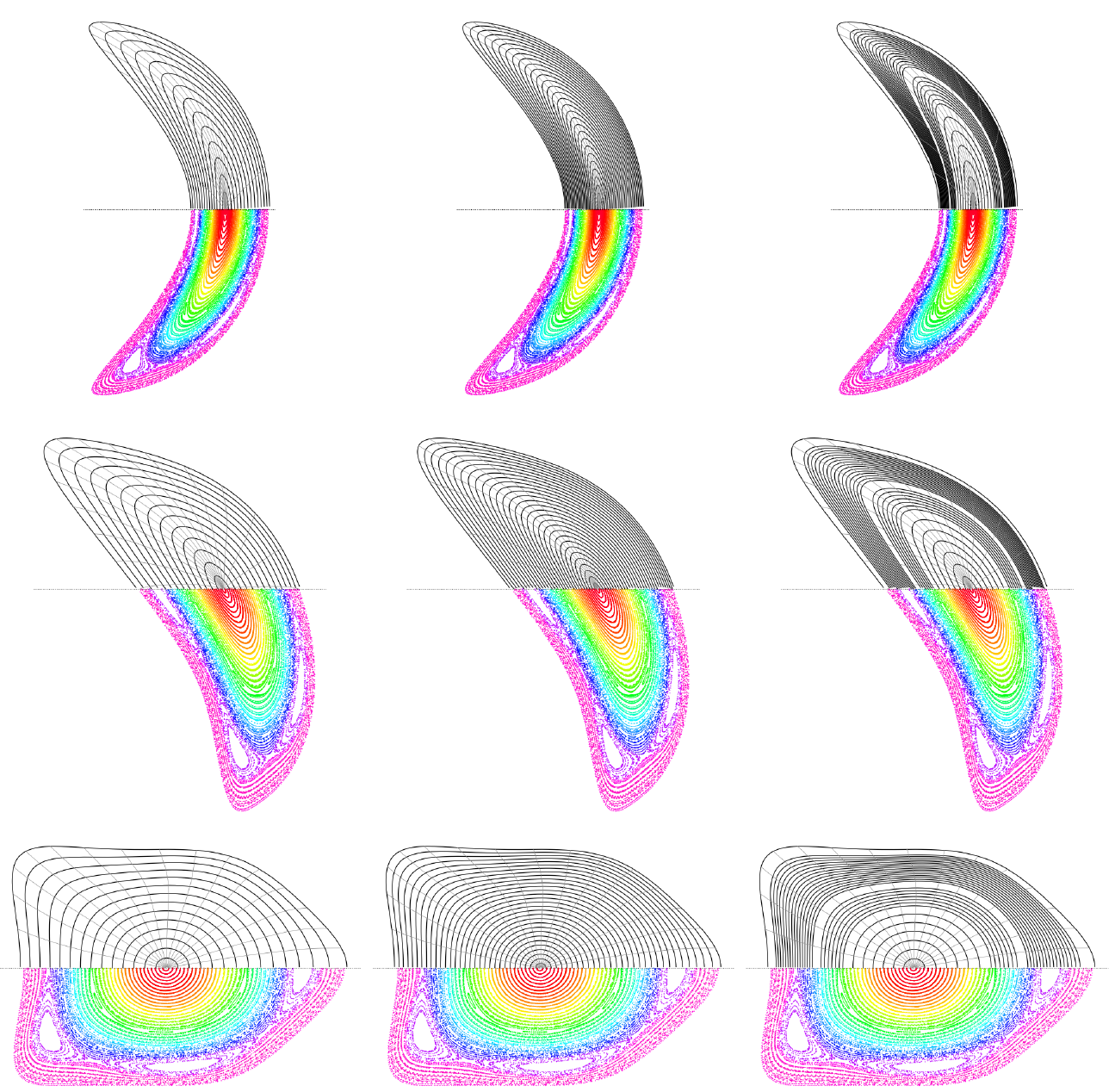}
\caption{Different meshes that can be used in MRX at $\zeta = $ const. slices.
For illustration, we also show Poincaré sections of a reconnected field on these meshes, see Section~\ref{sec:reconnection}.
}
  \label{fig:mesh_refinement}
\end{figure}

We show some of the grids employed in MRX in the illustrative Figure~\ref{fig:mesh_refinement}.

\subsection{Hodge decomposition}
\label{subsec:hodge_decomposition}
The $L^2$-orthogonal Hodge--Helmholtz decomposition expresses a vector field as the sum of a gradient field, a divergence-free field, and a harmonic component.
In fact, vector fields in $L^2(\Domain)$ admit two $L^2$-orthogonal Hodge--Helmholtz decompositions~\citep[Section 7.5.5]{abraham_manifolds_1988}, corresponding to the decomposition of $V^1$ and $V^2$ introduced in Section~\ref{subsec:derham_complex}:
\begin{align}
    \label{eq:helmholtz_decomposition}
    L^2(\Domain) = \grad H^1(\Domain) \overset{\perp}{\oplus} \curl \Hcurl_0(\Domain) \overset{\perp}{\oplus} \mathfrak{H}^1(\Domain) = \curl \Hcurl(\Domain) \overset{\perp}{\oplus} \grad H^1_0(\Domain) \overset{\perp}{\oplus} \mathfrak{H}^2(\Domain).
\end{align}
The orthogonality relations here are understood with respect to the $L^2$ inner product.
In terms of differential and co-differential, the Hodge decomposition reads
\begin{align}
    \label{eq:hodge_decomposition}
    L^2(\Domain) = \range \rd^{k-1} \oplus \range \rd^*_{k+1} \oplus \mathfrak{H}^k, \quad \mathfrak{H}^k = \ker \rd^k \cap \ker \rd^*_k.
\end{align}
The two decompositions of vector fields in $L^2(\Domain)$ in particular correspond to 1-forms and 2-forms.

\begin{definition}[Harmonic forms]
    \label{def:harmonic_forms}
    The spaces $\mathfrak{H}^k(\Domain)$, $k \in \{ 0, 1, 2, 3 \}$, are called harmonic spaces, elements $\mathfrak h_k$ of $\mathfrak{H}^k(\Domain)$ are harmonic forms and the $L^2$-projection onto $\mathfrak H^k(\Domain)$ is denoted by $\Pi^{{\mathfrak H}^k}$.
    The harmonic spaces of the de~Rham complex with homogeneous boundary conditions are denoted by $\mathfrak H^k_0(\Domain)$.
\end{definition}

\begin{lemma}[{\citep[Corollary 7.5.4]{abraham_manifolds_1988}}]
    \label{lem:harmonic_spaces}
    $\grad \mathfrak{h}_0 = 0$ in $\Domain$: the elements of $\mathfrak H^0(\Domain)$ are the functions constant on every connected component of $\Domain$.
    The elements of $\mathfrak{H}^1(\Domain)$ and $\mathfrak{H}^2(\Domain)$ are vector fields with $\curl \mathfrak h = \dvg \mathfrak h = 0$, with zero normal trace in $\mathfrak{H}^1(\Domain)$ and zero tangential trace in $\mathfrak{H}^2(\Domain)$, and $\mathfrak{H}^3(\Domain) = \{ 0 \}$.
    Moreover $\dim {\mathfrak H}^k(\Domain) = \dim {\mathfrak H}^{3-k}_0(\Domain)$, the two spaces being isomorphic.
\end{lemma}

All harmonic forms lie in the kernel of the Hodge--Laplace operator with corresponding boundary conditions.

\begin{definition}[Hodge--Laplace operator]
    \label{def:hodge_laplacian}
    The Hodge--Laplacian $\mathcal L^k$ on $V^k$ is
    \begin{align}
        \label{eq:hodge_laplacian}
        \mathcal L^k = \rd^*_{k+1} \rd^k + \rd^{k-1} \rd^*_k,
    \end{align}
    with domain $\{ u \in V^k \cap V^k_* \,:\, \rd^k u \in V^{k+1}_*,\ \rd^*_k u \in V^{k-1} \}$.
    
    For the de~Rham complex~\eqref{eq:derham_complex}, $\mathcal L^0$ and $\mathcal L^3$ are the scalar Laplacian $-\dvg \grad$, while $\mathcal L^1$ and $\mathcal L^2$ are the vector Laplacians $\curl \curl - \grad \dvg$.
    The boundary conditions differ due to the domains of the co-differentials.
\end{definition}

By construction $\mathcal L^k$ is self-adjoint and positive semi-definite, $\ip{\mathcal L^k u}{u} = \norm{\rd^k u}^2 + \norm{\rd^*_k u}^2$, its kernel is $\mathfrak H^k$ by definition, and it is invertible on $(\mathfrak{H}^k)^\perp$ by the Hodge decomposition~\eqref{eq:hodge_decomposition}.

As a consequence of de~Rham's theorem~\citep[Section 7.5]{abraham_manifolds_1988}, the dimensions of the harmonic spaces are the Betti numbers of the domain.
That is, $\dim \mathfrak{H}^0(\Domain)$ is the number of connected components of $\Domain$, $\dim \mathfrak{H}^1(\Domain)$ is the number of tunnels (or handles) in $\Domain$, and $\dim \mathfrak{H}^2(\Domain)$ is the number of cavities.

For a solid torus, the form of all domains in this work, $\dim \mathfrak{H}^0(\Domain) = \dim \mathfrak{H}^1(\Domain) = 1$ and $\dim \mathfrak{H}^2(\Domain) = 0$.
In the plasma physics community, elements of $\mathfrak{H}^2_0$ are usually referred to as vacuum fields. They describe magnetic fields with zero normal boundary trace that induce no current.
For an axisymmetric toroidal domain, $\mathfrak h_1 \propto R^{-1} e_\phi$.

\subsection{Discretization}
\label{subsec:discretization}

MRX is built on B-spline basis functions defined on the logical domain $\hat \Domain$, mapped to the physical domain $\Domain$ via $\Mapping$.   

\paragraph{Finite element exterior calculus}
The finite element framework we will follow, known as Finite Element Exterior Calculus (FEEC), crucially retains important features of the de~Rham complex introduced in Section~\ref{sec:framework} after discretization such as the identities $\curl \grad = \dvg \curl = 0$.
These relations hold to machine precision at any discretization resolution, they are topological (incidence) operations.
Furthermore, this approach comes with certain guarantees about numerical stability, in particular, the inf-sup stability criterion for various saddle point problems that arise is satisfied~\citep[Theorem 3.8]{arnold_finite_2010} and the approximation spaces can be decomposed via a discrete Hodge--Helmholtz decomposition.

Various excellent references exist on the theory behind FEEC and its connection to cohomology theory~\citep{arnold_finite_2006, arnold_finite_2010}.
The exposition we provide here for the sake of self-containedness will be presented on the level of the matrix-vector equations that arise after discretization.

The $k$-forms ($k \in \{0, 1, 2, 3 \}$) are represented as piece-wise polynomial functions valued in $\bR$ (for $k \in \{0, 3\}$) and $\bR^3$ (for $k \in \{1,2\})$.
We denote these discrete function spaces by $V^k_h, k \in \{0, 1, 2, 3\}$. 

These spaces are constructed such that the images of the (continuous) $\grad$, $\curl$ and $\dvg$ operators are contained in the subsequent space: $f_h \in V^0_h$ implies $\grad f_h \in V^1_h$, $A_h \in V^1_h$ implies $\curl A_h \in V^2_h$, and $B_h \in V^2_h$ implies $\dvg B_h \in V^3_h$.

\begin{example}[Vector spaces that do not form a complex]
    This property does not hold for every FE method.
    Consider, for example the space $V^0_h$ consisting of piece-wise linear functions and $V^{\mathrm{vec}}_h = \mesh{ V^0_h}{ V^0_h}{ V^0_h}$, a vector-valued approximation space where every component is an element of $V^0_h$.
    For $f \in V^0_h$, $\grad f$ is a discontinuous function, hence $\grad f \not \in V^{\mathrm{vec}}_h$.
\end{example}

The spaces with (essential) homogeneous boundary conditions are denoted $V^k_0 \; \forall k \in \{0,1,2,3\}$.
The basis functions that span $V^k$ are denoted $\{ \Lambda^k_i \}_i$.
We denote the $L^2$-projection operator to the space $V^k$ by $\Pi^k : \Pi^k u = u_h = \sum_i \mathtt{u}_i \, \Lambda^k_i$ {where} $\sum_j \mathtt{u}_j \ip{\Lambda_i^k}{\Lambda_j^k} = \ip{u}{\Lambda^k_i} \; \forall i$.   

\paragraph{Spline finite elements}
In this work, the spaces $\{ V^k \}_{k=0}^3$ are spanned by tensor products of $B$-splines.
$B$-splines are defined by their degree $p$ and a vector of knot points.
We use the same notation for spline degree and pressure.
Splines are piece-wise polynomial functions, $C^\infty$ between knot points and $C^{p-1}$ at simple knots.
In particular, given degrees $(p_r, p_\theta, p_\zeta) \in \mathbb N^3$, resolutions $(n_r, n_\theta, n_\zeta) \in \mathbb N^3$ such that $n_c> p_c$, $c \in \{r, \theta, \zeta\}$, let $\{ N^0_{c, i} \}_{i=1}^{n_c}$ denote one-dimensional B-splines of degree $p_c$.
The basis functions $\{ \Lambda^0_i \}_{i=1}^{n_r n_\theta n_\zeta}$ are then defined via tensor products and a flattening of their indices via $\mathrm{vec}: \mathbb N^3 \to \mathbb N : (j,k,l) \mapsto i$.
Evaluated at the point $\hat x = (r, \theta, \zeta)$, this yields $\Lambda^0_i(\hat x) |_{i = \mathrm{vec}(j,k,l)} = N^0_{r, j}(r) \, N^0_{\theta, k}(\theta) \, N^0_{\zeta, l}(\zeta)$.   
When we denote the space spanned by one-dimensional B-splines of degree $p$ by $S^p$,
\begin{align}
    V^0_h = \bigotimes_{c \in \{r, \theta, \zeta\}} S^{p_c}, \quad
    V^1_h = \begin{bmatrix} 
        S^{p_r-1} \otimes S^{p_\theta} \otimes S^{p_\zeta} \\ 
        S^{p_r} \otimes S^{p_\theta-1} \otimes S^{p_\zeta} \\
        S^{p_r} \otimes S^{p_\theta} \otimes S^{p_\zeta-1}
    \end{bmatrix}, \quad
    V^2_h = \begin{bmatrix} 
        S^{p_r} \otimes   S^{p_\theta-1} \otimes S^{p_\zeta-1} \\ 
        S^{p_r-1} \otimes S^{p_\theta} \otimes   S^{p_\zeta-1} \\
        S^{p_r-1} \otimes S^{p_\theta-1} \otimes S^{p_\zeta}
    \end{bmatrix}, \quad
    V^3_h = \bigotimes_{c \in \{r, \theta, \zeta\}} S^{p_c-1}.
\end{align}

Here $n_c$ is the number of splines of $V^0$ in direction $c$: in the periodic angles this is also the number of cells, while the clamped radial direction has $n_r - p_r$ cells. Throughout this work, $h_r = 1 / (n_r - p_r)$, $h_\theta = 1 / n_\theta$ and $h_\zeta = 1 / n_\zeta$ denote the knot spacings, the widths of the cells of the logical grid, and $\vavg{g_{rr}}^{1/2} h_r$ the physical length of a radial cell edge.

For more detailed references regarding spline finite elements in computational electromagnetics, we refer to~\citep{buffa_isogeometric_2010, campos-pinto_broken-feec_2024}.

\begin{remark}
    In the interest of avoiding notational clutter, we will drop the subscript $_h$ in the following.
\end{remark}

\paragraph{Polar splines}
The problems we consider occur naturally in toroidal geometry.
As a result, the mapping $\Mapping$ is singular at $r = 0$.
We modify the tensor-product structure of the spline basis functions because regularity requirements on the axis imply linear dependence of some of the splines with support at $r = 0$: The pull-back of any (physical) 0-form $f$ will lead to $\hat f: \partial_\theta \hat f |_{r = 0} = 0$.
The requirement that all elements of $V^0$ satisfy this relation leads to a number of linear constraints, analogously for $k>0$.
A detailed reference for these constructions is~\citep[Chapter 5]{holderied_struphy_2022}.

The number of splines affected depends on the desired regularity around the axis.
Polar splines are introduced in~\citep{toshniwal_multi-degree_2017, toshniwal_isogeometric_2021, patrizi_isogeometric_2025} and used for MHD simulations in~\citep{holderied_magneto-hydrodynamic_2022}.
In the interest of brevity, we refer to these references for further details.

\begin{example}[Continuously differentiable polar splines]
    For elements of $V^0$ and $V^0_{0}$, $C^0$ continuity demands that in every plane of constant $\zeta$, the coefficients of all $n_\theta$ splines that are clamped at $r = 0$, the inner ring of basis functions per poloidal domain, are equal.
    Hence, $\dim V^{\mathrm{polar},0} = ((n_r - 1) n_\theta + 1) n_\zeta$.
    To obtain $C^1$ regularity, the two inner rings of basis functions are affected, a total of $2 n_\theta n_\zeta$ functions.
    In this case, three $C^1$-compatible basis functions per poloidal plane are added to replace them, and $\dim V^{\mathrm{polar},0} = ((n_r - 2) n_\theta + 3) n_\zeta$.
\end{example}
In order to use polar splines in toroidal domains, it is sufficient that there exists a smooth diffeomorphism between $\Domain$ and a circular axisymmetric torus.    

This is the case for every geometry produced by a nested flux surface solver such as VMEC or GVEC.
These solvers represent the domain as a family of poloidal cross-sections swept around the torus, each the smooth image of the unit disc, and their nested-surface ansatz requires $\det D\Mapping$ not to change sign for $r > 0$, so $\Mapping$ is a local diffeomorphism away from the axis.
It is injective because distinct flux surfaces do not intersect and the plasma does not intersect itself.

All maps used in this work are represented by the same splines that are used to span $V^0$.

\paragraph{Maps and initial fields from nested flux surface solvers}
\label{par:maps_initial_fields}
Both GVEC and VMEC supply $(r, \theta, \zeta) \mapsto (R, Z)(r, \theta, \zeta)$ as a discretized function: a Fourier expansion in $\theta, \zeta$ whose coefficients are B-splines in $r$ (GVEC) or values on a radial grid in the normalized toroidal flux $s$ (VMEC), which MRX interpolates by splines in $r = \sqrt{s}$.

The spline coefficients of the MRX map are the $L^2$ projection of this series onto the map's tensor-product spline space, a cheap operation as it decouples by direction.
The map $\Mapping$ is then assembled as in~\eqref{eq:stellarator_map}.

The initial magnetic field is obtained as the discrete curl of a vector potential, $B_0 = \curl A_0$.
The nested-surface solver provides the toroidal and poloidal fluxes $r \mapsto \Psi_{\mathrm t}(r), \Psi_{\mathrm p}(r)$, and the angle stream function $(r, \theta, \zeta) \mapsto \lambdaup(r, \theta, \zeta)$. From these, we obtain the closed-form reference 1-form
\begin{align}
    \label{eq:vmec_vector_potential}
    \hat A_{0}(r, \theta, \zeta) 
    = 2\pi\left( -\frac{\lambdaup \, \partial_r \Psi_{\mathrm t}(r)}{2\pi},  \Psi_{\mathrm t}(r), -  \frac{\Psi_{\mathrm p}(r)}{\nfp{}} \right),
\end{align}
in the logical frame. Its curl is the equilibrium field in the Clebsch form of the nested-surface solver.

The discrete field is $B_0 = \curl A_0$, so $B_0$ is divergence-free to machine precision.
The potential $A_0 \in V^1$ is fixed by prescribing the line integrals of $\hat A_{0}$ over one set of cell-sized intervals (Greville histopolation).
The toroidal flux is the circulation of $A_0$ along the wall, a sum of these prescribed line integrals.

\subsection{Implementation}
\label{subsec:implementation}
We now give a brief overview of some implementation choices.
\paragraph{Matrices}
By rescaling the splines of lower order in $V^k$, $k>0$, it is possible to ensure that the degree-of-freedom to degree-of-freedom representations of grad, curl, and div operators on the tensor-product spaces are incidence matrices, i.e., they only hold entries $\pm 1$, independent of the geometry. The polar extraction adds fixed, geometry-independent corrections to grad and curl near the axis.
We will denote these matrices by $\mathbb D_k$ ($k=0$: grad, $k=1$: curl, $k=2$: div).
It holds that $\mathbb D_{k+1} \mathbb D_k = 0$, independent of geometry and mesh size.

Furthermore, we denote the $k$-form mass matrix by $\mathbb M_k$ with entries $(\mathbb M_k)_{ij} = \ip{\Lambda^k_i}{\Lambda^k_j}$.
The mass matrices depend on the metric through the pull-back operation, as they must.
The integrals over $\Domain$ in this definition are pulled back to integrals over $\hat \Domain$ and evaluated using Gauss--Legendre quadrature in every knot interval (cell) of the spline spaces.
The quadrature grid thus mimics the tensor-structure of the basis functions.

All derivative operators in this work can be expressed as combinations of these basic matrices.
For example, stiffness-like matrices are given by $\mathbb D_k^T \mathbb M_{k+1} \mathbb D_k$.

For $k=0$, this represents the bilinear form $\ip{\grad \Lambda^0_i}{\grad \Lambda^0_j}$, for $k=1$ it is a curl--curl operator, and for $k=2$, it is the bilinear form corresponding to $-$grad--div.

A weak derivative operator that maps $k$-form degrees of freedom to $k-1$-form degrees of freedom is expressed by $(-1)^k \mathbb M_{k-1}^{-1} \mathbb D_{k-1}^T \mathbb M_{k}$. These are denoted with a tilde, $\widetilde{\grad}$, $\widetilde{\curl}$, $\widetilde{\dvg}$ for the three weak derivatives.
It is called weak as the divergence/curl/gradient only holds in a weak $L^2$ sense, while the incidence relations (gradient of a 0-form is a 1-form, curl of a 1-form is a 2-form, divergence of a 2-form is a 3-form) hold point-wise.   

The full Hodge--Laplace operators combine weak and strong derivatives, reflecting the differential/co-differential formulation of the continuous Hodge Laplacian $\mathcal L^k = \rd^*_{k+1} \rd^k + \rd^{k-1} \rd^*_k$:
\begin{align}
    \mathbb L_k = \mathbb D_k^T \mathbb M_{k+1} \mathbb D_k + \mathbb M_{k} \mathbb D_{k-1} \mathbb M_{k-1}^{-1} \mathbb D_{k-1}^T \mathbb M_{k}, \label{eq:hodge_laplacian_matrix}
\end{align}
where it is understood that terms with indices outside $\{0, 1, 2, 3\}$ are omitted.

\paragraph{Matrix-free implementation}
None of these matrices are ever assembled.
They are applied through the tensor-product structure of the spline bases: a $k$-form is evaluated on the $(p+1)^3$ Gauss--Legendre quadrature points of every cell by three one-dimensional contractions, multiplied by the metric coefficients stored on the quadrature grid, and contracted back.
The cost of one operator apply is therefore proportional to the number of quadrature points, and the memory footprint to the geometry stored on them, a few floats per point from the symmetric metric tensor (6 unique entries), its inverse, and the Jacobian determinant.

\paragraph{Solvers and preconditioners}
Every linear system is solved by a Krylov method on the matrix-free operators: preconditioned conjugate gradients for symmetric positive definite operators, MINRES for indefinite symmetric operators. 

We apply a custom-built preconditioner exploiting the sequence properties and tensor structure of our approximation space.
This preconditioning strategy will be explained in a forthcoming publication.

The stopping criterion of every solve is the residual measured in the norm of its space, where the mass matrix preconditioner serves as a cheap proxy for the inverse mass-weighted norm.    

The Hodge--Laplacian is only invertible on the subspace orthogonal to harmonic forms.
We hence first compute the harmonic forms for every geometry and then deflate them from every solve, i.e., project both the right-hand side and solution of $\mathcal L^k u = f$ to be orthogonal to every harmonic form.

\paragraph{Precision}
The code runs in three modes.
In float64 every operation is in double precision.
In float32, used for computational efficiency and needed for running on specialized architectures like TPUs, the Krylov iterations run in single precision with a tolerance of $10^{-5}$. It still reaches the same force balance values as float64 at less than half the computational cost, see~\ref{subsec:further_newton}, Figure~\ref{fig:newton_tol_sweep}.
In mixed mode, the Krylov iterations run in float32 while the residual of the outer iterative refinement is formed in float64 and the correction is accumulated in float64, which reaches double precision tolerances at a cost between those of float32 and float64.

Default tolerances are $10^{-10}$ in float64, $10^{-8}$ in mixed precision and
$10^{-5}$ in float32, the code's default.

\begin{remark}
    All timings cited are on a single H100 GPU using the mixed precision setting.
    Note that setup times involve host-side compilation, which might be slowed by concurrent jobs taking up resources.
\end{remark}

\section{Vacuum numerics}
\label{sec:vacuum_numerics}

The first verification experiment we perform concerns the computation of a harmonic (vacuum) field in stellarator geometry. 
We begin by defining the diagnostics used throughout this work in Section~\ref{subsec:diagnostics}.
We then verify the computations against an analytical solution in Section~\ref{subsec:manufactured_vacuum} and one computed from VMEC in Section~\ref{subsec:vmec_convergence}.
Section~\ref{subsec:shape_optimization} demonstrates automatic differentiation and shape-optimization capabilities by tuning a vacuum configuration for quasi-axisymmetry.

\begin{figure}[tbp]
  \centering
  \includegraphics[trim={0cm 0.6cm 0cm 0.6cm},clip,width=0.5\linewidth]{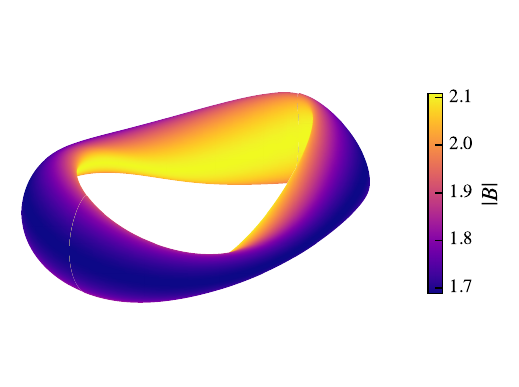}
  \caption{Vacuum case: Magnitude $|B|$ of the vacuum magnetic field on the boundary of the QA device.
  Spline resolution $(\mesh{32}{64}{32})$, degree $p = 3$, on the map of the high-resolution VMEC equilibrium of Section~\ref{subsec:vmec_convergence} (201 flux surfaces, 16/12 modes). 
  The seam after one field period is a plotting artifact.
  }
  \label{fig:vacuum_qa_bmag}
\end{figure}

Our numerical experiments in this section are run on the geometry of the Landreman--Paul precise quasi-axi\-sym\-met\-ric (QA) device~\cite{LandremanPaul2022}, shown in Figure~\ref{fig:vacuum_qa_bmag}.\footnote{The original low-res input file for this configuration can be found at: \url{https://github.com/hiddenSymmetries/simsopt/blob/master/tests/test_files/input.LandremanPaul2021_QA_lowres}} 

\subsection{Diagnostics}
\label{subsec:diagnostics}

The following diagnostics are central to our numerical experiments.

\paragraph{Normalizations}
Errors against a reference solution $u^{\star}$ are reported relative to it, $\normrel{u - u^{\star}} = \norm{u - u^{\star}} / \norm{u^{\star}}$. 
The magnetic pressure is $\tfrac12 |B(x)|^2$ and the magnetic energy $\Energy(B) = \tfrac12 \norm{B}^2$. 
Every integral, norm and average is taken over the computational domain $\Domain$, which is one field period of the device (with stellarator symmetry, half a period with doubled weights, see~\ref{sec:stellarator_symmetry}).
The volume of the domain is denoted $|\Domain|$.
In every relaxation run the initial field is normalized to $\norm{B} = 1$, so that $\Energy = \tfrac12$ and $|\Domain| \vavg{|B|^2} = 1$. 
Plasma $\beta$ is the ratio of the kinetic to the magnetic pressure: $\betavol{} = \vavg{p} / \vavg{\tfrac12 |B|^2} = 2 \ip{p}{1} / \norm{B}^2$.
We report $p$ normalized by the magnetic pressure, i.e., $p_{\mathrm{norm}}(x) = p(x) / \vavg{\tfrac12 |B|^2}$.

\paragraph{Force residual}
Following GVEC's convention, we normalize the force balance by the magnetic pressure gradient.
\begin{align}
\label{eq:force_residual}
    \normnorm{F}^2 &= \frac{\norm{F}^2}{\norm{ \grad ( \tfrac12 |B|^2 )}^2 }, \qquad F = J \times B - \grad p.
\end{align}
Here $p$ is defined by~\eqref{eq:pressure_problem}, so that $\dvg F = 0$.

\paragraph{Poincaré plots}
To compute integral curves we integrate $\rd x_t / \rd t = B(x_t)$ with $x_{t=0} = x_0$.
This is done in logical coordinates: the integral curve there is $\hat x_t = \Mapping^{-1} \circ x_t$, so $\tfrac{\rd}{\rd t} \hat x_t = D\Mapping^{-1}(x_t)\, \tfrac{\rd}{\rd t} x_t = D\Mapping^{-1}(x_t)\, B(x_t)$ and $\hat x_{t=0} = \Mapping^{-1}(x_0)$.
Using the definition of $B \in V^2$ through the logical representative $\hat B$,
\begin{align}
    \frac{\rd }{\rd t} \hat x_t = \big( \det D\Mapping(\hat x_t) \big)^{-1} \hat B(\hat x_t).
\end{align}
Since we want intersections of the integral curves with planes $\zeta = \mathrm{constant}$, we re-parametrize by $\zeta$, i.e.\ we divide by the $\zeta$-component of the right-hand side:
\begin{align}
    \frac{\rd }{\rd \zeta} \hat x = \frac{\hat B(\hat x)}{\hat B^{\zeta}(\hat x)},
\end{align}
where $\hat B^{\zeta}$ is the $\zeta$-component of the logical vector $\hat B$.
The $\zeta$-component of the right-hand side is then unity, so $\zeta$ is the independent variable and we record crossings at fixed-$\zeta$ planes, stepping forward period by period.
In our numerical experiments field lines are integrated with a fixed-step fifth-order explicit (Tsitouras') Runge--Kutta scheme from the {diffrax}~\citep{kidger_neural_2022} package.

\begin{remark}
    This is in contrast to, e.g., the SIMSOPT suite, where field lines are traced in Cartesian coordinates and plane crossings are detected in a secondary post-processing step via root-finding.
    We have found the re-parametrization with $\zeta$ to be significantly faster.
\end{remark}

\paragraph{Rotational transform}
We use two definitions here, they coincide for the case of nested flux surfaces.

\begin{definition}[Rotational transform]
    \label{def:rotational_transform}
    The rotational transform is defined as the fraction of poloidal versus toroidal rotations along a field line.

    We denote the rotational transform computed by field line tracing by $\iota$.
    Following the flow $\tfrac{\rd}{\rd t} x_t = B(x_t)$, we unwrap the toroidal angle $\zeta$ and the poloidal angle about the magnetic axis in the logical cross-section.
    The rotational transform $\iota$ is $\nfp{}$ times the slope of unwrapped poloidal against toroidal angle, determined by a least-squares fit.
    The traced $\iota$ is a property of a single field line and is defined on islands and on chaotic lines as well.
    
    The flux-ratio rotational transform $\iotabar$, the contained normalized toroidal flux $s$, and the mean rotational transform are
    \begin{align}
        \label{eq:flux_ratio_iota}
        \iotabar(r) = \nfp{} \, \frac{\langle \hat B^\theta \rangle_{\theta, \zeta}(r)}{\langle \hat B^\zeta \rangle_{\theta, \zeta}(r)}, 
        \qquad
        s(r) = \frac{\int_0^r \langle \hat B^\zeta \rangle_{\theta, \zeta}(r') \, \rd r'}{\int_0^1 \langle \hat B^\zeta \rangle_{\theta, \zeta}(r') \, \rd r'}, \qquad 
        \langle \iotabar \rangle_s = \int_0^1 \iotabar \, \rd s = \nfp{} \, \frac{\int_0^1 \langle \hat B^\theta \rangle_{\theta, \zeta} \, \rd r}{\int_0^1 \langle \hat B^\zeta \rangle_{\theta, \zeta} \, \rd r}.
    \end{align}
    The flux ratio $\iotabar$ requires no field-line tracing, but needs the logical surfaces $r = \mathrm{const.}$ to coincide with flux surfaces.
\end{definition}

\paragraph{Island size}
We measure the width of an island chain at $\iota = \nfp{} \, n / m$ on the traced field lines, where $n$ counts the chain's toroidal turns per field period, so that the toroidal mode number is $\nfp{} n$. 
Among the traced lines whose fitted rotational transform lies within $2\times 10^{-3}$ of the rational, we take the largest excursion $\max r - \min r$ of the logical radius over all their crossings on all planes.
A line inside the island near its separatrix spans the full island width. The field lines of MRX do not necessarily lie on the logical surfaces $r = \mathrm{const}$, the flux surfaces of the fixed VMEC map, so even a line on a nested surface varies in $r$. We quote this variation as a baseline.
Widths are given in logical $r \in [0, 1]$.

Island size is predicted by the pendulum estimate, see~\citep[Section 2.11]{helander_theory_2014} and \citep[Equations (44--45)]{cary_simple_1991}:
\begin{align}
    \label{eq:pendulum_estimate}
    w = \sqrt{ \frac{8 a \, \nfp}{\pi m \, |\partial_r \iota(\res{m}{n})|} },
\end{align}
where $a = |\delta \hat B^r_{mn}(\res{m}{n})| / \langle \sqrt{g} B^\zeta \rangle_{\theta, \zeta}(\res{m}{n})$ is the amplitude of the resonant normal field. This resonant field is the $(m, n)$ harmonic of $\sqrt g \, \delta B^r$ at the rational surface $\res{m}{n}$ divided by the surface-averaged toroidal field. The harmonic is computed via Fourier transform in the straight field line angle $\theta^* = \theta + \lambdaup$ with $\lambdaup$ from the VMEC input file.

\begin{figure}[t]
\centering
\begin{minipage}[t]{0.48\linewidth}
\centering
\resizebox{\linewidth}{!}{\includegraphics{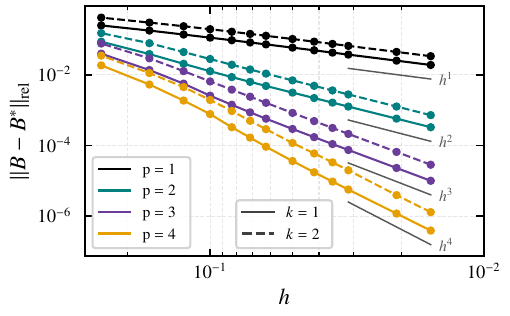}}
\caption{Convergence of the discrete vacuum field to the analytic solution $B^{\star}$ with respect to resolution: $k=1$ is the scalar-potential route $B = \grad \varphi + \tau \mathfrak{h}_1$, $k=2$ the vector-potential route $B = \curl A$.
The gray lines indicate the theoretical convergence rate $h^p$, which is observed for all cases.}
\label{fig:analytic_vacuum}
\end{minipage}\hfill
\begin{minipage}[t]{0.48\linewidth}
\centering
\resizebox{\linewidth}{!}{\includegraphics{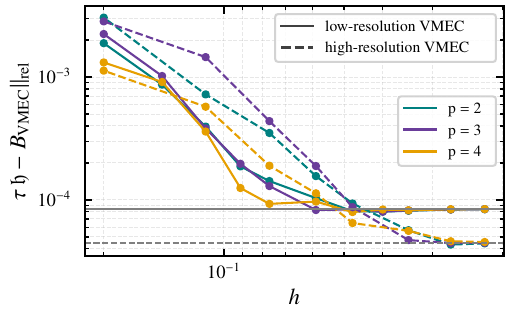}}
\caption{Agreement $\norm{B_{\mathrm{VMEC}} - \tau \, \mathfrak{h}_2} / \norm{B_{\mathrm{VMEC}}}$ of the discrete harmonic $2$-form with the high-resolution VMEC vacuum field.
The discrepancy decreases with resolution at every $p$ and saturates at $4.5 \times10^{-5}$, the truncation of the reference.
Against the low-resolution reference the difference plateaus earlier, at $8.4 \times 10^{-5}$.}
\label{fig:vmec_vacuum}
\end{minipage}
\end{figure}

\subsection{Manufactured vacuum solution}
\label{subsec:manufactured_vacuum}
We consider the Landreman--Paul precise QA design with $\nfp{} = 2$ and aim to solve the following problem: Find $B$ such that $\dvg B = \curl B = 0$ in $\Domain$ with boundary data given by $B_{\mathrm{vac}}$ on $\partial \Domain$. The two routes below impose its normal and its tangential trace, respectively.
The idea here is that we can specify an analytic vacuum field as a boundary condition, and validate $(h,p)$ convergence of our code to the analytic vacuum field in the whole volume. 
The field $B_{\mathrm{vac}}$ is chosen as
\begin{align}
    B_{\mathrm{vac}}(R, \phi, Z) = \frac{e_\phi}{R} + \grad \left( R^2 \cos (\nfp{} \phi) \right)
\end{align}
in cylindrical coordinates with $\phi \in [0, 2\pi]$.
All computations take place in the toroidal coordinates $(r, \theta, \zeta)$.

There are two distinct ways to solve this problem, corresponding to the two Hodge decompositions of the vector-valued $L^2(\Domain)$ space, i.e., expressing the magnetic field as a 1-form $H$ or a 2-form $B$.

\paragraph{Solution for $k=1$}
When we approximate the magnetic field as $H \in V^1$, then it holds that
\begin{align}
    H = \tau \mathfrak{h}_1 + \grad \varphi
\end{align}
where $\mathfrak{h}_1$ denotes the harmonic field, i.e., the normalized basis element of $\mathfrak{H}^1$, $\varphi \in V^0$, and $\tau \in \bR$ is a scale parameter.
Note that $\mathfrak{h}_1 \perp_{L^2} \grad \varphi$ by the discrete Hodge--Helmholtz decomposition property of the FEEC spaces.

To satisfy $\dvg H = 0$ and the normal boundary conditions, we can solve the Neumann problem
\begin{align}
    \ip{\grad \varphi}{\grad f} = \ip{B_{\mathrm{vac}}}{\grad f} \quad \text{for all} \quad f \in V^0.
\end{align}
This constitutes the inversion of a $k=0$ Hodge--Laplacian operator with natural boundary conditions.
The scale $\tau$ is set by matching the circulation along a closed toroidal curve, which in logical coordinates is obtained by following the vector field $(0,0,1)$.

\paragraph{Solution for $k=2$}
When approximating the vacuum field as $B \in V^2$, there is no explicit harmonic contribution as $\mathfrak{H}^2 = \{0\}$.
Instead $B = \curl A$, where the vector potential $A \in V^1$ solves the curl-curl problem
\begin{align}
    \ip{\curl A}{\curl u} = \ip{B_{\mathrm{vac}}}{\curl u} \quad \text{for all} \quad u \in V^1,
\end{align}
i.e., a $k=1$ Hodge--Laplacian problem with natural boundary condition $n \times (\curl A - B_{\mathrm{vac}}) = 0$ on $\partial \Domain$.
The toroidal flux is carried by the boundary circulation of $A$ and there is no harmonic contribution.

Results are presented in Figure~\ref{fig:analytic_vacuum}.
The meshes have $\mesh{n_r}{2 n_r}{n_r}$ splines with $n = n_r - p$ radial cells, $h_r = 1/n$. The $1$-form space then has $6n_r^3 - 12n_r^2 + 5n_r$ and the $2$-form space $6n_r^3 - 12n_r^2 + 2n_r$ degrees of freedom, from $475$ ($n = 4$, $p = 1$) to $1{,}830{,}000$ ($n = 64$, $p = 4$).

\paragraph{Additional symmetries}     
The map~\eqref{eq:stellarator_map} builds in field-period symmetry. The computational domain covers only one $\nfp{}$th of the device and hence reduces the number of discretization points by a factor $\nfp{}$.

Most stellarators are in addition stellarator symmetric, invariant under $(\theta, \zeta) \mapsto (-\theta, -\zeta)$. Exploiting this symmetry further reduces the computational cost, as half of the degrees of freedom are duplicates (potentially with reversed sign) of the other half.

The QA device is stellarator symmetric, so the same computation can be done on the whole torus, on one field period, or on half a period (\ref{sec:stellarator_symmetry}).
Table~\ref{tab:symmetry_vacuum} compares the three models at $n = 48$ radial cells and $p = 2$. The two solves take $1.8$ times less time on one field period than on the whole torus, and another $1.9$ times less on half a period, close to the factor two each model saves. The setup, dominated by compilation, shrinks by $1.7$ times at each step.

\begin{table*}[htb]
% 2026-09-23 (Tobias): only n = 48 cells, where the speed-up shows; records outputs/symmetry_vacuum_2026-09-18/{torus,
% period,half}_50 of the newton worktree (jobs 18742181-3). The half-period DoFs are the symmetric half of the period's.
% The n = 24 and 32 rows are in attic_2026-09-23/pre_ad_rewrite/symmetry_vacuum_table_jcp.tex.
\centering
\footnotesize
\setlength{\tabcolsep}{4pt}
\begin{tabular}{l l r r r r c c}
\toprule
model & splines & DoFs of $V^1$ & setup [s] & $k=1$ solve [s] & $k=2$ [s] & error $k=1$ & error $k=2$ \\
\midrule
whole torus, no symmetry & $\mesh{50}{100}{100}$ & $1{,}440{,}500$ & 2529 & 1428 & 245 & $6.22\cdot10^{-4}$ & $1.10\cdot10^{-3}$ \\
one field period & $\mesh{50}{100}{50}$ & $720{,}250$ & 1484 & 811 & 106 & $6.22\cdot10^{-4}$ & $1.10\cdot10^{-3}$ \\
half a period, stellarator symmetry & $\mesh{50}{100}{50}$ & $360{,}125$ & 870 & 416 & 62 & $6.22\cdot10^{-4}$ & $1.10\cdot10^{-3}$ \\
\bottomrule
\end{tabular}

\caption{Solving the manufactured vacuum test case. Exploiting symmetry in the computation saves substantial computation time. All three models use the stellarator-symmetric potential $R^2 \sin(\nfp{} \phi)$ in place of $R^2 \cos(\nfp{} \phi)$.}
\label{tab:symmetry_vacuum}
\end{table*}

\subsection{Convergence to a VMEC equilibrium}
\label{subsec:vmec_convergence}
Next, we compare the discrete harmonic $2$-form $\mathfrak{h}_2 \in \mathfrak{H}^2_0$ against the vacuum field $B_{\mathrm{VMEC}}$ of a VMEC equilibrium of the same QA design.
To match the scale of $B_{\mathrm{VMEC}}$, we compute the circulation $\tau \in \bR$ and report $\normrel{B_{\mathrm{VMEC}} - \tau \, \mathfrak{h}_2} = \norm{B_{\mathrm{VMEC}} - \tau \, \mathfrak{h}_2}  \,/\, \norm{B_{\mathrm{VMEC}}}$.

Unlike the manufactured case there is no analytical solution. We measure agreement with another computed equilibrium and are ultimately limited by the accuracy of the independent solve.
Against a low-resolution {VMEC} reference ($75$ flux surfaces, $8$/$8$ poloidal/toroidal modes) the difference saturates at $\approx 8.4 \times 10^{-5}$.
Recomputing the reference at higher resolution (201 flux surfaces, 16/12 modes) lowers that floor to $\approx 4.5 \times 10^{-5}$, which all three degrees ($p = 2, 3, 4$) reach at the finest meshes (Figure~\ref{fig:vmec_vacuum}).
The solution has essentially nested flux surfaces, consistent with a similar calculation in SPECTRE~\citep[Figure 5b]{balkovic2026spectre}.
We expect further convergence of the floor will be limited both by resolution and by the assumption of nested flux surfaces in VMEC. 

\subsection{Shape optimization for quasi-axisymmetry}
\label{subsec:shape_optimization}
There are many optimization, stability, and control problems to solve for nuclear fusion devices, many of which depend on the MHS solution.
In the context of stellarator optimization, the computation of MHD equilibria comprises a constraint within an outer optimization loop.
In particular, when $Q \colon \Hdiv_0(\Domain) \to \bR$ is a given function that measures the quality of a magnetic field configuration for the sake of some application (e.g., quality of particle confinement, engineering feasibility), then the full optimization problem reads
\begin{align}
    \min_{B \in \Hdiv_0(\Domain)} Q(B) \quad \text{s.t.} \quad \curl B \times B = \grad p, \; \dvg B = 0.
\end{align}
Given a parametrized mapping $\Mapping_\parameter \colon \hat \Domain \to \Domain_\parameter$, a $C^1$ diffeomorphism for every parameter vector $\parameter$, we can also pose the problem
\begin{align}
    \label{eq:shape_optimization}
    \min_{\parameter} Q \big( (\Mapping_\parameter)^2_* \hat B_\parameter \big) \quad \text{s.t.} \quad  \curl \big( (\Mapping_\parameter)^2_* \hat B_\parameter \big) \times (\Mapping_\parameter)^2_* \hat B_\parameter = \grad p \; \text{and} \; \dvg (\Mapping_\parameter)^2_* \hat B_\parameter = 0,
\end{align}
where $\hat B_\parameter$ is the equilibrium for the parameter $\parameter$.
To solve this shape optimization problem, we require the derivative of $(\Mapping_\parameter)^2_* \hat B_\parameter$ with respect to $\parameter$.
This derivative is obtained by the implicit function theorem together with the automatic differentiation tools within JAX.

In this example, we perform this optimization for vacuum fields, $B = \mathfrak h_2$, and optimize the map for quasi-axisymmetry.
The starting point is a strong deformation of the QA device and the task for the optimization is to return to it.

\paragraph{Derivative of the vacuum field}
The vacuum field is computed from an arbitrary first guess. We pick as this the field $\mathfrak s \in V^2_0$, the histopolation of the logical 2-form $(0, 0, 1)$, which is divergence-free in any geometry by Lemma~\ref{lem:pushforward_properties}. The following lemma crucially depends on $\mathfrak{H}^1_0 = \{0\}$. Its proof and that of the subsequent proposition are given in~\ref{subsec:shape_proofs}.
\begin{lemma}[Harmonic 2-form computation]
    \label{lem:harmonic_from_seed}
    Let $\mathfrak s \in V^2_0$ with $\dvg \mathfrak s = 0$, and let $A \in V^1_0$ solve
    \begin{align}
        \label{eq:harmonic_from_seed}
        \widetilde{\curl} \curl A = \widetilde{\curl} \, \mathfrak s.
    \end{align}
    Then $\mathfrak h_2 = \mathfrak s - \curl A$ is a nonzero harmonic 2-form, $\mathfrak h_2 \in \mathfrak H^2_0 \setminus \{ 0 \}$.
\end{lemma}

\begin{proposition}[Shape derivative]
    \label{prop:shape_gradient}
    Let the map $\parameter \mapsto \Mapping_\parameter$ be $C^1$ and denote the single harmonic 2-form on $\Domain_\parameter$ by $\mathfrak h_2(\parameter)$. Let the quality criterion $Q(\mathfrak h_2, \parameter)$ be differentiable. Then,
    \begin{align}
        \label{eq:shape_gradient}
        \frac{\rd}{\rd \parameter} Q\big( \mathfrak h_2(\parameter), \parameter \big) = \partial_\parameter Q + \partial_\parameter \langle \curl U, \mathfrak h_2 \rangle_{\Domain_\parameter}, \quad \text{where $U \in V^1_0$ solves} \quad
        \widetilde{\curl} \curl U = - \widetilde{\curl} \, \nabla_{\mathfrak h_2} Q.
    \end{align}
    Partial derivatives $\partial_\parameter$ are taken at fixed degrees of freedom of $U$ and $\mathfrak h_2$, and $\nabla_{\mathfrak h_2} Q \in V^2$ is the $L^2$ gradient of $Q$.
\end{proposition}
In the implementation, we wrap the solve of Section~\ref{subsec:implementation} in JAX's \texttt{custom\_linear\_solve} to differentiate the matrix-vector product. The preconditioner is held fixed at the initial geometry.

\paragraph{Parametrization}
The parameter $\parameter$ collects the spline coefficients of the discretizations of $R, Z: \hat x \mapsto x$ from ~\eqref{eq:stellarator_map}, denoted by $\parameter = (\mathtt R, \mathtt Z)$. Stellarator symmetry is enforced at all times.

We will drop the explicit dependence on $\parameter$ in the following to avoid notational clutter.

\paragraph{Quality criterion}
A vacuum field is quasi-axisymmetric if the following two-term residual vanishes~\citep{helander_theory_2014, LandremanPaul2022}:
\begin{align}
    \label{eq:qa_residual}
    Q_{\mathrm{QA}}(B) = \frac{1}{|B|^3} \bigg[ \bigg( \frac{\nfp{}}{2 \pi} \int_{\hat \Domain} B_\zeta \, \rd \hat x \bigg) \, B \cdot \grad |B| - \iotabar \, (B \times \grad \psi) \cdot \grad |B| \bigg],
\end{align}
with $\psi$ the toroidal flux divided by $2\pi$. A third term, proportional to the toroidal current, vanishes in vacuum.

When $r = \mathrm{const.}$ are flux surfaces, it holds that $2 \pi \grad \psi = \langle \hat B^\zeta \rangle_{\theta, \zeta}(r) \grad r$ and we can compute $\iotabar$ from the flux ratio~\eqref{eq:flux_ratio_iota}.

The total minimization objective is, with $\mathfrak h_2$ the harmonic 2-form ($\norm{\mathfrak h_2} = 1$),
\begin{align}
    \label{eq:qa_criterion}
    Q = \langle Q_{\mathrm{QA}}(\mathfrak h_2)^2 \rangle_{\Domain \cap \{r \geq h_r\}} \quad \text{s.t.} \quad \langle \iotabar(\mathfrak h_2) \rangle_s = \iotabar^\star, \; \text{and} \;
    P(\mathfrak h_2) = \left \| \mathfrak h_2 \cdot  \frac{\grad r}{|\grad r|} \right \|_{\Domain}^2 \leq 10^{-8}. 
\end{align}
The objective $Q_{\mathrm{QA}}$ is averaged over the domain without the first radial knot span, $r < h_r$, where the polar map is close to singular.
We re-scale the device after every step to retain volume and aspect ratio.

The constraints pin the mean rotational transform~\eqref{eq:flux_ratio_iota} to its value $\iotabar^\star = 0.41931$ on the VMEC map of the device of~\citep{LandremanPaul2022}, and keep logical surfaces $r = \mathrm{const.}$ close to flux surfaces. 
The latter is necessary for the definition of $Q_{\mathrm{QA}}$ and $\iotabar$ to be well-posed.
How to generalize these criteria to the non-nested case is an interesting question for future work. For instance, when $B \cdot n=0$, the computational boundary is a prescribed flux surface where quasi-symmetry and rotational transform can be targeted~\cite{nies2022adjoint}.

\paragraph{Setup}
Both constraints are imposed by an augmented Lagrangian method (\ref{subsec:shape_constraints}).
Optimization is done using a L-BFGS-B method~\citep[Chapter 7]{nocedal_numerical_2006} from \texttt{scipy.optimize} with a jitted value and gradient function.

\begin{figure}[tbp]
\centering
\begin{minipage}[t]{0.48\linewidth}
\centering
\resizebox{\linewidth}{!}{\includegraphics{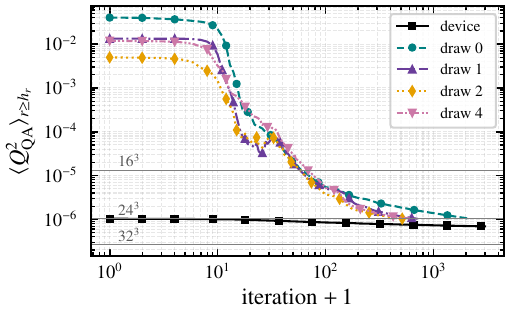}}
\caption{Quasi-axisymmetric shape optimization at $\mesh{24}{48}{24}$, $p = 3$: $\langle Q_{\mathrm{QA}}^2 \rangle_{r \geq h_r}$ at every L-BFGS-B iteration, from the interpolated VMEC map of the device and from four starts with random $10$~mm RMS boundary displacements. Gray: the criterion on the interpolated device at $\mesh{16}{32}{16}$, $\mesh{24}{48}{24}$ and $\mesh{32}{64}{32}$.}
\label{fig:shape_optimization}
\end{minipage}\hfill
\begin{minipage}[t]{0.48\linewidth}
\centering
\resizebox{\linewidth}{!}{\includegraphics{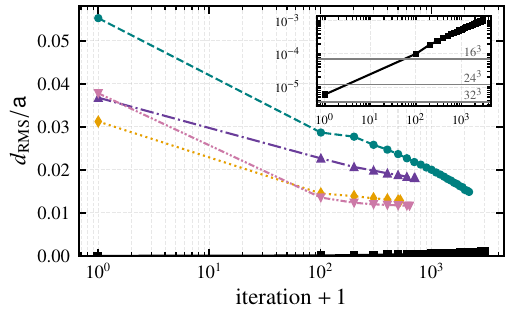}}
\caption{The runs of Figure~\ref{fig:shape_optimization}, in its line styles: the RMS distance $d_{\mathrm{RMS}}$ of the boundary to that of the interpolated device over the minor radius, every $100$ iterations. Inset: the run from the device on a logarithmic axis, and in gray the same distance of the interpolated boundary to the VMEC boundary.}
\label{fig:shape_errors}
\end{minipage}
\end{figure}

As a baseline we take the criterion on the VMEC map of the device, interpolated as in Section~\ref{subsec:vmec_convergence} on meshes $\mesh{n}{2n}{n}$ with $p = 3$.
For the interpolated solution, $\langle Q_{\mathrm{QA}}^2 \rangle_{r \geq h_r} = 1.3 \times 10^{-5}$, $1.0 \times 10^{-6}$ and $2.7 \times 10^{-7}$ for $n = 16$, $24$, and $32$ respectively, these are the gray lines in Figure~\ref{fig:shape_optimization}.
The observed rates are $6.3$ ($16 \to 24$) and $4.7$ ($24 \to 32$).

\paragraph{Optimization}
We perform the optimization at resolution $\mesh{24}{48}{24}$ ($55{,}296$ parameters to be optimized) and $p=3$. 
The parametrization consists of (i) the boundary degrees of freedom, whose deformation is extended smoothly to the interior by harmonic extension, and (ii) the remaining, interior degrees of freedom.
The vacuum field $\mathfrak h_2$, and hence the QA criterion, only depends on the boundary shape, so the latter are mostly in charge of satisfying the $P(\mathfrak h_2) \leq 10^{-8}$ constraint.

Running the optimization we see that the interpolated device is not a minimum of the discrete problem~\eqref{eq:qa_criterion}, see Figure~\ref{fig:shape_errors}.
L-BFGS-B lowers the criterion to $0.67$ times the baseline in $3{,}000$ iterations and moves the boundary by $d_{\mathrm{RMS}} / \mathsf a = 1.0 \times 10^{-3}$, meeting the constraints on $\langle \iotabar(\mathfrak h_2) \rangle_s$ and $P$.
We measure the change of the boundary by $d_{\mathrm{RMS}} = \langle \mathrm{dist}(\cdot, \partial \Domain_{\mathrm{VMEC}})^2 \rangle_{\theta,\zeta}^{1/2} \big |_{r=1}$, the distance of the boundary to the boundary of the interpolated device. The minor radius of the device is denoted by $\mathsf a$.

The initial configurations of these runs are random perturbations of the boundary of the interpolated map with modes $m, |n| \leq 4$ per field period. 
Their amplitudes decay as $1 / (1 + m^2 + n^2)$ (\ref{subsec:shape_constraints}) to enforce regularity and are scaled so their RMS is $6\%$ of $\mathsf a$. 

Four such random perturbations are applied, starting out at $d_{\mathrm{RMS}} / \mathsf a = 0.031$ to $0.055$.
They break the QA criterion ($4.8 \times 10^3$ to $3.9 \times 10^4$ times the baseline) and move $\langle \iotabar \rangle_s$ off $\iotabar^\star$ by $0.009$ to $0.031$.
We then run the optimization until the QA criterion hits the baseline value $1.0 \times 10^{-6}$. 
One evaluation of $Q$ and its gradient takes about $3$~s on one H100 GPU.
The gradient from autodifferentiation agrees with central finite differences to $10^{-7}$.

\paragraph{Results}
All curves in Figure~\ref{fig:shape_optimization} fall, so the gradient drives every start to a quasi-axisymmetric configuration, at fixed mean rotational transform, volume and aspect ratio. The Lagrange multipliers allow us to determine the cost of the constraints.
Moving $\iotabar^\star$ by $10^{-3}$ changes the criterion by $0.24\%$ of the baseline, and loosening the bound on $P$ by $10^{-8}$ lowers it by $0.15\%$.

The final boundaries differ from the device by one to two percent of its minor radius. 
Quasi-axisymmetry together with these three scalars does not determine the boundary shape. 
This is no surprise. 
Large databases of optimized stellarators contain families of nearby quasi-axisymmetric devices with a similarly low quasi-symmetry error at the same aspect ratio and rotational transform~\citep{giuliani_comprehensive_2024}.

\begin{figure}[tbp]
  \centering
  \includegraphics{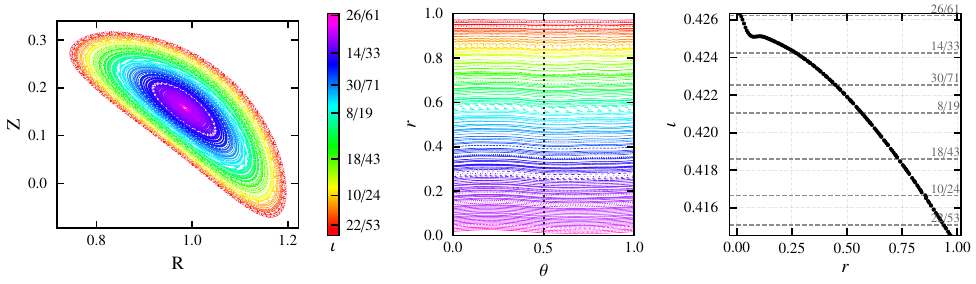}
  \caption{Vacuum field of an optimized boundary of Section~\ref{subsec:shape_optimization}. Left: Poincar\'e section at $\zeta = 0.25$ in the physical frame, field lines colored by rotational transform $\iota$. Center: the same section in the logical frame. Right: the rotational transform of every traced line, one marker per line, against the logical radius $r$ at which the line crosses $\theta = 0.5$ (the dotted line in the center plot).}
  \label{fig:vacuum_qa_poincare}
\end{figure}

Lastly, we verify $\langle \iotabar \rangle_s$ against $\iota$ computed by field line tracing, see Figure~\ref{fig:vacuum_qa_poincare}. 
The mean rotational transform agrees with $\langle \iotabar \rangle_s$ to $4 \times 10^{-5}$ in all runs.

These results demonstrate the automatic differentiation and shape optimization capabilities of MRX and conclude the discussion of vacuum configurations.

\section{Magnetic Relaxation}
\label{sec:magnetic_relaxation}

We now move on to describing finite-$\beta$ MHS solutions obtained by magnetic relaxation.
We present the context of previous work in Section~\ref{subsec:previous_work} and present relaxation as a constrained minimization problem in Section~\ref{subsec:constrained_minimization}. Treatment of pressure is discussed in Section~\ref{subsec:pressure}. The discrete computation scheme itself is presented in Section~\ref{subsec:discrete_relaxation}. Lastly, we explain how to accelerate the relaxation computations using Newton's method in Section~\ref{subsec:newton}.

\subsection{Relation to previous work}
\label{subsec:previous_work}
The fact that magnetic equilibria are stationary points of the energy with respect to admissible variations has been known for many decades.

\paragraph{Energy principle}
In~\cite{grad_hydromagnetic_1958} and~\cite{bernstein_energy_1958}, two different derivations of admissible variations are given.
The latter authors credit Lundquist~\cite{lundquist_stability_1951} with the development of this energy principle.
The same papers also derive the stability conditions of equilibria based on the eigenvalues of $\delta^2 \Energy$.
Alternative derivations of the stability condition are given in~\cite{hain_zur_1957,merkel_holonome_1977}.

\paragraph{Helicity}
The observation that the helicity is conserved under admissible variations is usually credited to Woltjer~\cite{woltjer_theorem_1958}, who also pointed out that the fields $B: \curl B = \lambda B$ with constant $\lambda \in \bR$ are minima of the magnetic energy subject to the constraint of constant helicity.
The term helicity was coined by Moffatt~\cite{moffatt_degree_1969}, who also connected it to the degree of entanglement of field lines.
The term magnetic relaxation was also popularized by the fluid dynamics community, where the same process is used to study stationary solutions of Euler's equations~\cite{moffatt_magnetostatic_1985, moffatt_magnetostatic_1986}.
The textbook~\cite{arnold_topological_2021} gives an overview of magnetic relaxation.

\paragraph{Numerics}
The first works to attempt a constrained minimization of $\Energy$ as a means to compute magnetic equilibria numerically were, to our knowledge, the works by Chodura and Schlüter~\cite{chodura_3d_1981} and the BETA code~\cite{bauer_computational_1978, bauer_magnetohydrodynamic_1984}, see Table~\ref{tab:codes}.
A more recent application of this idea is the SIESTA code~\cite{hirshman_siesta_2011}, which MRX is in many ways similar to.

\subsection{Constrained minimization}
\label{subsec:constrained_minimization}

Consider a magnetic field $B \in \Hdiv_0(\Domain)$ with $\dvg B = 0$.
\begin{definition}[Energy functional]
\label{def:energy}
    The magnetic energy $\Energy: \Hdiv_0(\Domain) \rightarrow \bR$ of the field $B$ is given by
    \begin{align}
    \label{eq:energy}
        \Energy(B) = \tfrac12 \norm{B}^2.
    \end{align} 
\end{definition}

\paragraph{Admissible variations}
Ideal perturbations of $B$ conserve the magnetic topology.
They are defined as follows:
\begin{definition}[Admissible variations]
\label{def:admissible_variations}
    Given $v \in \Hdiv_0 (\Domain)\cap C^1(\Domain)$, we define admissible variations $v \mapsto \delta_v B$ by:
    \begin{align}
        \delta_v B = \curl (v \times B).
    \end{align}
\end{definition}

\begin{proposition}[First variation of the energy]
    \label{prop:first_variation}
    For $B \cdot n = v \cdot n = 0$ on $\partial \Domain$ and $J = \curl B$,
    \begin{align}
        \label{eq:first_variation}
        \delta_v \Energy(B) = \ip{B}{\delta_v B} = - \ip{J \times B}{v} .
    \end{align}
\end{proposition}
\begin{proof}
    From the definition of $\delta_v B$ and integration by parts, $\delta_v \Energy(B) = \ip{B}{\curl (v \times B)} = \ip{\curl B}{v \times B}$.
    The boundary term vanishes since $\big( (v \times B) \times B \big) \cdot n = (v \cdot B) (B \cdot n) - |B|^2 (v \cdot n) = 0$.
    Finally, $\ip{J}{v \times B} = - \ip{J \times B}{v}$.
\end{proof}

\paragraph{Arnold's inequality}
To be useful, magnetic relaxation schemes should avoid relaxing to the vacuum equilibrium $B \propto \mathfrak h$.
To construct such a magnetic relaxation scheme, we will recall the following lower bound on the energy, often referred to as Arnold's theorem~\cite{arnold_asymptotic_1974}. 
\begin{definition}[Generalized helicity]
    \label{def:helicity}
    Let $A \in \Hcurl_0(\Domain)$ with $\curl A = (\Id - \Pi^{{\mathfrak H}^2_0}) B$ and $\dvg A = 0$.
    The generalized helicity of $B$ is
    \begin{align}
        \label{eq:helicity}
        \Helicity(B) = \ip{A}{B + \Pi^{{\mathfrak H}^2_0} B}.
    \end{align}
    Its value does not depend on the gauge of $A$, and the gauge $\dvg A = 0$ is fixed for concreteness.
    It is the relative helicity of~\cite{berger_topological_1984} with the harmonic part $\Pi^{{\mathfrak H}^2_0} B$ as the reference field.
    When $\mathfrak H^2(\Domain) \neq \{ 0 \}$, this is no longer the case and the definition must be adjusted.
\end{definition}

\begin{proposition}[Helicity bounds energy]
\label{prop:helicity_bounds_energy}
    When $\mathfrak H^2(\Domain) = \{ 0 \}$, the magnetic energy is bounded from below by the generalized helicity, with $\lambda_\Domain$ a constant that depends only on $\Domain$:
    \begin{align}
        \label{eq:arnold_inequality}
        \frac {4 \Energy(B)} {\sqrt{\lambda_\Domain}}  \geq |\Helicity(B)|.
    \end{align}
\end{proposition}

\begin{proof}
    Write $B_{\mathfrak H} = \Pi^{{\mathfrak H}^2_0} B.$
    The Poincaré inequality~\citep[Proposition 4.1]{ern_discrete_2024} and Hodge--Helmholtz decomposition~\eqref{eq:helmholtz_decomposition} give
    $\norm{A}^2 \leq \norm{\curl A}^2 / \lambda_\Domain \leq \norm{B}^2 / \lambda_\Domain$.
    Hence,
    \begin{align}\notag
        \sqrt{\lambda_\Domain} |\Helicity| = \sqrt{\lambda_\Domain} |\ip{A}{B + B_{\mathfrak H}}| \leq \sqrt{\lambda_\Domain} \norm{A} \left( \norm{B} + \norm{B_{\mathfrak H}} \right) \leq 2 \sqrt{\lambda_\Domain} \, \norm{A} \norm{B} \leq 2 \norm{B}^2  = 4 \Energy(B).
    \end{align}
    Here, we use the Poincaré and Cauchy--Schwarz inequalities and finally $\norm{\curl A} = \norm{B - B_{\mathfrak H}} \leq \norm{B}$ and $\norm{B_{\mathfrak H}} \leq \norm{B}$.
\end{proof}

\begin{figure}
    \centering
    \resizebox{0.5\linewidth}{!}{\includegraphics{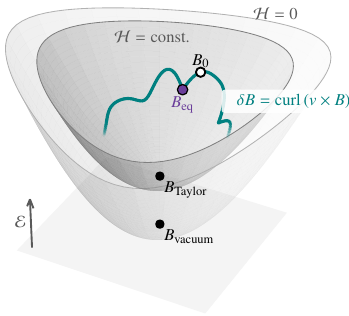}}
\caption{
A sketch of magnetic relaxation as a constrained minimization problem. 
Field configuration energy $\Energy$ is displayed on the $z$-axis.
Constant helicity $\Helicity$ sheets are plotted in gray. 
The lowest energy states at fixed helicity are Taylor states, $\curl B_{\mathrm{Taylor}} = \lambda \, B_{\mathrm{Taylor}}$.
By Arnold's inequality~\eqref{eq:arnold_inequality}, the energy within one sheet is bounded below by a multiple of $|\Helicity|$ and the vacuum field $\curl B_{\mathrm{vacuum}} = 0$ is the lowest energy state.
The orbit generated by admissible variations (teal) satisfies the stricter constraint of constant field topology within one helicity sheet. 
Relaxation from $B_0$ descends along this orbit to the nearest minimum, $B_{\mathrm{eq}}$. 
Smooth local minima of the energy along this orbit satisfy $\ip{J \times B}{v} = 0$. 
The orbit is plotted with kinks to demonstrate the regularity issues discussed in Section~\ref{subsec:rational_surfaces}: at a kink the energy is not differentiable along the orbit, and a minimum there need not be force-free.
}
\label{fig:landscape}
\end{figure}

Figure~\ref{fig:landscape} sketches the resulting picture of relaxation as a constrained minimization.
Conserving the generalized helicity provides a barrier to relaxing to the trivial equilibrium. 
Resistive dissipation can be used to purposefully induce topological changes, but it is important to control any change in helicity that occurs.
We justify below why admissible variations allow for sufficient control over this generalized helicity loss. 

\begin{proposition}[Helicity preservation]
\label{prop:helicity_preservation}
    Admissible variations of the magnetic field preserve $\dvg B = 0$ and the harmonic component $\Pi^{{\mathfrak H}^2_0} B$ and they leave the generalized helicity of Definition~\ref{def:helicity} unchanged: with $A \in \Hcurl_0(\Domain)$, $\curl A = B - \Pi^{\mathfrak H^2_0} B$,
    \begin{align}
        \delta_v \Helicity(B) &= \delta_v \int_\Domain A \cdot (B + \Pi^{{\mathfrak H}^2_0} B) \, \rd x = 0.
    \end{align}
\end{proposition}

\begin{proof}
    Let $B_{\mathfrak H} = \Pi^{{\mathfrak H}^2_0} B$.
    First, $\dvg B = 0$ is preserved since $\dvg \curl = 0$ and $\curl (v \times B) \perp \mathfrak H^2_0$ by the Hodge--Helmholtz decomposition.
    Second, for the vector potential, $\delta_v A = v \times B + \grad f$ for some $f \in \Hone(\Domain)$ depending on the gauge.
    Hence,
    \begin{align}
        \delta \Helicity(v) 
        = \ip{\delta_v A}{B + B_{\mathfrak H}} + \ip{A}{\delta_v B}
        = \ip{v \times B + \grad f}{B + B_{\mathfrak H}} + \ip{A}{\curl (v \times B)}.
    \end{align}
    We now integrate by parts and use $\curl A = B - B_{\mathfrak H}$, $B \times B = 0$, and $\grad f \perp_{L^2(\Domain)} \{ B, B_{\mathfrak H} \}$ for all $f \in \Hone(\Domain)$ to find
    \begin{align}
        \delta \Helicity(v) = \ip{v \times B + \grad f}{B + B_{\mathfrak H}} + \ip{B - B_{\mathfrak H}}{v \times B} = \ip{v \times B}{B_{\mathfrak H}} - \ip{B_{\mathfrak H}}{v \times B} = 0.
    \end{align}
\end{proof}

For more information on generalized helicities, we refer to~\citep[Section III.7]{arnold_topological_2021} and~\citep{mactaggart_magnetic_2019, he_topology-preserving_2025}.

In fact, helicity preservation is merely a consequence of the preservation of magnetic topology. We consider the special case of divergence-free velocity fields, anticipating the next section. The proof for this case is given in~\ref{sec:proofs}, the general one for example in~\citep[Lemma C.1]{enciso_obstructions_2025}.

\begin{lemma}[Admissible flows]
    \label{lem:admissible_flows}
    Let $\mathcal F_t$ be the flow of a velocity field $v_t$, $\tfrac{\rd}{\rd t} \mathcal F_t(x) = v_t(\mathcal F_t(x))$, $\mathcal F_0(x) = x$, with $\dvg v_t = 0$. Then $\partial_t B_t = \curl (v_t \times B_t)$ with $\dvg B_0 = 0$ implies $B_t(\mathcal F_t(x)) = D\mathcal F_t(x) B_0(x)$.
\end{lemma}

\begin{proposition}[Topology preservation]
\label{prop:topology_preservation}
    Admissible variations of the magnetic field map magnetic field lines bijectively onto magnetic field lines.
\end{proposition}

\begin{proof}
    Denote by $\Gamma_0$ a field line of $B_0$, i.e., $\tfrac{\rd}{\rd \ell} \Gamma_0(\ell) = B_0 \circ \Gamma_0(\ell)$. Let $\Gamma_t = \mathcal F_t \circ \Gamma_0$. We show that $\Gamma_t$ is a field line of $B_t$:
    \begin{align}
        \frac{\rd}{\rd \ell} \Gamma_t(\ell) = D \mathcal F_t( \Gamma_0(\ell)) \frac{\rd}{\rd \ell} \Gamma_0(\ell) = D \mathcal F_t( \Gamma_0(\ell)) B_0( \Gamma_0(\ell) ) = B_t(\mathcal F_t \circ \Gamma_0(\ell))
    \end{align}
    by Lemma~\ref{lem:admissible_flows}. Hence $\rd \Gamma_t(\ell) / \rd \ell = B_t \circ \Gamma_t(\ell)$, so $\Gamma_t$ is a field line of $B_t$. Since $\mathcal F_t$ is a diffeomorphism of $\Domain$, the map $\Gamma_0 \mapsto \mathcal F_t \circ \Gamma_0$ is a bijection of the field lines of $B_0$ onto those of $B_t$.
\end{proof}

\begin{remark}
    It has been conjectured in~\citep[Section I.9]{arnold_topological_2021} and shown in~\citep{enciso_helicity_2016} that helicity is the only integral invariant of volume-preserving maps given certain regularity assumptions.   
\end{remark}

\subsection{Pressure}
\label{subsec:pressure}
Clearly, we need to generalize this picture to include finite pressure. We want to satisfy $J\times B = \grad p$ at equilibrium, not $J \times B = 0$.

\paragraph{Leray projection}
Finite-pressure equilibria in MRX are computed by restricting the velocity $v$ to be divergence-free, or, equivalently, restricting the flows $\mathcal F_t$ generated by $v$ to be volume-preserving.

\begin{definition}[Leray projection]
    \label{def:leray}
    The Leray projection of $u \in \Hdiv_0(\Domain)$ is
\begin{align}
\label{eq:pressure_problem}
    \Leray u = u - \grad p, \text{ where } - \dvg \grad p = - \dvg u,
\end{align}
    with homogeneous Neumann boundary conditions on $p$. By construction, $\dvg \Leray u = \dvg (u - \grad p) = 0$.
\end{definition}

\begin{corollary}[Force balance at a stationary point]
    \label{cor:force_balance}
    For $v = \Leray (J \times B)$,
\begin{align}
    \delta_v \Energy(B)
    = - \norm{J \times B - \grad p}^2.
    \label{eq:force_balance}
\end{align}
    Formally, a stationary point of this descent satisfies $J \times B = \grad p$.
\end{corollary}
To see this, note that $\delta_v \Energy(B) = -\ip{J \times B}{J \times B - \grad p}$. Add and subtract $\grad p$ to obtain
\begin{align}
    \delta_v \Energy(B) = - \norm{J \times B - \grad p}^2 - \ip{\grad p}{J \times B - \grad p}. \notag
\end{align}
The second term vanishes after integration by parts as $J \times B - \grad p$ is divergence-free and hence $L^2$-orthogonal to $\grad p$.

More generally, the Hodge--Helmholtz decomposition gives: 
\begin{corollary}[Decomposition of the Lorentz force]
    \label{cor:force_decomposition}
    By the Hodge--Helmholtz decomposition~\eqref{eq:helmholtz_decomposition}, the Lorentz force $J \times B$ decomposes as
\begin{align}
    \label{eq:force_decomposition}
    J \times B = \grad p + \curl \omega + \tau \mathfrak h,
\end{align}
    with $\curl \mathfrak h = \dvg \mathfrak h = 0$, $\norm{\mathfrak h} = 1$ and $\tau \in \bR$.
    The three components are $L^2$-orthogonal to one another, hence $\norm{J \times B - \grad p} \leq \norm{J \times B}$.
\end{corollary}
We hence obtain a different path to computing $J \times B - \grad p$.
\begin{corollary}[Potential velocity]
    \label{cor:potential_velocity}
    It holds that
\begin{align}
    \label{eq:potential_velocity}
    J \times B - \grad p =  \curl \, (\curl \curl)^{-1} \curl (J \times B) + \ip{J \times B}{\mathfrak h} \, \mathfrak h.
\end{align}
\end{corollary}
We found this second route more computationally efficient in the numerical experiments, see~\ref{sec:additional_numerics}. It also generalizes very well to the Newton method introduced in Section~\ref{subsec:newton}.

\begin{proposition}[Force under stellarator symmetry]
    \label{prop:no_harmonic_force}
    When $\Domain$ and $B$ are stellarator symmetric, $\ip{J \times B}{\mathfrak h} = 0$.
\end{proposition}
The proof is in~\ref{sec:stellarator_symmetry}.

\paragraph{Pressure profiles}
For Grad--Shafranov solvers and for 3D nested equilibrium solvers that solve for the flux function $\psi$, the profile functions $[0,1] \to \bR$ are inputs.
The physical fields $p(x)$ and $\iota(x)$ are then given by $p(x) = p_{\mathrm{profile}}(\psi(x))$ and $\iota(x) = \iota_{\mathrm{profile}}(\psi(x))$.
With enough experience, one can use the inputs $p_{\mathrm{profile}}$ and $\iota_{\mathrm{profile}}$ to shape the outputs $p$ and $\iota$.
Even in this case, the final pressure in physical space is also an output of the numerical method, since it will vary as $\psi$ varies.
In three-dimensional configurations, this procedure breaks down as $B$ is no longer determined by $\psi$ and the toroidal flux value alone.

In our case, the pressure $p$ is not treated as a dynamical variable and function of the density as in e.g.~\cite{chodura_3d_1981, hender_calculation_1985, hirshman_siesta_2011}, but rather as a Lagrange multiplier.
We prescribe an initial pressure and rotational transform only through the initial condition $B_0$.
At each step, it is the solution to the elliptic problem~\eqref{eq:pressure_problem} with homogeneous Neumann boundary conditions.
In other words, the final pressure profile will depend on the initial condition.     
\begin{remark}
    This choice is common among relaxation methods in fluid dynamics but to our knowledge has not been used in previous magnetic relaxation codes. 
\end{remark}

For practical purposes, one way to replicate the profile design is to prescribe a $p$ and $\iota$ profile using a code that assumes nested flux surfaces and then use the output of this code as an input to a relaxation run, potentially with a small perturbation to escape the (local) minimum with nested flux surfaces.
As the relaxation process starts close to an equilibrium, we can expect that $\norm{B - B_0}$ is small, where $B$ denotes the state of the magnetic field throughout the relaxation.
The same is true for $\norm{p - p_0}$ and  $\norm{\iota - \iota_0}$, giving us control of $p$ and $\iota$ through the initial condition.
In this case, the lack of direct control over the profiles might actually be a useful feature. 
The prescribed $p_{\mathrm{profile}}$ and $\iota_{\mathrm{profile}}$ are roughly respected, and some modifications and flattening of the profiles can be achieved where islands and chaos open up in the solution (Section~\ref{subsec:seeding}).

\subsection{Discrete relaxation}
\label{subsec:discrete_relaxation}
We now provide the discrete counterparts to the continuous relaxation.

\paragraph{Regularization}

There are a priori only very weak regularity guarantees on $J \times B$ and this can lead to grid-scale oscillations in $v$ in a numerical scheme.
For this reason, it is beneficial to apply a smoothing operation to $v$, as has been done in~\cite{chodura_3d_1981} and also is done in the analytic treatment of the relaxation problem in ~\citep[Equation 2.1]{beekie_moffatts_2022}.

We can let $v = (\Id - \varepsilon \dvg \grad)^{-\gamma} (J \times B - \grad p)$ ($\varepsilon > 0$) in order to dampen high-frequency oscillations.
We note the following solvability result from ~\citep[Section 6.2.3]{arnold_finite_2010}:
\begin{lemma}[Smoothing divergence-free fields]
    \label{lem:smoothing_divfree}
For any $F \in \Hdiv_0(\Domain)$ satisfying $F \perp {\mathfrak H}^2_0$ and $\dvg F = 0$, there exists $u \in \Hdiv_0(\Domain)$ such that $\curl \curl u = F$, $\dvg u = 0$, and $u \perp {\mathfrak H}_0^2$.
\end{lemma}

Since $\dvg \curl = 0$, every application of $(\Id + \varepsilon \curl \widetilde{\curl})^{-1}$ preserves the divergence, so $v$ is divergence-free and
\begin{align}
    \delta_v \Energy(B)
    = -\ip{(\Id - \varepsilon \dvg \grad)^{-\gamma} (J \ttimes B - \grad p)}{J \ttimes B}    
    = - \| J \times B - \grad p \|^2_{H_{0,\varepsilon}^{-\gamma}(\Domain)} = - \| v \|^2_{H_{0,\varepsilon}^{\gamma}(\Domain)},
\end{align}
i.e., the energy dissipation equals an $\varepsilon$-weighted Sobolev norm of order $-\gamma$ of the force residual. We only look at the case $\gamma = 1$ in the numerical experiments.

\paragraph{Algorithm}
Relaxation is performed by solving the following system of equations:
\begin{definition}[Relaxation step]
    \label{def:relaxation_step}
    Let $B^n \in V^2_{0}$, $\delta t, \, \varepsilon \in \bR_{> 0}$.
    The relaxation step from $B^n$ to $B^{n+1}$ is the solution of
\begin{align}
    \label{eq:relaxation_step}
    J^n = \widetilde{\curl} \, B^{n}, \qquad E^n = {\Pi}^1_0 (v^n \times B^n), \qquad B^{n+1} = B^n + \delta t \curl E^n.
\end{align}  
The operations $\widetilde{\grad}$, $\widetilde{\curl}$, $\widetilde{\dvg}$ are the weak derivatives as described in Section~\ref{subsec:implementation}.
They are the discrete version of the co-differential operators from~\eqref{eq:codifferential}.   

The velocity $v^n$ is defined in one of two ways. The gradient descent approach with smoothing sets
\begin{align}
    v^n &= (\Id + \varepsilon \curl \widetilde{\curl})^{-1} (J^n \times B^n - \widetilde{\grad} \, p^n),
    \label{eq:velocity_gradient_descent}
\end{align}
where either $p^n = (\dvg \widetilde{\grad})^{-1} \dvg (J^n \times B^n)$ (Leray projection, \eqref{eq:pressure_problem}), or $J^n \times B^n - \widetilde \grad p^n = \curl \, \omega^n + \ip{J^n \times B^n}{\mathfrak h} \, \mathfrak h$ with $\omega^n = (\widetilde \curl \curl)^{-1} \widetilde \curl \, (J^n \times B^n)$ (potential velocity, \eqref{eq:potential_velocity}).

The Newton approach sets
\begin{align}
    v^n &= \curl \omega^n + \tau^n \mathfrak h
    \label{eq:velocity_newton}
\end{align}
with $(\omega^n, \tau^n)$ the solution of the Newton system~\eqref{eq:newton_system} at $B^n$, which will be discussed in detail in Section~\ref{subsec:newton}.

The inversions in the gradient descent step are to be understood as solving the corresponding Hodge--Laplace saddle problems, whose discrete well-posedness is shown in~\citep[Sections 6.2.2--4]{arnold_finite_2010}.
\end{definition}

\paragraph{Discrete helicity preservation}
We now discuss how the continuous helicity preservation from Proposition~\ref{prop:helicity_preservation} reflects in the discrete setting.
\begin{definition}[Discrete vector potential and helicity]
    \label{def:discrete_helicity}
    The discrete vector potential $A \in V^1_0$ of $B \in V^2_0$ is the solution of the Hodge--Laplace problem for $k = 1$ with essential boundary conditions~\citep[Section 6.2.2]{arnold_finite_2010},
\begin{align}
    \label{eq:discrete_vector_potential}
    \widetilde{\curl} \curl A - \grad q = \widetilde{\curl} \, B, \quad q = \widetilde{\dvg} \, A, \quad A \perp_{L^2} {\mathfrak H}^1_0,   
\end{align}
    which gives $\curl A = B - \Pi^{\mathfrak H^2_0} B$. For the problems in this work, ${\mathfrak H}^1_0 = \{ 0 \}$.
    The discrete helicity is that of Definition~\ref{def:helicity}, $\Helicity = \ip{A}{B + \Pi^{\mathfrak H^2_0} B}$.
\end{definition}

The following identity holds for any $E^n \in V^1_0$ when $B^{n+1} = B^n + \delta t \curl E^n$, the proof is given in~\ref{sec:proofs}.

\begin{lemma}[Discrete helicity balance]
\label{lem:helicity_balance}
    Let $B^n \in V^2_0$, $E \in V^1_0$ and $B^{n+1} = B^n + \delta t \curl E^n$.
    Then
    \begin{align}
        \label{eq:helicity_balance}
        \Helicity^{n+1} - \Helicity^{n} &= 2 \, \delta t \, \ip{E^n}{B^{n}} + \delta t^2 \, \ip{E^n}{\curl E^n}.
    \end{align}
\end{lemma}

The continuous flow conserves the helicity because $E = v \times B$ is point-wise orthogonal to $B$ and the second term, time-integration error of the explicit step, vanishes.
Discretely, the projection of $v \times B$ onto $V^1_0$ to obtain $E$ leads to a residual, the product of the projection errors of $v \times B$ and of $B$ (Corollary~\ref{cor:helicity_remainder}).
This residual can be removed by a rank-one correction of $E$.

\begin{corollary}[Discrete helicity remainder]
\label{cor:helicity_remainder}
    For $E_\sigma = \Pi^1_0 (v^n \times B^n) - \sigma \, \Pi^1_0 B^n$, $\sigma \in \bR$,
    \begin{align}
        \frac{\Helicity^{n+1} - \Helicity^{n}}{2 \, \delta t}
        &= \ip{\Pi^1_0 (v^n \times B^n) - v^n \times B^n}{\; B^n - \Pi^1_0 B^n} - \sigma \norm{\Pi^1_0 B^n}^2 + \frac{\delta t}{2} \, \ip{E^n_\sigma}{\curl E^n_\sigma} .
        \label{eq:helicity_remainder}
    \end{align}
\end{corollary}

The right-hand side of \eqref{eq:helicity_remainder} is a quadratic function of $\sigma$. For small $\delta t$ it has a real root close to that of its linear part, and this root makes it vanish, restoring exact discrete helicity preservation. An auxiliary-variable formulation \cite{he_topology-preserving_2025} with an implicit midpoint step achieves the same end. We discuss and demonstrate this in Table~\ref{tab:helicity_preservation},~\ref{subsec:further_gradient_descent}.

We emphasize here that the preservation of global helicity is not a constraint that is as strong as the preservation of field line topology that is present in the continuous problem.
There is an argument that the auxiliary-variable formulation is a more faithful discretization of the continuous problem.
The reason we do not follow it here is that it requires a vanishing tangential trace of $B$ on the domain boundary, which is far from true in the settings we are investigating.

\paragraph{Discrete energy evolution}
The following result is proven in~\ref{sec:proofs}.

\begin{proposition}[Discrete energy evolution]
    \label{prop:energy_evolution}
    Solutions to~\eqref{eq:relaxation_step} satisfy
    \begin{align}
        \frac{\Energy^{n+1} - \Energy^{n}}{\delta t} = - \| v^n \|^2_{\mathcal A} + \frac{\delta t}{2} \norm{\curl E^n}^2,
    \end{align}
    where $\mathcal A v^n = J^n \times B^n - \widetilde \grad \, p^n$, tested with divergence-free fields, $\dvg v^n = 0$, $\| v^n \|^2_{\mathcal A} = \ip{v^n}{\mathcal A v^n}$, and $\mathcal A$ is either the operator $\Id + \varepsilon \curl \widetilde{\curl}$ from~\eqref{eq:velocity_gradient_descent} or the penalized Hessian of \eqref{eq:velocity_newton}. The energy decreases for $0 < \delta t < 2 \| v^n \|^2_{\mathcal A} / \norm{\curl E^n}^2$.
\end{proposition}

\begin{lemma}[Exact line search]
    \label{lem:line_search}
    Since the magnetic energy is quadratic in the increment $\delta B^n = \curl E^n$, the line search $\min_{\delta t} \Energy(B^n + \delta t \curl E^n)$ has the closed-form solution
\begin{align}
    \label{eq:line_search}
    \delta t_* = -\frac{\ip{B^n}{\curl E^n}}{\norm{\curl E^n}^2} .
\end{align}
\end{lemma}

We clip this time step so that the velocity advects the field by at most a fraction $C$ of a cell in one step: with $v^i = \hat v^i / \det D\Mapping$ the contravariant components of $v$ and $h_i$ the knot spacing of direction $i \in \{r, \theta, \zeta\}$.
\begin{align}
\label{eq:cfl}
    \delta t = \min \Big\{ \delta t_*, \; \frac{C}{\max_{\hat x \in \hat \Domain,\, i} |v^i(\hat x)| / h_i} \Big\}
\end{align}
with $C < 1$.
We pick $C = 0.5$ in all numerical experiments.
Empirically, the cap binds only in the first steps of a relaxation from a rough initial field.

\paragraph{Conjugate gradients}
Chodura and Schl\"uter~\cite{chodura_3d_1981} accelerate the descent with a conjugate-gradient (Polak--Ribi\`ere) update of the velocity.
We found this method to require many restarts in our setting, as consecutive gradients fail to remain sufficiently orthogonal. Our acceleration of choice is Newton's method (Section~\ref{subsec:newton}). We also note that Chodura and Schl\"uter find that smoothing $v$ is beneficial for convergence.

\paragraph{Implementation choices}
The relaxation runs in compiled chunks of steps, called from an outer loop in pure Python.
Between outer loop steps, any callback routine can run (e.g., computing and storing diagnostics).
As a result, host-device synchronization happens once per chunk rather than once per step, offering a good compromise between speed and flexibility.
The fastest performance can be achieved running the entire relaxation in one chunk, but the overhead from the outer loop is typically small (1--5\% of wall time). 

The full descent state is written to disk at every chunk and can be restarted from any checkpoint. This allows a relaxation to be branched off or resumed at a later time.  

\subsection{Newton's method}
\label{subsec:newton}
Well above the floor of the force residual (Section~\ref{subsec:rational_surfaces}) and close to an equilibrium, the optimization problem is well-behaved.
We can make use of second-order information of the energy functional.

\paragraph{Second variation}
The Hessian used in Newton's method must be aware of the curvature of the constraint manifold.
We now derive the expression for the second admissible variation.
\begin{proposition}[Second variation of the energy]
    \label{prop:second_variation}
    Along the flow $\mathcal F_t$ generated by $v$, with $\delta_v B = \curl(v \times B)$ and $\delta^2_{u,v} B = \curl(u \times \delta_v B)$, the second variation of the energy is
\begin{align}
\label{eq:Hessian}
    &\delta^2_{u,v} \Energy(B) = \ip{\delta_u B}{\delta_v B} + \tfrac12 \ip{B}{\curl(v \times \delta_u B)} + \tfrac12 \ip{B}{\curl(u \times \delta_v B)} .
\end{align}
\end{proposition}
\begin{proof}
    Expand the push-forward $B_t = \exp(\mathcal F_t) B$ in $t$ to obtain $B_t = B + t \, \delta_v B + \tfrac12 t^2 \delta^2_{v,v} B + O(t^3)$ and substitute into $\Energy(B_t) = \tfrac12 \norm{B_t}^2$.
    The second-order contribution to the energy is $\tfrac12 t^2 \big( \ip{\delta_v B}{\delta_v B} + \ip{B}{\delta^2_{v,v} B} \big)$, adding the contributions from $u$ and $v$ gives~\eqref{eq:Hessian}.
\end{proof}
The Newton step minimizes the quadratic approximation of $\Energy(B_t)$: find the velocity $v$ with $\dvg v = 0$ such that
\begin{align}
    \delta^2_{u,v} \Energy(B) = \ip{J \times B - \grad p}{u} \quad \text{for all} \quad u : \dvg u = 0 .
\end{align}
The constraint $\dvg v = 0$ can easily be incorporated by formulating the problem in terms of $(\omega, \tau): \curl \omega + \tau \mathfrak h = v$.

Testing with the same space, $u = \curl \omega' + \tau' \mathfrak h$, we obtain the system to solve for $(\omega, \tau) \in V^1_0 \times \bR$:
\begin{align}
    \label{eq:newton_system}
    \begin{pmatrix}
        \widetilde{\curl} \, \delta^2 \Energy(B) \, \curl
        & {\widetilde{\curl} \, \delta^2 \Energy(B) \, \mathfrak h} \\
        {\ip{\curl \, \cdot \,}{\delta^2 \Energy(B) \, \mathfrak h}}
        & {\ip{\mathfrak h}{\delta^2 \Energy(B) \, \mathfrak h}}
    \end{pmatrix}
    \begin{pmatrix} \omega \\ \tau \end{pmatrix}
    =
    \begin{pmatrix} \widetilde{\curl} \, (J \times B) \\ {\ip{J \times B}{\mathfrak h}} \end{pmatrix} ,
\end{align}
where $\delta^2 \Energy(B)$ denotes the operator of the bilinear form from~\eqref{eq:Hessian}.

In the case of stellarator symmetry, $\tau \equiv 0$ for any $B$ by Proposition~\ref{prop:no_harmonic_force} and the system reduces to
\begin{align}
    \widetilde{\curl} \, \delta^2 \Energy(B) \, \curl \, \omega = \widetilde{\curl} \, (J \times B). \notag
\end{align}

\paragraph{Solution strategy}

It remains to efficiently solve~\eqref{eq:newton_system} at every relaxation step.
\begin{lemma}[The Newton system]
    \label{lem:newton_system}
    The system~\eqref{eq:newton_system} is symmetric and singular.
    Every gradient $\grad f$, $f \in V^0_0$, lies in its kernel and the right-hand side is orthogonal to these gradients.
    
    The Gauss--Newton part $\ip{\delta_u B}{\delta_v B}$ of~\eqref{eq:Hessian} has the further kernel of the velocities with $\curl(v \times B) = 0$, the symmetries of $B$, on which the right-hand side vanishes as well by Proposition~\ref{prop:first_variation}.
\end{lemma}

\begin{proof}
    This system is symmetric since $\ip{\delta^2 \Energy(B) \curl \omega}{\mathfrak h} = \ip{\omega}{\widetilde{\curl} \, \delta^2 \Energy(B) \, \mathfrak h}$ and $\widetilde{\curl} \, \delta^2 \Energy(B) \, \curl$ is symmetric. The fact that gradients lie in its kernel follows from the de~Rham property of the discretization. The right-hand side is orthogonal to gradients since $\ip{\widetilde{\curl} (J \times B)}{\grad f} = \ip{J \times B}{\curl \grad f} = 0$ and $\mathfrak{h}$ is orthogonal to gradients.
\end{proof}

We rely on an inexact Newton method with MINRES as the inner solver, the Newton-MR method of Liu and Roosta~\cite{liu_newton-mr_2022}. 
We found this to be superior in performance when compared to the more standard CG--Steihaug method (see~\citep[Chapter 7]{nocedal_numerical_2006}). 

The condition number of $\widetilde{\curl} \, \delta^2 \Energy(B) \curl$ is estimated at $\mathcal O(10^5)$ using Lanczos iterations in the configurations tested here, so to achieve anything with a fixed budget of MINRES iterations, we require a decent preconditioner. 
The second variation~\eqref{eq:Hessian} has a null space that the force does not see: a velocity $v_\parallel = f B$ along the field has $v_\parallel \times B = 0$ and hence $\delta_{v_\parallel} B = 0$, and it is divergence-free wherever $B \cdot \grad f = 0$, on every flux surface.
Discretely these flows leave a continuum of very small eigenvalues and render Newton's method numerically unstable.
We therefore add to the operator a Levenberg--Marquardt term on the parallel component only,
\begin{align}
    \label{eq:parallel_penalty}
    \delta^2 \Energy + \kappa \, \mathcal M_\parallel, \qquad \ip{u}{\mathcal M_\parallel v} = \int_\Domain \chi_\parallel(r) \, \frac{(u \cdot B)(v \cdot B)}{|B|^2} \, \rd x , \; \text{where} \;
    \chi_\parallel = \frac{\langle B^\theta \rangle_{\theta, \zeta}^2 \, \langle |\partial_\theta B|^2 \rangle_{\theta, \zeta} + \langle B^\zeta \rangle_{\theta, \zeta}^2 \, \langle |\partial_\zeta B|^2 \rangle_{\theta, \zeta}}{\langle B^\theta \rangle_{\theta, \zeta}^2 + \langle B^\zeta \rangle_{\theta, \zeta}^2},
\end{align}
is the strain of $B$ along the field direction.
The constant $\kappa = 3$ is a hyperparameter, determined by a sweep shown in~\ref{subsec:further_newton}, Table~\ref{tab:newton_penalty}.

The preconditioner approximates the Gauss--Newton term $(u,v) \mapsto \ip{\curl(v \times B)}{\curl(u \times B)} + \kappa \, \ip{u}{\mathcal M_\parallel v}$ and exploits the tensor-product structure of the approximation space using fast diagonalization.
We will provide a detailed description of our preconditioning strategy in a forthcoming paper.

\paragraph{Comparison with SIESTA}
Overall, our procedure is similar to that of SIESTA, described in~\citep[Section IX.A]{hirshman_siesta_2011}.

SIESTA linearizes the same force operator, but assembles the resulting matrix column by column from finite differences of the force. 
This matrix is block-tridiagonal (due to the radial coupling) and factorized with BCYCLIC~\citep{hirshman_bcyclic_2010}. 
This factorization is the preconditioner of a GMRES iteration on the linearized force balance. In MRX, by contrast, the Hessian is never assembled. It is only ever applied
 at the cost of three mass matrix solves, and we use MINRES to solve~\eqref{eq:newton_system}, with a budget of $200$ iterations per step (Table~\ref{tab:newton_inner}).

In SIESTA, the plasma is compressible, with a pressure that is transported and compressed by the displacement. There is an additional term proportional to $(v,u) \mapsto \ip{\dvg v}{\dvg u}$ in their Hessian as a result. In MRX, displacements are divergence-free and this term vanishes.

\section{Relaxation numerics}
\label{sec:relaxation_numerics}
We now formulate a robust group of experiments to test the numerics, hyperparameters, computational efficiency, and convergence of the proposed methodology.
We introduce a benchmark example in Section~\ref{subsec:ncsx} and use the diagnostics of Section~\ref{subsec:diagnostics}. Results for the gradient descent approach are shown in Section~\ref{subsec:gradient_descent} and for Newton's method in Section~\ref{subsec:newtons_method}. The discretization of a finite-$\beta$ equilibrium is verified against an exact solution in~\ref{sec:landreman_verification}, Table~\ref{tab:landreman_convergence}.

The exact specifications of all numerical experiments that are reported in this work are given in~\ref{subsec:experiment_parameters}.
As before, all reported timings are for a single H100 GPU.

\subsection{NCSX configuration} \label{subsec:ncsx}
The National Compact Stellarator Experiment (NCSX) was a fusion device at Princeton Plasma Physics Laboratory, designed to determine whether a compact QA stellarator can combine ideal confinement properties of tokamaks with the functionality of stellarators~\cite{ncsx2001physics, Zarnstorff2001}.
Although the project was not fully realized, this configuration remains a useful tool for analysis of high-$\beta$ design possibilities. 

The configuration we build on is the \texttt{li383} NCSX reference equilibrium computed by {VMEC} with $49$ flux surfaces, $m_{\mathrm{pol}} = 9$ poloidal and $n_{\mathrm{tor}} = 5$ toroidal modes, three field periods, major radius $1.42\,$m, minor radius $0.33\,$m, on-axis field $1.47\,$T and a volume-averaged $\beta$ of $4.25\%$.
The rotational transform rises from $0.393$ on the axis to $0.655$ at the edge, crossing both the $\iota = \tfrac12$ and $\iota = \tfrac35$ surfaces.
This is the geometry from Figure~\ref{fig:ncsx_sections}.\footnote{The high-res data file for this configuration can be found at \url{https://github.com/landreman/regcoil/blob/master/equilibria/wout_li383_1.4m.nc}}

The reference run is at $(n_r, n_\theta, n_\zeta) = \mesh{32}{64}{64}$, $p = 2$, mixed precision. The gradient descent runs and the parameter sweeps of \ref{sec:additional_numerics} are at $\mesh{16}{32}{32}$.
The detailed numerical configurations for all results are listed in Table~\ref{tab:runs}, the default values in Table~\ref{tab:hyperparameters}.

\begin{remark}
    The seconds per step quoted in the tables are the steady stepping rate over compiled chunks after the first, hence excluding just-in-time compilation.
    Wall-time axes and numbers include this first chunk. 
    The diagnostics evaluated between chunks (helicity, pressures, check-pointing) are excluded from both.
    This overhead is small (1--5\%) compared to the full run time.
\end{remark}

\subsection{Gradient descent}
\label{subsec:gradient_descent}

We now show a number of different configurations that the relaxation code can be run in, and demonstrate the impacts of the different hyperparameter choices.
A comprehensive list of choices available to the user is shown in~\ref{subsec:code_settings}.   

\paragraph{Smoothing} 

The hyperparameter $\varepsilon$ that controls the strength of the smoothing is set to $\varepsilon = 0.07 \, \vavg{g_{rr}} \, h_r^2$. The factor $h_r^2$ offsets the grid-dependence of the Laplacian spectrum, while $\vavg{g_{rr}}$ absorbs the physical size of the device.
The constant $0.07$ was determined by a sweep. For constants between $0.007$ and $0.7$ the squared force residual after $2{,}000$ steps stays within a factor two of its value at $0.07$.

\begin{figure}
\centering
\begin{minipage}[t]{0.48\linewidth}
\centering
\resizebox{\linewidth}{!}{\includegraphics{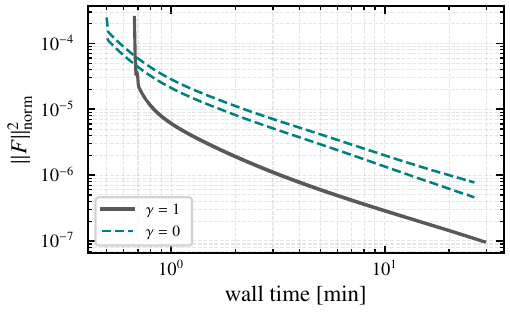}}
\caption{
Impact of velocity smoothing on gradient descent at comparable wall time: $\gamma = 1$ (grey, solid, $5{,}000$ steps) and $\gamma = 0$ (teal, dashed, $7{,}000$ steps).
Without smoothing, the descent falls into a two-step cycle on a grid-scale mode. We draw odd and even steps as two separate lines.
}
\label{fig:gamma_sweep}
\end{minipage}\hfill
\begin{minipage}[t]{0.48\linewidth}
\centering
\resizebox{\linewidth}{!}{\includegraphics{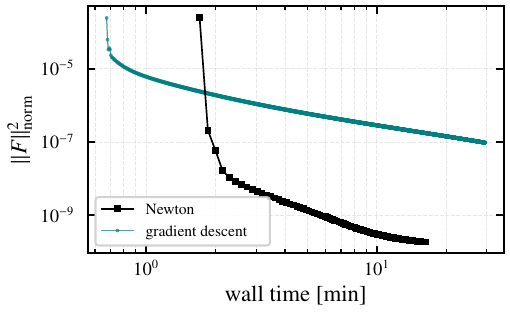}}
\caption{
Squared force residual against wall time for gradient descent (teal, circular markers) and Newton's method (black, square markers) at the same parameters, $\mesh{16}{32}{32}$, $p=2$.
Every marker location corresponds to one relaxation step.
Newton's method reaches $10^{-8}$ in $2.4$ minutes and $10^{-9}$ in $5.8$~minutes. Gradient descent reaches $10^{-7}$ after $28$ minutes.
}
\label{fig:newton}
\end{minipage}
\end{figure}

Figure~\ref{fig:gamma_sweep} illustrates why smoothing the velocity before advecting the magnetic field is beneficial.
Without smoothing ($\gamma = 0$), the squared force alternates between two branches on every pair of steps, with step sizes $4.1 \times 10^{-5}$ and $6.8 \times 10^{-5}$.
We identify this as a two-step cycle of steepest descent with an exact line search~\cite{akaike_successive_1959}. The iteration alternates between two directions in the plane of the largest and the smallest mode.
In that limit the two step sizes satisfy $1/\delta t_1 + 1/\delta t_2 = \lambda_{\max} + \lambda_{\min}$. 
The measured value, $3.9 \times 10^{4}$, is the largest eigenvalue of the discrete second variation and corresponds to a grid-scale mode.

The smoothed descent ($\gamma = 1$) exhibits the same cycle with branches only $1\%$ apart.
Its step sizes give $\lambda_{\max} + \lambda_{\min} = 2.1 \times 10^{3}$. 
The smoothing acts as a preconditioner that lowers the effective stiffness of the grid-scale modes.
At comparable wall time the squared force residual on the lower branch is $1.1 \times 10^{-7}$ against $4.6 \times 10^{-7}$ for the unsmoothed run.

The need for regularization of magnetic relaxation dynamics is well understood in the continuous case~\cite{beekie_moffatts_2022}.
To our knowledge, the only work that systematically studied this numerically so far is~\cite{yeates_limitations_2022}, but only for a one-dimensional toy model.

Numerical results on helicity balance and a comparison of the Leray projection to the potential velocity formulation are shown in~\ref{subsec:further_gradient_descent}, Tables.~\ref{tab:helicity_preservation} and~\ref{tab:velocity_choices}.

\subsection{Newton's method}
\label{subsec:newtons_method}
\begin{figure}
\centering
\resizebox{0.725\linewidth}{!}{\includegraphics{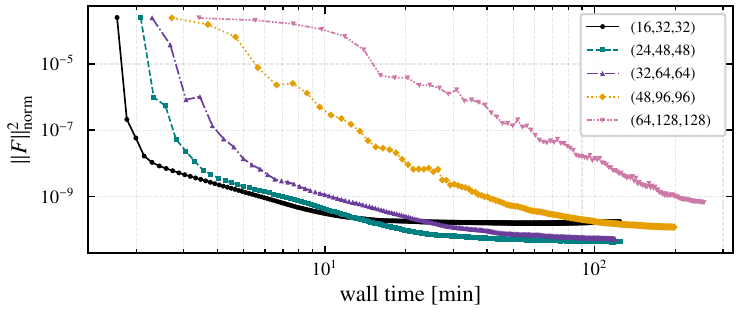}}
\caption{Newton's method at different resolutions. The cost per step grows as $n_r^{1.92}$ over this range (a fit to the s/step of Table~\ref{tab:newton_convergence}), and the time to $10^{-8}$ from $2.4$ minutes at $n_r = 16$ to $20$ minutes at $n_r = 48$.
Every step is indicated with a marker symbol.
Within two hours all resolutions but the two largest reach a floor between $4 \times 10^{-11}$ and $1.5 \times 10^{-10}$.
Past the floor the residual stays flat to within a few percent while the energy (not shown) continues to decrease.}
\label{fig:newton_h_sweep}
\end{figure}

\begin{figure}[tb]
    \centering
    \resizebox{\linewidth}{!}{\includegraphics{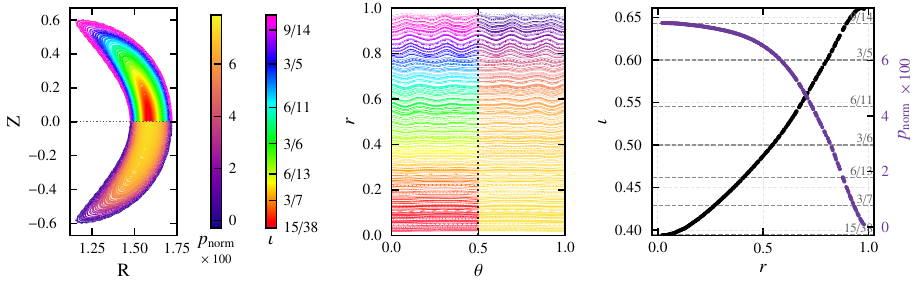}}
    \caption{Converged ideal equilibrium of the NCSX case at $\mesh{48}{96}{96}$, Poincar\'e section at $\zeta = 0$ after $200$ Newton steps.
    The tiles are as in Figure~\ref{fig:vacuum_qa_poincare}.
    Pressure is normalized, $p_{\mathrm{norm}} = p / \vavg{\tfrac12 |B|^2}$. The pressure profile is plotted with error bars that show the spread of $p_{\mathrm{norm}}$ along a field line. These error bars are not visible here, as $B \cdot \grad p$ is very small in this state. 
    The diagnostics can be read from Table~\ref{tab:newton_convergence}: $\normnorm{F}^2 = 1.2 \times 10^{-10}$, $\betavol = 4.24\%$ ($4.26\%$ for the VMEC initial condition at this resolution, $4.25\%$ as reported by VMEC).}
    \label{fig:islands_equilibrium}
\end{figure}

The addition of Newton's method speeds up the method significantly and allows us to reach the discretization floor with normalized square force residuals of order $10^{-10}$.
Results are shown in Table~\ref{tab:newton_convergence}, Figure~\ref{fig:newton} and Figure~\ref{fig:newton_h_sweep}. 
The minimum squared force residual of the runs that reach their floor lies between $4 \times 10^{-11}$ and $1.5 \times 10^{-10}$ and does not move systematically with the resolution. The relative helicity drift stays below $1.05 \times 10^{-5}$ and shrinks with the resolution as $n_r^{-1.8}$, a fit to Table~\ref{tab:newton_convergence}.
A Poincar\'e section for one of the relaxed states is shown in Figure~\ref{fig:islands_equilibrium}.
In~\ref{subsec:further_newton}, we show Poincar\'e sections of the relaxed states at three resolutions.
Only grid-scale features differ between them, while profiles of $p$ and $\iota$ and force residuals are in agreement across resolutions.

\begin{table}[htb]
\caption{Comparing relaxation performance for gradient descent and Newton's method at different resolutions using stellarator symmetry, see also Figure~\ref{fig:newton_h_sweep}.
Newton's method reaches $\normnorm{F}^2 < 10^{-8}$ within minutes at modest resolutions. 
The two highest resolutions tested reach $\normnorm{F}^2 < 10^{-8}$ in under two hours and are still descending at the wall time limit.
Helicity drift is small and decreases with increasing resolution. 
For the Newton runs, plasma $\betavol$ drifts by less than $4 \times 10^{-3}$ relative to VMEC's value of $4.252\%.$}
%% JCP (single-column) version of the Newton convergence table, generated by
%% outputs/table4_2026-09-18/make_table4.py (worktree newton) from the runs of launch.sh; do not edit
\centering
\footnotesize
\setlength{\tabcolsep}{3pt}
\begin{tabular*}{\textwidth}{@{\extracolsep{\fill}}l l r r r c r r r r r r r r@{}}
\toprule
method & resolution & DOF count & steps & s/step & $\normnorm{F}^2$
  & \multicolumn{3}{c}{\makebox[0pt]{steps to $\normnorm{F}^2$}}
  & \multicolumn{3}{c}{\makebox[0pt]{time to $\normnorm{F}^2$ [min]}}
  & $\Delta \Helicity / \Helicity$ & $\Delta \betavol / \betavol^{\mathrm{VMEC}}$ \\
\cmidrule(lr){7-9}\cmidrule(lr){10-12}
 & & & & [s] & floor
  & $10^{-7}$ & $10^{-8}$ & $10^{-9}$
  & $10^{-7}$ & $10^{-8}$ & $10^{-9}$
  & $\times 10^{-6}$ & $\times 10^{-3}$ \\
\midrule
gradient & $\mesh{16}{32}{32}$ & 20{,}998 & 5000 & 0.34 & $9.62 \times 10^{-8}$ & 4813 & -- & -- & 28.3 & -- & -- & -7.3 & 15.0 \\
\midrule
Newton & $\mesh{16}{32}{32}$ & 20{,}998 & 820 & 8.98 & $1.54 \times 10^{-10}$ & 3 & 6 & 29 & 2.0 & 2.4 & 5.8 & 10.4 & -0.4 \\
 & $\mesh{24}{48}{48}$ & 74{,}886 & 500 & 14.85 & $4.15 \times 10^{-11}$ & 4 & 7 & 21 & 2.8 & 3.5 & 7.0 & -9.4 & -2.0 \\
 & $\mesh{32}{64}{64}$ & 182{,}278 & 300 & 23.52 & $5.18 \times 10^{-11}$ & 6 & 9 & 23 & 4.2 & 5.4 & 10.8 & -3.5 & -2.5 \\
 & $\mesh{48}{96}{96}$ & 631{,}302 & 200 & 58.88 & $\leq$ $1.18 \times 10^{-10}$ & 12 & 19 & 38 & 13.5 & 20.4 & 39.1 & -1.8 & -2.7 \\
 & $\mesh{64}{128}{128}$ & 1{,}515{,}526 & 120 & 126.84 & $\leq$ $6.43 \times 10^{-10}$ & 27 & 51 & 93 & 58.3 & 108.8 & 197.6 & -1.0 & -3.5 \\
% \midrule
% fit $\propto n_r$ & & $ n_r^{3.09}$ & & $ n_r^{1.92}$ &  & $ n_r^{1.57}$ & $ n_r^{1.51}$ & $ n_r^{0.82}$ & $ n_r^{2.38}$ & $ n_r^{2.67}$ & $ n_r^{2.50}$ & $ n_r^{-1.80}$ &  \\
\bottomrule
\end{tabular*}

\label{tab:newton_convergence}
\end{table}

\section{Reconnection}
\label{sec:reconnection}

Magnetic reconnection allows the field line topology to change by opening and closing magnetic islands near resonant surfaces and current sheets.
Reconnection happens either on grid-scale through finite-resolution effects or can be actively enabled by adding resistivity, denoted by $\eta > 0$. 
This section discusses how current sheets form near rational surfaces in Section~\ref{subsec:rational_surfaces} and how we change the topology deliberately by seeding in Section~\ref{subsec:seeding}.

\subsection{Rational surfaces}
\label{subsec:rational_surfaces}

We now review the physical mechanism of current sheets forming on rational surfaces, what is known about the continuous problem from the analysis side, and how these difficulties lead to finite resolution effects. 

\paragraph{Current sheets}
Ideal relaxation in the continuous limit keeps the topology of the field lines, so the equilibrium it tends to retains the rotational transform of the initial field on every surface.
On a surface with rational transform $\iota = \nfp{} n/m$ the field lines close.
Force balance across such a surface has no smooth solution in general. 
The perpendicular current $B \times \grad p / |B|^2$ has a divergence that the parallel Pfirsch--Schlüter current must close, since $\dvg J = 0$ holds.
For the $(m, n)$ harmonic the closing current is proportional to $1 / (\iota m - \nfp{} n)$~\citep{loizu_magnetic_2015}.

The ideal displacement that would align the field is discontinuous across the surface for the same reason.
At rational surfaces, $B$ is tangentially discontinuous, hence the current $\curl B$ is a singular sheet. Figure~\ref{fig:current_sheets} in~\ref{subsec:further_newton} shows the current of our relaxed states with and without seeded islands.
This mechanism is what the Grad conjecture~\citep{grad_toroidal_1967, constantin_flexibility_2021} and the Parker problem~\citep{parker_spontaneous_1994, moffatt_magnetostatic_1985} are concerned with.

\paragraph{The continuous problem}
The well-posedness of the magnetic relaxation is largely an open problem in analysis.
It is known that the regularized relaxation is globally well-posed on the flat torus for $\gamma > 5/2$ and its velocity tends to zero~\citep{beekie_moffatts_2022, bae_local_2025}. Local well-posedness holds for all $\gamma \geq 0$.
However, the limit need not be smooth and generic fields cannot relax to a smooth equilibrium under their topological constraint~\citep{cieliebak_note_2017, enciso_obstructions_2025}. 

\paragraph{Discretization floor}
For any discretization, a real current sheet would be a sub-grid quantity.
After a fast initial descent, any relaxation run will form them and the force residual stalls near the discretization floor where the forming current sheets can no longer be resolved.
In our experiments, the value of this floor does not decrease under $h$ or $p$ refinement (Table~\ref{tab:newton_convergence}, Figure~\ref{fig:newton_p_sweep}): the sheets remain sub-grid at every resolution tested.

This floor is not a stationary point of the discrete energy, at which the projected force would vanish.
The optimization landscape becomes very flat along the directions of modifying the sheet-scale structures.
At this stage any progress in decreasing the energy comes at the price of numerical reconnection.
We observe that this can lead to a slight increase in force residual after the initial floor is reached.

The rise of the force residual can be pictured as taking long steps along a narrow, curved valley of the energy landscape.
The quadratic model is accurate along the valley floor, yet any misalignment sends the step up the valley's sides, where the gradient (force residual) grows even though the energy still decreases.

\subsection{Island seeding}
\label{subsec:seeding}

The relaxation is ideal: apart from grid-scale reconnection, which it does not control, it cannot change the topology of the initial field. 
Reaching a state with prescribed islands therefore requires the topology to be modified explicitly.
In MRX, this is done by seeding islands at rational surfaces following an energy criterion.

\paragraph{Perturbations}
The seed is a resonant perturbation $\dB = \curl \dA$.
We chose the same form of perturbation as SIESTA~\citep[Equation (8.2)]{hirshman_siesta_2011}, namely
\begin{align}
    \dA_{mn}(r, \theta, \zeta) = \frac{B}{|B|} a_{mn}(r) \cos\big(2 \pi (m \theta - n \zeta)\big) 
\end{align}
for the rational surface $(m,n)$ located at $r_{mn}$, in the logical coordinates, where $\zeta \in [0, 1]$ spans one field period and the toroidal mode number is $\nfp{} n$. 
The total perturbation is then given by $\dB = \sum_{(m,n)} \curl \, \dA_{mn}$ for all rational surfaces considered.
The profile $a_{mn}(r)$ determines the shape and amplitude.
We restrict $a_{mn}(r) = 0$ for $r=0$, $r=1$, and all $r: |r - r_{mn}| > 1/m$, so the perturbation is local around $r_{mn}$ and respects the boundary conditions at $\partial \Domain$ and the coordinate axis.
The locality criterion follows from minimizing the energy of a single perturbation, $\norm{\curl \dA_{mn}}^2$, which has the form $\exp(-m|r - r_{mn}|)$.

\paragraph{Energy criterion}
Different choices of $r \mapsto a_{mn}(r)$ lead to different topologies and thus determine the equilibria that can be reached after reconnection.
We need a criterion to choose these profiles in a principled manner.

\begin{figure}[tb]
\centering
\begin{minipage}[t]{0.57\linewidth}
\vspace{0pt}
\centering
{\small\setlength{\tabcolsep}{3pt}
\centering
\begin{tabular}{l c c c c c c}
\toprule
$(m, n)$ & $\iota$ & $r_{mn}$ & $\delta \hat B^r_{mn}$ & $w_{\mathrm{pred}}$ & $w_{\mathrm{pred}} / h_r$ & $w_{\mathrm{meas}}$ \\
\midrule
% 2026-09-24: delta B^r_mn without the (1 + d_theta lambda) weight (job 18753633, outputs/seed_width_fix_2026-09-24)
$(5, 1)$ & $\tfrac35$ & $0.794$ & $5.16 \cdot 10^{-3}$ & $0.125$ & $3.76$ & $0.142 \pm 0.028$ \\
\addlinespace[0.15em]
$(6, 1)$ & $\tfrac12$ & $0.544$ & $5.38 \cdot 10^{-4}$ & $0.047$ & $1.41$ & $0.055 \pm 0.023$ \\
\addlinespace[0.15em]
$(11, 2)$ & $\tfrac{6}{11}$ & $0.674$ & $1.27 \cdot 10^{-5}$ & $0.005$ & $0.14$ & -- \\
\addlinespace[0.15em]
$(7, 1)$ & $\tfrac37$ & $0.269$ & $2.46 \cdot 10^{-6}$ & $0.004$ & $0.11$ & -- \\
\addlinespace[0.15em]
$(14, 3)$ & $\tfrac{9}{14}$ & $0.882$ & $7.13 \cdot 10^{-6}$ & $0.003$ & $0.09$ & -- \\
\addlinespace[0.15em]
$(15, 2)$ & $\tfrac25$ & $0.112$ & $1.29 \cdot 10^{-6}$ & $0.002$ & $0.07$ & -- \\
\addlinespace[0.15em]
$(19, 3)$ & $\tfrac{9}{19}$ & $0.453$ & $2.90 \cdot 10^{-6}$ & $0.002$ & $0.06$ & -- \\
\addlinespace[0.15em]
$(21, 4)$ & $\tfrac47$ & $0.734$ & $5.49 \cdot 10^{-6}$ & $0.002$ & $0.06$ & -- \\
\bottomrule
\end{tabular}}
\captionof{table}{Seed selection on the converged NCSX equilibrium, $\mesh{32}{64}{64}$, $p = 2$, $h_r = 0.033$.
Every rational surface in the range of $\iota$ is seeded with the profile $\aopt(r)$ following the energy criterion~\eqref{eq:seed_amplitude}.
We consider the $28$ resonances with $m \leq n_\theta / 2$ and show the eight largest.
The predicted width $w_{\mathrm{pred}}$ of the seeded island with $\aopt$ by the pendulum estimate~\eqref{eq:pendulum_estimate} is contrasted with the width $w_{\mathrm{meas}}$ measured in the Poincar\'e section of the final state shown in Figure~\ref{fig:islands_seeded32}.
The $\pm$ value is the measurement baseline: the width measured at the same resonance in the unseeded equilibrium.}
\label{tab:seed_selection}
\end{minipage}\hfill
\begin{minipage}[t]{0.40\linewidth}
\vspace{0pt}
\centering
\resizebox{\linewidth}{!}{\includegraphics{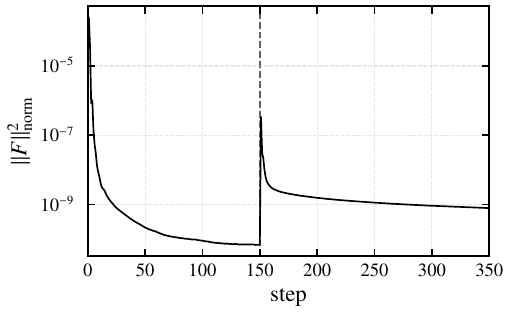}}
\caption{Force residual over the relax--seed--relax sequence (Poincar\'e sections of the state before seeding in Figure~\ref{fig:reconnection_before}, top, and of the final state in Figure~\ref{fig:islands_seeded32}). After reaching an initial floor, the residual spikes as the seed is applied. The force residual floor in the topologically different state is one order of magnitude larger.}
\label{fig:seed_trace}
\end{minipage}
\end{figure}

The exact shape of $r \mapsto a_{mn}(r)$ for all $(m,n)$ are determined using an energy criterion. Considering $\Energy(B + \dB) = \tfrac12 \norm{B + \dB}^2$, the perturbation changes the magnetic energy by
\begin{align}
    \label{eq:seed_energy_criterion}
    \Energy(B + \dB) = \Energy(B) + a \cdot \nabla_a \Energy(B) + \tfrac12 a^T \mathbb G a \quad \text{with} \quad (\nabla_a \Energy(B))_i = \ip{J}{\dA_i} \quad \text{and} \quad \mathbb G_{ij} = \ip{\curl \dA_i}{\curl \dA_j}.
\end{align}
The matrix $\mathbb G$ is symmetric positive definite, so there is a unique value $\aopt$ that minimizes the magnetic energy under the given constraints, namely
\begin{align}
  \label{eq:seed_amplitude}
  \aopt = - \mathbb G^{-1} \nabla_a \Energy(B) .
\end{align}
The minimizer defines, for every $(m, n)$ resonance, a profile $a_{mn}$ localized around $r_{mn}$. 
The amplitudes the criterion returns span four orders of magnitude between resonances. 
Most of them produce islands narrower than a single radial cell, see Table~\ref{tab:seed_selection}.
The criterion is exact for the seeded field, whose energy is quadratic in the amplitudes, but it ignores the relaxation that follows, so the true decrease in energy after relaxing the seeded state is larger.

\begin{figure}[tb]
    \centering
    \includegraphics{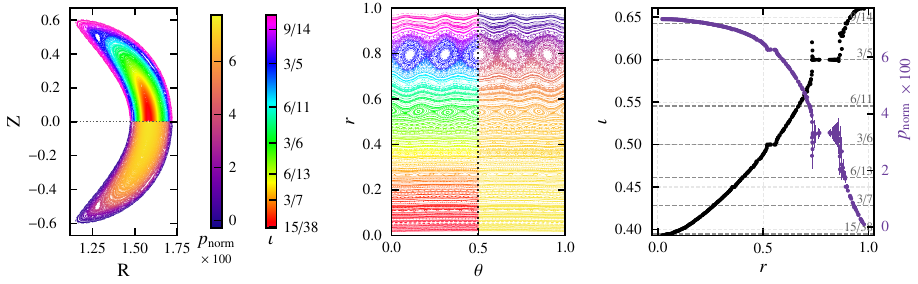}
    \caption{The NCSX case after relaxation (150 Newton steps), island seeding, and relaxation (200 Newton steps). 
    The Poincar\'e section $\zeta = 0$ is shown.
    The final state has $\normnorm{F}^2 = 7.9 \times 10^{-10}$, $\Helicity = 5.0120 \times 10^{-3}$ ($5.0122 \times 10^{-3}$ before seeding), and $\betavol = 4.24\%$ ($\betavol$ changes by an absolute $4 \times 10^{-6}$ compared to the nested equilibrium state).
    The $\iota$ estimation is noisy in the stochastic region around the $\iota = \tfrac35$ island. Pressure is not constant along these field lines either, as can be seen by the error bars indicating the standard deviation of $p$ along them.}
    \label{fig:islands_seeded32}
\end{figure}

\paragraph{Results}
Figure~\ref{fig:seed_trace} shows the force residual over the whole run. It initially relaxes from the VMEC initial condition to the nested equilibrium at $\mesh{32}{64}{64}$. After island seeding (dashed vertical line), the force residual spikes, and following relaxation brings it to the island configuration shown in Figure~\ref{fig:islands_seeded32}.

\begin{figure}[tb]
    \centering
    \includegraphics{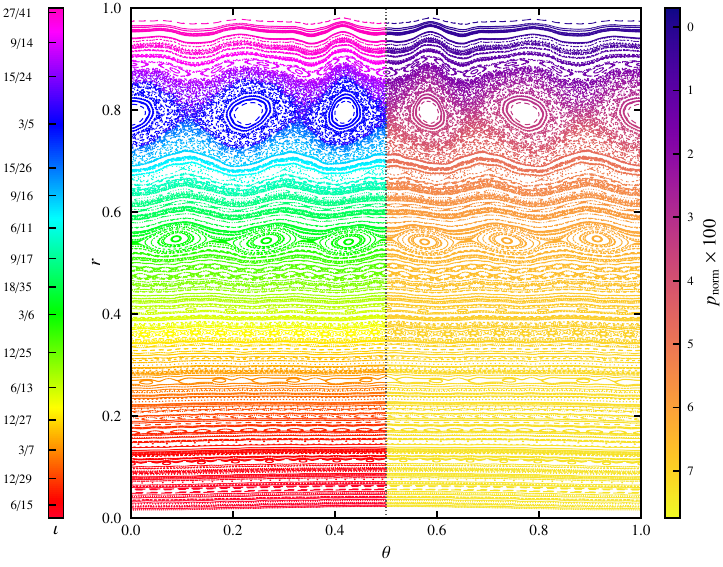}
    \caption{The final state of Figure~\ref{fig:islands_seeded32} in the logical coordinates $(\theta, r)$ at $\zeta = 0.5$. The crossings are coloured by the rotational transform of their field line for $\theta < \tfrac12$ and by its pressure for $\theta \geq \tfrac12$.}
    \label{fig:islands_seeded32_logical}
\end{figure}

A Poincar\'e section at $\zeta = 0$ of the final relaxed state is shown in Figure~\ref{fig:islands_seeded32}. Two large island chains at $\iota \in \{\tfrac12, \tfrac35\}$ are visible. The $\iota = \tfrac35$ chain in particular is surrounded by a stochastic region. Several smaller, sub-grid island chains at $\iota \in \{\tfrac37, \tfrac{6}{11}, \tfrac{9}{14}, \dots\}$ are also visible. Figure~\ref{fig:islands_seeded32_logical} shows the same state in the logical coordinates at $\zeta = 0.5$.

The same relax--seed--relax sequence was repeated in single precision (float32 solves and residuals, inner solves at $10^{-5}$) and in the mixed precision of the reference run, each with 100 Newton steps per relaxation at $\mesh{32}{64}{64}$. Single precision takes 13.5\,s per step against 23.5\,s in mixed precision. Its force residual floors at $\normnorm{F}^2 = 2.6 \times 10^{-10}$ after 76 steps. The energy criterion returns the same seed: $\delta \hat B^r_{51} = 5.17 \times 10^{-3}$ against $5.16 \times 10^{-3}$ and $\delta \hat B^r_{61} = 5.15 \times 10^{-4}$ against $5.38 \times 10^{-4}$, the predicted widths within 1\,\%. After the second relaxation the measured island widths are the same in both precisions.

\paragraph{Island pressure}
\label{par:island_pressure}
Typically, pressure profiles are flattened near the O-point of a magnetic island~\cite{fitzpatrick_helical_1995}, although this depends on the width of the island: smaller islands are better able to maintain a pressure gradient across them. 

If we look at the pressure profiles in Figure~\ref{fig:islands_seeded32}, we see that the spread of $p$ along field lines (the error bars in these plots) is largest for those field lines that lie in magnetic islands and chaotic regions.
This spread of $p$ is a sign that $p$ retains a spatial gradient in $r$.

This is expected.
A discrepancy $p |_{r_\mathrm{in}} - p |_{r_\mathrm{out}}$ between the field line's inner (high pressure) and outer (low pressure) edge leads to a parallel force residual, as $B \cdot F = - B \cdot \grad p$.
If we denote by $\ell$ the length a field line needs to move from $r_\mathrm{in}$ to $r_\mathrm{out}$, this parallel force residual is of order $(p |_{r_\mathrm{in}} - p |_{r_\mathrm{out}}) / \ell$.
The length $\ell$ is large, tens of toroidal transits in our examples.
This is in line with the Cary--Hanson formula~\citep[Equations (44--45)]{cary_simple_1991}, which predicts one full rotation around an O-point every $4/(m|\partial_r \iota|w)$ toroidal transits, with $m$ the poloidal mode number and $w$ the island width.
For this reason, the parallel force is small, and its impact on the volume force residual is smaller still by the island's volume fraction.

\section{Resistive relaxation}
\label{sec:resistive}

The relaxation of Sections~\ref{sec:magnetic_relaxation}--\ref{sec:reconnection} is ideal and, up to grid-scale reconnection, preserves the topology of the field it is given.
Another way to break topology is by introducing finite resistivity $\eta > 0$. We discuss the impact of resistivity and how we can guide relaxation with external current in Section~\ref{subsec:resistive_variations} and provide a numerical example in Section~\ref{subsec:reconnection_demo}.

\subsection{Resistive variations}
\label{subsec:resistive_variations}
Occasionally, it is useful in a magnetic relaxation scheme to induce some reconnection via magnetic diffusion. An external current $J^\star$ can be used to guide the state to a given target state. 

\paragraph{Resistive variations}
This modifies the variations to
\begin{align}
    \delta_v B = \curl (v \times B - \eta (J - J^\star))
\end{align}
with $J = \curl B \in \Hcurl_0(\Domain)$ as before. Our approach of guiding relaxation using an external current follows Huang et al.~\cite{huang_voigt_2025}. The external electric field $-\eta J^{\star}$ keeps the solution from decaying to a vacuum field.

\begin{lemma}[Resistive variations]
    \label{lem:resistive_variations}
    For $\delta_v B = \curl (v \times B - \eta (J - J^\star))$ with $J^\star \in \Hcurl_0(\Domain)$,
    \begin{align}
        \delta_v \Energy(B) = - \ip{J \times B}{v} - \eta \ip{J}{J - J^\star},
        \quad \text{and} \quad
        \delta_v \Helicity(B) = - 2 \eta \, \ip{J - J^\star}{B} .
    \end{align}
\end{lemma}
See~\ref{sec:proofs} for the proof.

While magnetic diffusion can drastically change the small-scale topology of the field, the global helicity can persist through much longer time scales, see~\citep[Remark 7.19]{arnold_topological_2021}.

\subsection{Reconnection between distinct relaxed states}
\label{subsec:reconnection_demo}

We demonstrate the crossing of topological barriers using magnetic diffusion by a numerical experiment.

\begin{figure}[tb]
    \centering
    \includegraphics{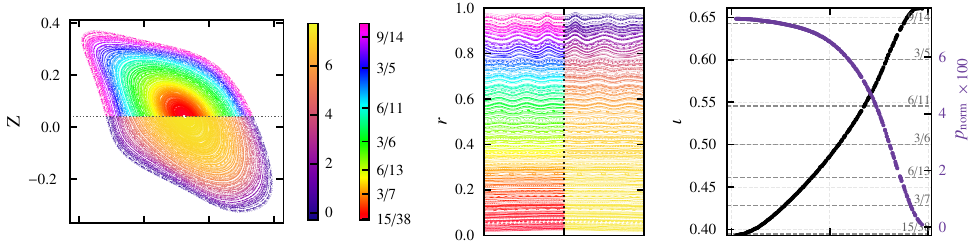} \\
    \includegraphics{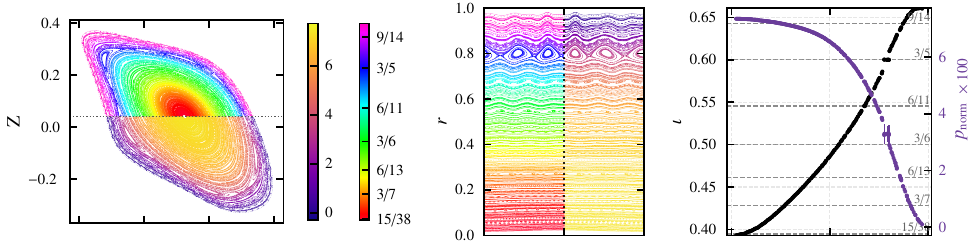} \\
    \includegraphics{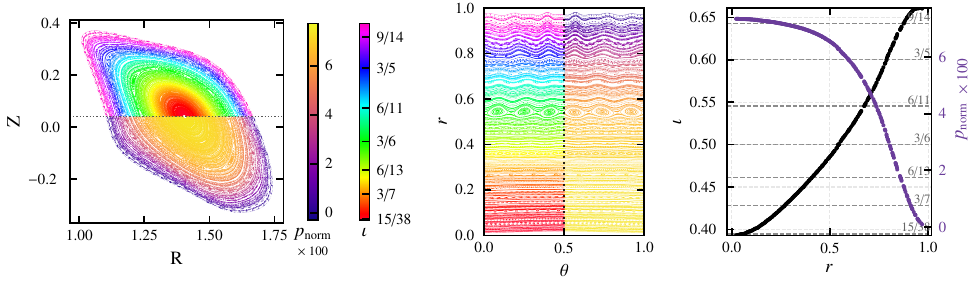}
    \caption{The three converged equilibria before the resistive phase, Poincar\'e sections at $\zeta = 0.375$: nested (top), seeded with the $\tfrac35$ chain (middle) and with the $\tfrac12$ chain (bottom).}
    \label{fig:reconnection_before}
\end{figure}

Three converged states in NCSX geometry are the starting points:
(i) the nested state $B_{\mathrm{nested}}$, the reference run at $\mesh{32}{64}{64}$ after 150 Newton steps (Figure~\ref{fig:reconnection_before}, top),
(ii) the same state, seeded at the $\tfrac35$ island chain alone at $0.25 \times \aopt$ (at the full $\aopt$, this seed produces a chain of width $0.14$ in a stochastic band, as in Figure~\ref{fig:islands_seeded32}), and
(iii) the same state, seeded at the $\tfrac12$ island chain alone at $\aopt$ determined by~\eqref{eq:seed_amplitude}.
All states are relaxed to a force residual floor with 200 Newton steps.

Each of the three is then run for $200$ steps with $\eta \, \delta t = 0.064 \, \vavg{g_{rr}} \, h_r^2 = 1.5 \times 10^{-5}$, followed by $100$ ideal Newton steps.
We pick $J^{\star} = \widetilde{\curl} \, B^\star = \widetilde{\curl} \, (\Id + \varepsilon \curl \widetilde{\curl})^{-1} B_{\mathrm{nested}}$ with $\varepsilon = 0.03 \, \vavg{g_{rr}} \, h_r^2 = 6.9 \times 10^{-6}$ as in the velocity smoothing from Section~\ref{subsec:discrete_relaxation}. 
As a result, all three states will evolve towards a common equilibrium state. 
Smoothing $B_{\mathrm{nested}}$ to obtain $B^\star$ means that the nested state can also open islands.
Without the smoothing step, we would simply have $\widetilde{\curl} \, B_{\mathrm{nested}} - J^{\star} = 0$ and no dynamics for case (i).
The state that the three runs approach is therefore set by $J^\star$, i.e.\ by $B_{\mathrm{nested}}$ and the smoothing $\varepsilon$, while $\eta \, \delta t$ only sets the rate at which it is approached.

\begin{figure}[tb]
    \centering
    \includegraphics{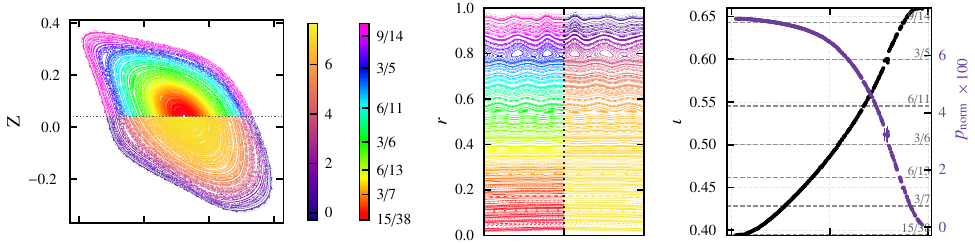} \\
    \includegraphics{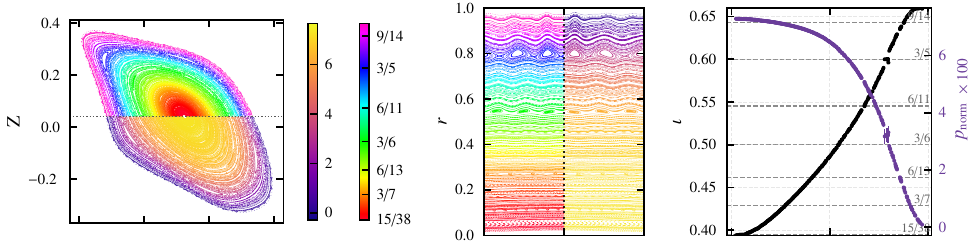} \\
    \includegraphics{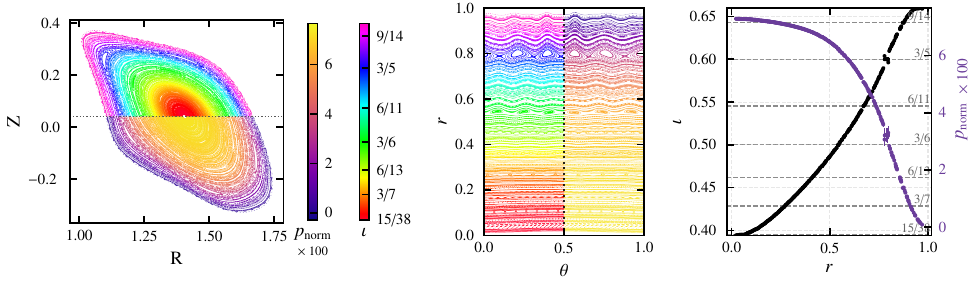}
    \caption{The states from Figure~\ref{fig:reconnection_before} after the resistive reconnection phase and subsequent relaxation to equilibrium, at $\zeta = 0.375$. All three runs have converged to essentially the same state.}
    \label{fig:reconnection_after}
\end{figure}

Table~\ref{tab:reconnection_demo} lists the force residual, $\betavol$, the helicity and the island widths at the two
resonances at the three stages of the three relaxation runs. Figure~\ref{fig:reconnection_before} and Figure~\ref{fig:reconnection_after}
show the Poincar\'e sections before the resistive phase and at the end of the run. 
They illustrate that the three cases all land on the same state. As some current is dissipated, helicity and plasma beta drop accordingly by $9 \times 10^{-4}$ and $1.4\%$ (both relative) in the resistive phase. The resistive phase thus allows the system to forget the initial condition.

\begin{table}[htb]
\caption{
    Diagnostics for the three reconnection runs. 
    In the resistive phase, the $\iota = \tfrac35$ chain converges to a width of $0.058$--$0.061$ in all three runs, opening where it was closed and shrinking where it was seeded. The $\iota = \tfrac12$ chain closes to the measurement baseline in all three runs. 
    The force residual after the resistive phase is set by $\eta$ and determined by $\curl( v \times B - \eta (J - J^\star) ) = 0$, hence $v \neq 0$.
    }

\centering
\begin{tabular}{l l c c c c c}
\toprule
seed & phase & $\normnorm{F}^2$ & $\betavol$ & $\Helicity \times 10^{3}$ & $w_{3/5}$ & $w_{1/2}$ \\
\midrule
none & after initial relaxation & $6.9 \times 10^{-11}$ & $0.0424$ & $5.0122$ & $0.028 \pm 0.028$ & $0.023 \pm 0.023$ \\
 & after resistive phase & $5.3 \times 10^{-10}$ & $0.0418$ & $5.0077$ & $0.058 \pm 0.028$ & $0.025 \pm 0.023$ \\
 & after final relaxation & $9.0 \times 10^{-11}$ & $0.0418$ & $5.0077$ & $0.061 \pm 0.028$ & $0.023 \pm 0.023$ \\
\midrule
$(5, 1)$, $\iota = \tfrac35$ & after initial relaxation & $1.1 \times 10^{-10}$ & $0.0424$ & $5.0122$ & $0.078 \pm 0.028$ & $0.021 \pm 0.023$ \\
 & after resistive phase & $5.4 \times 10^{-10}$ & $0.0418$ & $5.0076$ & $0.061 \pm 0.028$ & $0.025 \pm 0.023$ \\
 & after final relaxation & $9.4 \times 10^{-11}$ & $0.0418$ & $5.0076$ & $0.063 \pm 0.028$ & $0.023 \pm 0.023$ \\
\midrule
$(6, 1)$, $\iota = \tfrac12$ & after initial relaxation & $5.2 \times 10^{-11}$ & $0.0424$ & $5.0122$ & $0.030 \pm 0.028$ & $0.052 \pm 0.023$ \\
 & after resistive phase & $5.3 \times 10^{-10}$ & $0.0418$ & $5.0077$ & $0.059 \pm 0.028$ & $0.025 \pm 0.023$ \\
 & after final relaxation & $9.0 \times 10^{-11}$ & $0.0418$ & $5.0077$ & $0.061 \pm 0.028$ & $0.023 \pm 0.023$ \\
\bottomrule
\end{tabular}

\label{tab:reconnection_demo}
\end{table}

\section{Discussion and Outlook}
\label{sec:discussion}

In this work, we presented a magnetic relaxation method in the spirit of~\cite{chodura_3d_1981} and~\cite{hirshman_siesta_2011} in order to compute MHD equilibria without assuming nested flux surfaces.
Open questions and future work are discussed in Section~\ref{subsec:convergence} and Section~\ref{subsec:future_work}.

To summarize, this work has demonstrated:
\begin{itemize}
    \item A principled mathematical approach, based on the discrete Hodge decomposition on non-simply connected domains, to perform magnetic relaxation.
    \item Gradients of the vacuum field with respect to every coefficient of the map by automatic differentiation, used to optimize a stellarator for quasi-axisymmetry (Section~\ref{subsec:shape_optimization}).
    \item Very computationally efficient equilibrium calculations without the assumption of nested flux surfaces (a finite-$\beta$ NCSX configuration with $1.8 \times 10^5$ degrees of freedom can be computed to $10^{-8}$ normalized force balance in $5.4$ minutes, see Table~\ref{tab:newton_convergence}).
    \item A codebase that can be run on CPU, GPU, and TPU architectures and allows automatic differentiation for future incorporation into stellarator optimization routines.
\end{itemize}

\subsection{Open questions about convergence}
\label{subsec:convergence}
The continuous problem has no smooth limit in general (Section~\ref{subsec:rational_surfaces}), and on a mesh the descent stalls at a floor of the force residual that is not a stationary point of the discrete energy. Beyond this floor, the energy keeps decreasing while the residual stays flat or slowly increases (Figure~\ref{fig:newton_h_sweep}). We attribute this to grid-scale reconnection (cf.\ Figure~\ref{fig:newton_sections}).
The floor itself is robust, between $4 \times 10^{-11}$ and $1.5 \times 10^{-10}$ for $n_r = 16$ to $32$ (the two larger runs are still descending), with the helicity constant to $10^{-5}$ (Table~\ref{tab:newton_convergence}).

The second question concerns uniqueness and well-posedness.
The three states of Section~\ref{subsec:reconnection_demo} start from the same configuration, differ only in their island content, and all relax to the floor with $\betavol$ and helicity equal to four digits.
The initial topology decides which configuration is reached. 
This memory of the initial state or hysteresis can only be overcome with (resistive) reconnection (Figure~\ref{fig:reconnection_after}).

The amount and the form of the reconnection are therefore physics inputs of a non-nested equilibrium code, in the same way that the pressure and current profiles are inputs of a nested one.
Up to grid-scale reconnection, the initial condition determines the magnetic topology of the field.
Whether another input exists that is sufficient to fully determine the final state is open. 
This question would be of large interest when non-nested codes are run to match experimental data.

\subsection{Future work}
\label{subsec:future_work}
There are several useful directions for future work. 

\paragraph{Outer loop optimization}
Section~\ref{subsec:shape_optimization} differentiates the vacuum field with respect to the map. For finite $\beta$, the derivative of the relaxed state $\hat B_\parameter$ in~\eqref{eq:shape_optimization} follows from the implicit function theorem applied to the discrete force balance at an equilibrium reached from a given initial condition.
With this functionality in place, the presented code could be used as a back-end for equilibrium calculation in {SIMSOPT}~\cite{landreman2021simsopt} and other stellarator optimization suites.

\paragraph{Free boundary solves}
Free boundary solves with coils are the most robust check for designing the magnetic fields of a stellarator experiment.
We are able to replace the boundary condition $B \cdot n = 0$ by $(B - B_{\mathrm{vac}}) \cdot n = 0$ when we set $v = 0$ on the boundary.
This can be done with a penalty method and will allow us to move beyond the case where the computational boundary is a conducting wall or a flux surface~\cite{kaptanoglu_computational_2026}. 

\paragraph{Comparison of fixed-boundary codes for finite-$\beta$}
We noted in the introduction and Table~\ref{tab:codes} that HINT, SPECTRE, SIESTA are in use for fixed-boundary finite-$\beta$ stellarator equilibrium calculations.
A forthcoming publication will perform comparisons between these codes and MRX.  

\paragraph{More general maps}
All maps in this work are of the form from~\eqref{eq:stellarator_map}.
Stellarator symmetry is exploited as described in Section~\ref{subsec:manufactured_vacuum} (Table~\ref{tab:symmetry_vacuum}).
Future work will illustrate extension of the code to maps that are more general (e.g., devices where the magnetic axis is parallel to $e_Z$) and not dependent on an initial flux surface solution, using a re-implementation of map2disc~\citep{babin2025construction} in JAX.

\section*{Acknowledgments}
We would like to thank Omar Maj for pointing out the work of Chodura \& Schlüter and many enlightening discussions on the topic over the last years.
The authors would also like to thank Robert Babin, Martin Campos-Pinto, Mark Cianciosa, Yaman Güçlü, Christopher Hansen, Florian Hindenlang, Matt Landreman, Elizabeth Paul, Francesco Patrizi, Benjamin Peherstorfer, Stefan Possanner, Matteo Raviola, Georg Stadler, and Vlad Vicol for valuable discussions.

This work was supported through grants from the Simons Foundation under award 560651 and performed using computational resources sponsored by the U.S. Department of Energy’s Office of Critical Minerals and Energy Innovation and located at the National Laboratory of the Rockies. Additional computations were done using New York University's Greene and Torch HPC clusters.
We gratefully acknowledge the support of a Google TPU Research Award for providing funding and TPU computing resources.

\section*{Declaration of generative AI and AI-assisted technologies in the writing process}
Large language models developed by Anthropic PBC and OpenAI PBC were used throughout this work as coding assistants, for implementing and refactoring parts of the MRX code base and its test suite, for setting up, bookkeeping, and analyzing the numerical experiments, and for generating figures and tables from the run records. 
They were also used to proofread the manuscript and to check derivations. 
All code and text was reviewed and edited by the authors, who take full responsibility for the content of this article.

\section*{Declaration of competing interest}
The authors declare that they have no known competing financial interests or personal relationships that could have appeared to influence the work reported in this paper.

\section*{Data availability}
The MRX code is openly available at \url{https://github.com/ToBlick/mrx}.

% CRediT author statement
\section*{CRediT authorship contribution statement}
\textbf{Tobias Blickhan:} Conceptualization, Investigation, Methodology, Software, Writing.
\textbf{Julianne Stratton:} Data curation, Investigation, Writing.
\textbf{Alan A. Kaptanoglu:} Conceptualization, Funding acquisition, Writing.

\appendix
\makeatletter\@addtoreset{figure}{section}\@addtoreset{table}{section}\makeatother % floats fix

\section{Notation}
\label{sec:notation}
Tables~\ref{tab:notation_1}--\ref{tab:notation_2} list the notation of this work.
% Notation of mrx_jcp.tex, ALPHABETICAL (Tobias 2026-09-23): the conventions first, as the bare decorations (hat,
% star, super- and subscripts), then Latin a-z, A-Z, typewriter, blackboard bold, calligraphic, fraktur, Greek (lower,
% upper), operators. One row per family (Tobias: "if things are basically the same, summarize them"): variants that a
% convention explains (a subscript 0, a hat, a step n) get no row of their own, and a glyph with a second, unrelated
% meaning keeps one row with "also". Earlier versions in attic_2026-09-23/pre_qa_rewrite/.

\begin{table}[htbp]
\centering
\footnotesize
\setlength{\tabcolsep}{4pt}
\begin{tabular}{@{}l l l@{}}
\toprule
Symbol & Meaning & Defined \\
\midrule
\multicolumn{3}{@{}l}{\emph{Conventions}} \\
$\hat{\phantom{x}}$ & logical quantity, pulled back to $\hat \Domain$ (not written on $r$, $\theta$, $\zeta$, $\Mapping$, $\Lambda^k_i$) & \ref{subsec:domain_mapping} \\
${}^\star$ & reference solution, target, or optimal value & \ref{subsec:diagnostics} \\
${}^k$, ${}_k$ & form degree $k \in \{0, 1, 2, 3\}$ & \ref{subsec:derham_complex} \\
${}_0$ & homogeneous boundary conditions on a space or the initial state of a field & \ref{subsec:derham_complex}, \ref{par:maps_initial_fields} \\
${}_h$ & discrete quantity & \ref{subsec:discretization} \\
${}^i$, ${}_i$ & contravariant component, covariant component & \eqref{eq:covariant_contravariant} \\
$\widetilde{\phantom{xxxx}}$ & weak derivative, $(-1)^k \mathbb M_{k-1}^{-1} \mathbb D_{k-1}^T \mathbb M_k$ & \ref{subsec:implementation} \\
$\mathtt u$ & typewriter font: coefficients of a spline expansion, $u_h = \sum_i \mathtt u_i \Lambda^k_i$ & \ref{subsec:discretization} \\
$\mathbb M$ & blackboard font: matrices & \ref{subsec:implementation} \\
$\mathfrak h$ & harmonic forms & \ref{subsec:hodge_decomposition} \\
\addlinespace
\multicolumn{3}{@{}l}{\emph{Operators and brackets}} \\
$\grad$, $\curl$, $\dvg$ & strong derivatives & \eqref{eq:derham_complex} \\
$\ip{u}{v}$, $\norm{u}$, $|u|$ & $L^2$ inner product and norm, point-wise norm & \eqref{eq:inner_product} \\
$\vavg{u}$, $\langle u \rangle_{\theta, \zeta}$ & average over $\Domain$, over a logical surface & \eqref{eq:inner_product}, \eqref{eq:qa_criterion} \\
$\normrel{\cdot}$ & error relative to a reference & \ref{subsec:diagnostics}, \ref{subsec:discrete_relaxation} \\
$\nabla_{u} Q$ & $L^2$ gradient of $Q$ with respect to $u$ & Prop.~\ref{prop:shape_gradient} \\
\addlinespace
\multicolumn{3}{@{}l}{\emph{Latin, lower case}} \\
$a$ & amplitude of island seeds & \eqref{eq:pendulum_estimate}, \eqref{eq:seed_energy_criterion} \\
$\rd^k$, $\rd^*_k$ & differential and co-differential of the Hilbert complex & \eqref{eq:derham_complex_general}, \eqref{eq:codifferential} \\
$d_{\mathrm{RMS}}$ & RMS distance of a boundary to that of the interpolated device, $\langle \mathrm{dist}(\cdot, \partial \Domain_{\mathrm{VMEC}})^2 \rangle_{\theta, \zeta}^{1/2}$ at $r = 1$ & \ref{subsec:shape_optimization} \\
$e_i$ & unit vectors & \ref{subsec:hodge_decomposition} \\
$g_{ij}$, $g^{ij}$, $\sqrt g$ & metric $D\Mapping^T D\Mapping$, its inverse, $\det D\Mapping$ & \ref{subsec:domain_mapping} \\
$h$, $h_c$ & mesh size, knot spacing in direction $c \in \{r, \theta, \zeta \}$, $h_r = 1 / (n_r - p_r)$ & Fig.~\ref{fig:analytic_vacuum}, \ref{subsec:discretization} \\
$\ell$ & parameter along a field line & \ref{subsec:seeding} \\
$(m, n)$ & poloidal mode number and toroidal turns per field period, toroidal mode number $\nfp{} n$ & \ref{subsec:diagnostics} \\
$n$ & outward unit normal on $\partial \Domain$ & \ref{subsec:problem} \\
$n_c$, $\mesh{n_r}{n_\theta}{n_\zeta}$ & number of splines of $V^0$ per direction $c \in \{r, \theta, \zeta \}$ & \ref{subsec:discretization} \\
$\nfp$ & number of field periods & \eqref{eq:stellarator_map} \\
$p$, $p_{\mathrm{norm}}$ & pressure, $p$ over the mean magnetic pressure or the spline degree & \eqref{eq:pressure_problem}, \ref{subsec:discretization} \\
$p_{\mathrm{profile}}$, $\iota_{\mathrm{profile}}$ & profiles $[0,1] \to \bR$ prescribed to a nested-surface solver & \ref{subsec:special_cases} \\
$r$, $\theta$, $\zeta$, $\hat x$ & logical radius and angles, $r_{mn}$ the $(m, n)$ rational surface & \ref{subsec:domain_mapping}, \ref{subsec:seeding} \\
$s$ & normalized toroidal flux inside a logical surface & \eqref{eq:flux_ratio_iota} \\
$t$ & time, parameter of a flow and of a traced field line $x_t$ & \ref{subsec:diagnostics} \\
$v$, $v_t$, $v_\parallel$ & velocity, of a flow, along $B$ & Def.~\ref{def:admissible_variations}, \ref{subsec:newton} \\
$\mathrm{vec}$ & flattening $(j, k, l) \mapsto i$ of the tensor-product indices & \ref{subsec:discretization} \\
$w$ & island width & \eqref{eq:pendulum_estimate} \\
$x$, $\hat x$ & point in the physical/logical domain & \ref{subsec:problem}, \ref{subsec:domain_mapping} \\
\addlinespace
\multicolumn{3}{@{}l}{\emph{Latin, upper case}} \\
$A$ & vector potential, $\curl A = B - \Pi^{\mathfrak H^2_0} B$ & Def.~\ref{def:helicity}, \eqref{eq:vmec_vector_potential} \\
$B$, $H$ & magnetic field as a $2$-form, as a $1$-form & \eqref{eq:mhs}, \ref{subsec:manufactured_vacuum} \\
$C$, $C_\iota$, $C_P$ & CFL constant, also the scaled constraints of the shape optimization & \eqref{eq:cfl}, \ref{subsec:shape_constraints} \\
$E$, $E_\sigma$ & electric field $\Pi^1_0 (v \times B)$, helicity-corrected & \eqref{eq:relaxation_step}, \eqref{eq:helicity_remainder} \\
$F$, $\normnorm{F}$ & force $J \times B - \grad p$, normalized force residual & \eqref{eq:force_residual} \\
$\Hone$, $\Hcurl$, $\Hdiv$, $L^1$, $L^2$ & Sobolev spaces with weak $\grad$, $\curl$, $\dvg$, integrable and square-integrable fields & \ref{subsec:derham_complex} \\
$J$ & current density $\curl B$ & \eqref{eq:mhs} \\
$N^0_{c, i}$, $S^p$ & 1D B-spline $i$ in direction $c$, spline space of degree $p$ & \ref{subsec:discretization} \\
$P$ & normal-field fraction of a vacuum field on the logical surfaces & \eqref{eq:qa_criterion} \\
$Q$, $Q_{\mathrm{QA}}$, $q$ & quality criterion, quasi-axisymmetry residual, $Q$ over its value on the VMEC map & \eqref{eq:shape_optimization}, \eqref{eq:qa_criterion}, \eqref{eq:qa_residual} \\
$(R, \phi, Z)$ & cylindrical coordinates, $R$ and $Z$ also as functions of $\hat x$ & \ref{subsec:domain_mapping}, \eqref{eq:stellarator_map} \\
$U$ & adjoint potential in $V^1_0$ & Prop.~\ref{prop:shape_gradient} \\
$V^k$, $V^k_*$ & spaces of the (dual) de~Rham complex & \eqref{eq:derham_complex_general}, \eqref{eq:codifferential}, \ref{sec:stellarator_symmetry} \\
\addlinespace
\bottomrule
\end{tabular}
\caption{Notation (1/2).}
\label{tab:notation_1}
\end{table}

\begin{table}[htbp]
\centering
\footnotesize
\setlength{\tabcolsep}{4pt}
\begin{tabular}{@{}l l l@{}}
\toprule
Symbol & Meaning & Defined \\
\midrule
\multicolumn{3}{@{}l}{\emph{Calligraphic}} \\
$\mathcal A$, $\| v \|_{\mathcal A}$ & descent operator and induced norm & Prop.~\ref{prop:energy_evolution} \\
$\Energy$, $\nabla_a \Energy$ & magnetic energy $\tfrac12 \norm{B}^2$, gradient with respect to $a$ & \eqref{eq:energy}, \eqref{eq:seed_energy_criterion} \\
$\mathcal F_t$ & flow (of $v_t$) & Lem.~\ref{lem:admissible_flows} \\
$\Helicity$ & generalized helicity & \eqref{eq:helicity} \\
$\mathcal L^k$, $\mathbb L_k$ & Hodge--Laplacian on $V^k$, matrix discretization & \eqref{eq:hodge_laplacian}, \eqref{eq:hodge_laplacian_matrix} \\
$\mathcal M_\parallel$ & parallel-flow penalty operator & \eqref{eq:parallel_penalty} \\
$\mathcal S$, $\mathsf S$ & stellarator symmetry operators & \ref{sec:stellarator_symmetry} \\
\addlinespace
\multicolumn{3}{@{}l}{\emph{Greek, lower case}} \\
$\parameter$ & parameters (of the map $\Phi_\parameter$) & \ref{subsec:goals} \\
$\betavol$ & volume-averaged beta & \ref{subsec:diagnostics} \\
$\gamma$ & smoothing order & \ref{subsec:discrete_relaxation} \\
$\delta_v$, $\delta^2_{u, v}$ & first and second variation in direction $v$, $\delta_v B = \curl (v \times B)$ & Def.~\ref{def:admissible_variations}, \eqref{eq:Hessian} \\
$\delta t$, $\delta t_*$ & time step, exact line-search step & \eqref{eq:relaxation_step}, \eqref{eq:line_search} \\
$\dA_{mn}$ & seed perturbation at $(m,n)$ resonance, $\delta \hat B^r_{mn}$ its normal-field harmonic & \ref{subsec:seeding}, \eqref{eq:pendulum_estimate} \\
$\varepsilon$ & smoothing strength & \ref{subsec:discrete_relaxation} \\
$\eta$ & resistivity & \ref{subsec:resistive_variations} \\
$\theta^* = \theta + \lambdaup$ & straight field line angle & \ref{subsec:diagnostics} \\
$\iota$, $\iotabar$, $\langle \iotabar \rangle_s$ & rotational transform: traced, flux ratio, its mean over $s$ & Def.~\ref{def:rotational_transform} \\
$\kappa$ & parallel-flow penalty & \eqref{eq:parallel_penalty} \\
$\lambda$ & eigenvalue (Taylor state, Poincar\'e constant, etc.) & \ref{subsec:existing_codes}, \eqref{eq:arnold_inequality}, \ref{subsec:gradient_descent} \\
$\lambdaup$ & angle stream function of the nested-surface solver & \eqref{eq:vmec_vector_potential} \\
$\mu$, $\nu_\iota$, $\nu_P$ & penalty parameter and multipliers of the augmented Lagrangian & \eqref{eq:qa_augmented_lagrangian} \\
$\xi$ & random normal displacement of the boundary of a start & \ref{subsec:shape_constraints} \\
$\sigma$ & helicity correction & \eqref{eq:helicity_remainder} \\
$\tau$ & coefficient of the harmonic form $\mathfrak h$ in a field & \eqref{eq:force_decomposition}, \eqref{eq:newton_system} \\
$\chi_\parallel$ & strain of $B$ along the field & \eqref{eq:parallel_penalty} \\
$\psi$, $\Psi_{\mathrm t}$, $\Psi_{\mathrm p}$ & flux label, toroidal and poloidal flux over $2\pi$ & \ref{subsec:special_cases}, \eqref{eq:vmec_vector_potential}, \eqref{eq:qa_residual} \\
$\omega$ & potential of the velocity $\curl \omega + \tau \mathfrak h$ & \eqref{eq:force_decomposition}, \eqref{eq:newton_system} \\
\addlinespace
\multicolumn{3}{@{}l}{\emph{Greek, upper case}} \\
$\Gamma_t(\ell)$ & field line of $B_t$ with parameter $\ell$ & Prop.~\ref{prop:topology_preservation} \\
$\Lambda^k_i$ & basis functions of $V^k$ & \ref{subsec:discretization} \\
$\Pi^k$, $\Pi^{\mathfrak H^k}$, $\Leray$ & $L^2$ projection onto $V^k$, onto $\mathfrak H^k$, onto divergence-free fields & \ref{subsec:discretization}, Def.~\ref{def:harmonic_forms}, \eqref{eq:pressure_problem} \\
$\Mapping$, $D \Mapping$, $\Mapping^k_*$ & map $\hat \Domain \to \Domain$ ($\Mapping_\parameter$ for the parameters $\parameter$), Jacobian matrix, push-forward & \ref{subsec:domain_mapping}, \eqref{eq:pushforward} \\
$\hat \Domain$, $\Domain$, $\partial \Domain$, $|\Domain|$ & logical domain $[0,1]^3$, physical domain, its boundary, its volume & \ref{subsec:problem}, \ref{subsec:domain_mapping} \\
\addlinespace
\multicolumn{3}{@{}l}{\emph{Other}} \\
$\mathfrak H^k$, $\mathfrak h_k$ & harmonic space, harmonic $k$-form ($\norm{\mathfrak{h_k}} = 1$) & \eqref{eq:hodge_decomposition}, Def.~\ref{def:harmonic_forms} \\
$\mathtt R$, $\mathtt Z$ & coefficients of $R$ and $Z$ as functions of $\hat x$ & \ref{subsec:shape_optimization} \\
$\mathsf a$, $R_{\mathrm{major}}$ & minor and major radius of the device as in VMEC & \ref{subsec:shape_optimization} \\
$\mathbb D_k$ & incidence matrix of $\grad$, $\curl$, $\dvg$ for $k = 0, 1, 2$ & \ref{subsec:implementation} \\
$\mathbb G$ & Gram matrix & \eqref{eq:seed_energy_criterion} \\
$\mathbb K$ & curl--curl matrix $\mathbb D_1^T \mathbb M_2 \mathbb D_1$ & \ref{subsec:shape_proofs} \\
$\mathbb M_k$ & $k$-form mass matrix & \ref{subsec:implementation} \\
\bottomrule
\end{tabular}
\caption{Notation (2/2).}
\label{tab:notation_2}
\end{table}

\section{Proofs}
\label{sec:proofs}

\begin{proof}[Proof of Lemma~\ref{lem:pushforward_properties}]
    We use the push-forwards~\eqref{eq:pushforward} and~\eqref{eq:pushforward_eb}, $f = \hat f \circ \Mapping^{-1}$, $u \circ \Mapping = D\Mapping^{-T} \hat u$ and $v \circ \Mapping = (\det D\Mapping)^{-1} D\Mapping \, \hat v$, and the logical boundary conditions on $\{ r = 1 \}$: $\hat f = 0$ for $\Hone_0$, vanishing tangential components $\hat u_\theta = \hat u_\zeta = 0$ for $\Hcurl_0$, and a vanishing normal component $\hat v^r = 0$ for $\Hdiv_0$.
    Since $\Mapping^{-1}(\partial \Domain) = \{ r = 1 \}$, it holds for all $x\in\partial \Domain$ that $(\Mapping^0_* \hat f)(x) = \hat f(\Mapping^{-1}(x)) = 0$.
    Next, for $\hat x \in \{ r = 1 \}$, $D\Mapping(\hat x)^T u(\Mapping(\hat x)) = \hat u(\hat x) = (\hat u_r(\hat x), 0, 0)$.
    This implies $\partial_\theta \Mapping \cdot (u \circ \Mapping) = \partial_\zeta \Mapping \cdot (u \circ \Mapping) = 0$.
    Every tangent vector of $\partial \Domain$ is a combination of $\partial_\theta \Mapping$ and $\partial_\zeta \Mapping$, hence orthogonal to $u \circ \Mapping$.
    Lastly, for $\hat x \in \{ r = 1 \}$, $\hat v(\hat x) = (0, \hat v^\theta(\hat x), \hat v^\zeta(\hat x))$, hence ${v \circ \Mapping} = (\det D\Mapping)^{-1} \, D\Mapping \hat v = (\det D\Mapping)^{-1} ( \hat v^\theta \, \partial_\theta \Mapping + \hat v^\zeta \, \partial_\zeta \Mapping).$
    At the same time, the normal vector $n$ is proportional to $\partial_\theta \Mapping \times \partial_\zeta \Mapping$ and therefore orthogonal to $v \circ \Mapping$.
\end{proof}

\begin{proof}[Proof of Lemma~\ref{lem:admissible_flows}]
    The formula for $B_t$ holds at $t=0$. Write $x_t = \mathcal F_t(x)$ for short and take the time derivative
    \begin{align}
        \frac{\rd}{\rd t} \left( B_t(x_t) - D\mathcal F_t(x) B_0(x) \right) &= \partial_t B_t(x_t) + DB_t(x_t) \frac{\rd}{\rd t} x_t - \left( \frac{\rd}{\rd t} D\mathcal F_t(x) \right) B_0(x).
    \end{align}
    Differentiating the flow equation in $x$ gives $\frac{\rd}{\rd t} D\mathcal F_t(x) = Dv_t(x_t) \, D\mathcal F_t(x)$.
    The divergence of $B_t$ is conserved, $\partial_t \dvg B_t = \dvg \curl (v_t \times B_t) = 0$, so $\dvg B_t = 0$ for all $t$.
    With the vector identity $\curl (v_t \times B_t) = Dv_t B_t - DB_t v_t + v_t \dvg B_t - B_t \dvg v_t$ and $\dvg B_t = \dvg v_t = 0$, the first derivative above is $\frac{\rd}{\rd t} B_t(x_t) = \curl(v_t \times B_t) + DB_t v_t = Dv_t(x_t) B_t(x_t)$.
    Hence the difference $\Delta_t(x) = B_t(\mathcal F_t(x)) - D\mathcal F_t(x) B_0(x)$ satisfies the homogeneous linear equation
    \begin{align}
        \frac{\rd}{\rd t} \Delta_t(x) = Dv_t(x_t) \, \Delta_t(x), \qquad \Delta_0(x) = 0 ,
    \end{align}
    whose only solution is $\Delta_t \equiv 0$.
    The two expressions thus agree on every trajectory $x_t$, and since the flow is invertible, for all $x \in \Domain$.
\end{proof}

\begin{proof}[Proof of Lemma~\ref{lem:helicity_balance}]
    We use the discrete vector potential and helicity of Definition~\ref{def:discrete_helicity}, $A^n \in V^1_0$ with $\curl A^n = B^n - \Pi^{\mathfrak H^2_0} B^n$ and $\Helicity^n = \ip{A^n}{B^n + \Pi^{\mathfrak H^2_0} B^n}$, and $\ip{\curl u}{\mathfrak h} = \ip{u}{\widetilde{\curl} \, \mathfrak h} = 0$ for $u \in V^1_0$ and $\mathfrak h \in \mathfrak H^2_0$.
    Let $B_{\mathfrak H} = \Pi^{\mathfrak H^2_0} B^n$, constant in $n$ because $\curl E \perp \mathfrak H^2_0$.
    Then $\curl (A^{n+1} - A^n - \delta t E) = 0$ with $A^{n+1} - A^n - \delta t E \in V^1_0$.
    Hence this expression is a gradient: $A^{n+1} - A^n - \delta t E = \grad f$, $f \in V^0_0$ (as $\dim \mathfrak H^1_0 = 0$ in our case), and $\ip{\grad f}{B^{n+1} + B_{\mathfrak H}} = 0$ since both fields are divergence-free with vanishing normal trace.
    It follows that
    \begin{align}
    \notag
        \Helicity^{n+1} - \Helicity^{n}
        &= \ip{A^{n+1} - A^n}{B^{n+1} + B_{\mathfrak H}} + \ip{A^n}{B^{n+1} - B^n}
        = \delta t \, \ip{E}{B^{n+1} + B_{\mathfrak H}} + \delta t \, \ip{\curl A^n}{E} \notag \\
        &= \delta t \, \ip{E}{B^{n+1} + B_{\mathfrak H}} + \delta t \, \ip{B^n - B_{\mathfrak H}}{E}
        = \delta t \, \ip{E}{B^{n+1} + B^n}
    \end{align}
    by partial integration (with both $A^n$ and $E$ in $V^1_0$) and $\curl A^n = B^n - B_{\mathfrak H}$.
\end{proof}

\begin{proof}[Proof of Proposition~\ref{prop:energy_evolution}]
        We use the relaxation step~\eqref{eq:relaxation_step}, $J = \widetilde{\curl} \, B^n$, $E^n = \Pi^1_0 (v \times B^n)$ and $B^{n+1} = B^n + \delta t \curl E^n$, and the weak curl, $\ip{B}{\curl u} = \ip{\widetilde{\curl} \, B}{u}$ for $u \in V^1_0$.
        From $B^{n+1} = B^n + \delta t \curl E^n$, $\Energy^{n+1} - \Energy^n = \delta t \ip{B^n}{\curl E^n} + \tfrac{1}{2} \delta t^2 \norm{\curl E^n}^2$.
        Since $E^n \in V^1_0$, the definition of the weak curl operator gives $\ip{B^n}{\curl E^n} = \ip{J}{E^n}$.
        As $J \in V^1_0$ and $\Pi^1_0$ is $L^2$-orthogonal,
    \begin{align}
        \ip{J}{E^n} &= \ip{J}{v \times B^n}
        = - \ip{v}{J \times B^n}
        = - \ip{v}{J \times B^n - \widetilde{\grad} \, p},
    \end{align}
    since $v$ is divergence-free.
    Substitute $\ip{v}{J \times B^n - \widetilde{\grad} \, p} = \ip{v}{\mathcal A v}$, which holds for both choices of $\mathcal A$, and the claim follows.
\end{proof}

\begin{proof}[Proof of Lemma~\ref{lem:resistive_variations}]
    We use the first variation~\eqref{eq:first_variation}, $\delta_v \Energy(B) = \ip{B}{\delta_v B} = \ip{J}{v \times B}$ for $\delta_v B = \curl (v \times B)$, the helicity $\Helicity = \ip{A}{B + B_{\mathfrak H}}$ with $\curl A = B - B_{\mathfrak H}$, and $\ip{J}{v \times B} = - \ip{J \times B}{v}$.
    The first identity is Proposition~\ref{prop:first_variation} with $v \times B$ replaced by $v \times B - \eta (J - J^\star)$, since $\ip{J}{\eta (J - J^\star)} = \eta \ip{J}{J - J^\star}$.
    For the second, the harmonic component $B_{\mathfrak H}$ of $B$ is unchanged (the variation is the curl of a field in $\Hcurl_0$), so $\delta_v \Helicity = \ip{\delta_v A}{B + B_{\mathfrak H}} + \ip{A}{\delta_v B}$.
    
    With $\delta_v A = v \times B - \eta (J - J^\star) + \grad f$ for a gauge $f$, partial integration and $\curl A = B - B_{\mathfrak H}$ give $\ip{A}{\delta_v B} = \ip{\curl A}{\delta_v A} = \ip{B - B_{\mathfrak H}}{\delta_v A}$, since the tangential traces of $A$ and of $\delta_v A$ vanish.
    Hence $\delta_v \Helicity = 2 \ip{B}{\delta_v A}$, where $\grad f$ drops out against the divergence-free $B$ with vanishing normal trace. With $\ip{B}{v \times B} = 0$, we arrive at $\delta_v \Helicity = - 2 \eta \ip{J - J^\star}{B}$.
\end{proof}

\section{Stellarator symmetry}
\label{sec:stellarator_symmetry}

Let $\mathsf S$ denote the transformation $(R, \phi, Z) \mapsto (R, -\phi, -Z)$.
A configuration is stellarator symmetric when $\mathsf S \Domain = \Domain$ and the map~\eqref{eq:stellarator_map} satisfies $\Mapping(r, -\theta, -\zeta) = \mathsf S \Mapping(r, \theta, \zeta)$.
In this case half a field period determines the entire geometry.

Let $(\mathcal S u)(x) = \mathsf S \, u(\mathsf S x)$ for vector fields and $(\mathcal S f)(x) = f(\mathsf S x)$ for scalar fields. A field is even if $\mathcal S u = u$ and odd if $\mathcal S u = -u$.
Since $\mathsf S$ is orthogonal with unit Jacobian, the $L^2$ pairing is invariant, $\ip{\mathcal S u}{\mathcal S v} = \ip{u}{v}$, and therefore $\ip{u}{v} = 0$ whenever $u$ is even and $v$ is odd.

The magnetic field of a stellarator-symmetric equilibrium is odd: $\mathcal S B = -B$.
As $\mathsf S$ is a proper rotation, $\mathcal S$ commutes with $\grad$, $\curl$, and $\dvg$.
Furthermore, $\mathcal S (u \times v) = \mathcal S u \times \mathcal S v$.
Hence $J$ and $A$ are odd like $B$. 
The harmonic form $\mathfrak h$ is odd as well.
The Lorentz force $J \times B$, the pressure, the force $F = J \times B - \grad p$ and the descent velocity $v$ are even.
Note that $E = v \times B$ is odd (even $\times$ odd), as it has to be since $\partial_t B = \curl E$.

\begin{proof}[Proof of Proposition~\ref{prop:no_harmonic_force}]
    $J \times B$ and $F$ are even, $\mathfrak h$ is odd, hence $L^2$-orthogonality makes both pairings vanish at every step.
\end{proof}

Discretely, $(\theta, \zeta) \mapsto (-\theta, -\zeta)$ is a signed permutation of the degrees of freedom of every $V^k$.
Each space splits into its parity classes, $V^k = V^k_+ \oplus V^k_-$, for example $B \in V^2_-$ and $v \in V^2_+$. 
The parity classes hold half the degrees of freedom each, up to potential fixed points on the planes $\zeta = 0, 1/2$. 
Numerical quadrature need only cover half a field period.

Any fields without a definite parity cannot be represented in this setting, just as fields that do not respect the field periods cannot be represented by~\eqref{eq:stellarator_map}.

\section{Shape optimization}
\label{sec:shape_optimization_details}
This section provides additional details on the shape optimization performed in~Section~\ref{subsec:shape_optimization}.

\subsection{Proofs} We begin with the proof of the derivative of the optimization objective.
\label{subsec:shape_proofs}
\begin{proof}[Proof of Lemma~\ref{lem:harmonic_from_seed}]
    We use $\mathfrak h_2 = \mathfrak s - \curl A$ with
    \begin{align}
        \widetilde{\curl} \curl A = \widetilde{\curl} \, \mathfrak s, \qquad \dvg \mathfrak s = 0, \notag
    \end{align}
    the harmonic space $\mathfrak H^2_0 = \{ \mathfrak h \in V^2_0 \colon \dvg \mathfrak h = 0, \ \widetilde{\curl} \, \mathfrak h = 0 \}$, the kernel of the Hodge--Laplacian~\eqref{eq:hodge_laplacian}.
    Since $\dvg \curl = 0$, $\dvg \mathfrak h_2 = \dvg \mathfrak s = 0$, and by~\eqref{eq:harmonic_from_seed}, $\widetilde{\curl} \, \mathfrak h_2 = \widetilde{\curl} \, \mathfrak s - \widetilde{\curl} \curl A = 0$. Hence $\mathfrak h_2 \in \mathfrak H^2_0$.
    It is nonzero: by Stokes' theorem the flux of $\curl A$ through a cross-section $\zeta = \mathrm{const}$ is the circulation of $A$ along its boundary on $\partial \Domain$, which vanishes for $A \in V^1_0$. 
    Consequently, $\mathfrak h_2$ keeps the toroidal flux of $\mathfrak s$, which is nonzero.
\end{proof}

\begin{proof}[Proof of Proposition~\ref{prop:shape_gradient}]
    We use $\mathfrak h_2 = \mathfrak s - \curl A$ with $\widetilde{\curl} \curl A = \widetilde{\curl} \, \mathfrak s$ from Lemma~\ref{lem:harmonic_from_seed}.
    The degrees of freedom $\mathtt s$ of $\mathfrak s$ do not depend on $\parameter$.
    Let $\mathbb M_k$ be the mass matrix of $V^k_0$ on $\Mapping_\parameter$ and $\mathbb D_1$ the incidence matrix of $\curl \colon V^1_0 \to V^2_0$, which does not depend on $\parameter$.
    The weak curl on $V^2_0$, $\widetilde{\curl} = \mathbb M_1^{-1} \mathbb D_1^T \mathbb M_2$, does depend on $\parameter$. 
    In degrees of freedom,~\eqref{eq:harmonic_from_seed} reads $\mathbb K \mathtt a = \mathbb D_1^T \mathbb M_2 \mathtt s$ with $\mathbb K = \mathbb D_1^T \mathbb M_2 \mathbb D_1$, and $\mathtt h = \mathtt s - \mathbb D_1 \mathtt a$. 
    Differentiating with respect to $\parameter$ gives
    \begin{align}
        \mathbb K \, \partial_\parameter \mathtt a = \mathbb D_1^T (\partial_\parameter \mathbb M_2) (\mathtt s - \mathbb D_1 \mathtt a) = \mathbb D_1^T (\partial_\parameter \mathbb M_2) \, \mathtt h, \qquad \partial_\parameter \mathtt h = - \mathbb D_1 \, \partial_\parameter \mathtt a.
    \end{align}
    Let $\mathtt g = \partial Q / \partial \mathtt h$ and let $\mathtt u$ solve $\mathbb K \mathtt u = - \mathbb D_1^T \mathtt g$.
    This is the degree-of-freedom form of $\widetilde{\curl} \curl U = - \widetilde{\curl} \, \nabla_{\mathfrak h_2} Q$, with $\nabla_{\mathfrak h_2} Q = \mathbb M_2^{-1} \mathtt g$.
    Both systems are consistent, since their right-hand sides vanish on $\ker \mathbb K = \ker \mathbb D_1$.
    As $\mathbb K$ is symmetric,
    \begin{align}
        \mathtt g^T \partial_\parameter \mathtt h = - \mathtt g^T \mathbb D_1 \, \partial_\parameter \mathtt a = - (\mathbb D_1 \mathtt g)^T \partial_\parameter \mathtt a = ( \mathbb K \mathtt u )^T  \partial_\parameter \mathtt a = \mathtt u^T \mathbb K \, \partial_\parameter \mathtt a = \mathtt u^T \mathbb D_1^T (\partial_\parameter \mathbb M_2) \, \mathtt h = (\mathbb D_1 \mathtt u)^T (\partial_\parameter \mathbb M_2) \, \mathtt h,
    \end{align}
    which is $\partial_\parameter \ip{\curl U}{\mathfrak h_2}$ in degrees of freedom. 
\end{proof}

\subsection{Augmented Lagrangian and boundary perturbations}
\label{subsec:shape_constraints}
L-BFGS-B minimizes the augmented Lagrangian~\citep[Chapter 17]{nocedal_numerical_2006}
\begin{align}
    \label{eq:qa_augmented_lagrangian}
    \mathsf L_\mu = q + \nu_\iota C_\iota + \frac{\mu}{2} C_\iota^2 + \frac{\max(0, \nu_P + \mu C_P)^2 - \nu_P^2}{2 \mu}, \quad q = \frac{\langle Q_{\mathrm{QA}}(\mathfrak h_2)^2 \rangle_{\Domain \cap \{r \geq h_r\}}}{Q^2_{\mathrm{VMEC}}}, \quad C_\iota = \frac{\langle \iotabar \rangle_s - \iotabar^\star}{10^{-3}}, \quad C_P = \frac{P - 10^{-8}}{10^{-8}},
\end{align}
with $Q^2_{\mathrm{VMEC}}$ the value of $\langle Q_{\mathrm{QA}}(\mathfrak h_2)^2 \rangle_{\Domain \cap \{r \geq h_r\}}$ on the VMEC map.
Every $100$ iterations $\nu_\iota \leftarrow \nu_\iota + \mu C_\iota$ and $\nu_P \leftarrow \max(0, \nu_P + \mu C_P)$, starting from zero.
The parameter $\mu$ starts at $1$ and we take $\mu \leftarrow 10 \, \mu$ if a violation above its tolerance decreases by less than a factor four. 
At convergence, $-\nu_\iota$ and $-\nu_P$ are the derivatives of the optimal $q$ with respect to $\iotabar^\star$, per $10^{-3}$, and to the bound on $P$, per $10^{-8}$~\citep[Chapter 12]{nocedal_numerical_2006}.

For the initial perturbations, we displace the VMEC boundary along its normal by $\xi = \sum_{m \leq 4, |n| \leq 4} a_{mn} \cos 2 \pi (m \theta - n \zeta)$, $a_{mn} \sim \mathcal N(0, 1) / (1 + m^2 + n^2)$, scaled to the given RMS.
The perturbation is rejected if $\det D\Mapping$ falls below $0.2$ times that of the VMEC map or if the boundary crosses itself.

\section{Verification against an analytic finite-$\beta$ equilibrium}
\label{sec:landreman_verification}
The solution family of~\citep{landreman_analytic_2026} gives exact smooth equilibria with nested flux surfaces and finite pressure in closed form.
We take $\varepsilon = 2$, $S = 1$ and $k_b = 0.1$ with the free scale of the pressure set to one.
This leads to $\betavol = 1.79\%$ and $\iota$ between $2.2869$ on the axis and $2.2878$ on the boundary, with $p = 0$ on the boundary.
We use the closed-form flux surfaces as our radial mesh through the route of Section~\ref{par:maps_initial_fields}.
Table~\ref{tab:landreman_convergence} compares the projected exact field with the closed form on the meshes $\mesh{n}{2n}{2n}$.
The squared force residual of the projected field falls from $10^{-4}$ to $5 \times 10^{-9}$ at $p = 2$ and to $1.5 \times 10^{-9}$ at $p = 3$.

No relaxation is performed, this test is purely to verify the validity of our geometry import, initial field processing, and force evaluation.

\begin{table}[htb]
% Landreman's sheared exact equilibrium (arXiv:2609.26742; eps = 2, S = 1, k_b = 0.1, lambda = 1), the projected exact
% field without relaxation: outputs/landreman_verify_2026-09-24/partA_sheared/rung_*/result.json (B_err, resid);
% rates between consecutive meshes in h = 1/(n - p)
\centering
\setlength{\tabcolsep}{4pt}
\begin{tabular}{l cc cc cc cc}
\toprule
 & \multicolumn{4}{c}{$p = 2$} & \multicolumn{4}{c}{$p = 3$} \\
\cmidrule(lr){2-5} \cmidrule(lr){6-9}
mesh & $\normrel{B_h - B}$ & rate & $\normnorm{F}^2$ & rate & $\normrel{B_h - B}$ & rate & $\normnorm{F}^2$ & rate \\
\midrule
$\mesh{8}{16}{16}$ & $2.88 \cdot 10^{-2}$ & -- & $1.11 \cdot 10^{-4}$ & -- & $7.52 \cdot 10^{-3}$ & -- & $2.34 \cdot 10^{-5}$ & -- \\
$\mesh{12}{24}{24}$ & $9.38 \cdot 10^{-3}$ & $2.2$ & $1.46 \cdot 10^{-5}$ & $4.0$ & $2.25 \cdot 10^{-3}$ & $2.1$ & $6.48 \cdot 10^{-6}$ & $2.2$ \\
$\mesh{16}{32}{32}$ & $4.47 \cdot 10^{-3}$ & $2.2$ & $1.77 \cdot 10^{-6}$ & $6.3$ & $8.65 \cdot 10^{-4}$ & $2.6$ & $9.11 \cdot 10^{-7}$ & $5.3$ \\
$\mesh{24}{48}{48}$ & $1.81 \cdot 10^{-3}$ & $2.0$ & $6.72 \cdot 10^{-8}$ & $7.2$ & $1.82 \cdot 10^{-4}$ & $3.2$ & $2.16 \cdot 10^{-8}$ & $7.8$ \\
$\mesh{32}{64}{64}$ & $9.93 \cdot 10^{-4}$ & $1.9$ & $4.64 \cdot 10^{-9}$ & $8.6$ & $6.47 \cdot 10^{-5}$ & $3.2$ & $1.50 \cdot 10^{-9}$ & $8.3$ \\
\bottomrule
\end{tabular}

\caption{The projected exact field of the sheared equilibrium of~\citep{landreman_analytic_2026} against its closed form, without relaxation. The field converges at $h^p$ as expected.}
\label{tab:landreman_convergence}
\end{table}

\section{Additional numerical results}
\label{sec:additional_numerics}

\subsection{Further numerical results: gradient descent}
\label{subsec:further_gradient_descent}

This section shows further numerical results and hyperparameter sweeps for the gradient descent approach.

\paragraph{Helicity preservation}
We verify the helicity balance of Section~\ref{subsec:discrete_relaxation}.
Per step the plain explicit step on $B$ drifts by the remainder~\eqref{eq:helicity_remainder}, which can be compensated by solving for the correction $\sigma$.
This requires operating at high numerical precision, since the drift of a single-precision field is of the order of its own machine epsilon.

Table~\ref{tab:helicity_preservation} shows that in double precision the corrected step conserves the helicity to $3.5 \times 10^{-16}$, machine precision and ten orders of magnitude below the explicit step's $5.7 \times 10^{-6}$.
In mixed precision, where the field and its helicity are stored in single precision, the corrected step drifts by $9.3 \times 10^{-8}$ against $5.0 \times 10^{-6}$ for the plain step. The remainder is the rounding of the stored helicity.
Removed energy differs between the corrected and the explicit step by less than $2\%$ (relative).

\begin{table}[htb]
\centering
\begin{minipage}[t]{0.48\linewidth}
\centering
\caption{Helicity preservation, with and without correction term, throughout gradient descent for different choices of precision.
The correction term reduces the helicity drift to round-off in both precisions.}

\centering
\begin{tabular}{l c c}
\toprule
 & \multicolumn{2}{c}{$|\Delta \Helicity / \Helicity|$} \\
\cmidrule(lr){2-3}
precision & plain step & corrected step \\
\midrule
% 2026-09-22 rerun: reruns_2026-09-22/hel_{mixed,f64}_{B,corr} (half period, current code), float64 unchanged
mixed & $5.02 \times 10^{-6}$ & $9.29 \times 10^{-8}$ \\
float64 & $5.67 \times 10^{-6}$ & $3.46 \times 10^{-16}$ \\
\bottomrule
\end{tabular}

\label{tab:helicity_preservation}
\end{minipage}\hfill
\begin{minipage}[t]{0.48\linewidth}
\centering
\caption{Velocity formulations: potential and Leray projection.
The two formulations agree, but the potential route is faster and extends to Newton's method.}
\centering
\setlength{\tabcolsep}{3pt}
\begin{tabular*}{\linewidth}{@{\extracolsep{\fill}}l c c c c c c@{}}
\toprule
velocity & s/step & $\normnorm{F}^2$
  & \multicolumn{3}{c}{\makebox[0pt]{steps to $\normnorm{F}^2$}}
  & $\Delta \Helicity / \Helicity$ \\
\cmidrule(lr){4-6}
 & & floor $\times 10^{-8}$
  & $10^{-6}$ & $10^{-7}$ & $10^{-8}$
  & $\times 10^{-6}$ \\
\midrule
% 2026-09-22 rerun: potential = table4_2026-09-18/gradient_16 (Tab. 3's gradient row), Leray = reruns_2026-09-22/gd_leray
potential & 0.34 & 9.62 & 443 & 4813 & -- & -7.3 \\
Leray     & 0.49 & 9.63 & 445 & 4819 & -- & -7.1 \\
\bottomrule
\end{tabular*}
\label{tab:velocity_choices}
\end{minipage}
\end{table}

\begin{remark}
    We do, however, see that the effects here are much milder than those shown in~\citep[Section 4.2.2]{he_topology-preserving_2025}, where the scheme without auxiliary variables dissipated the helicity entirely in only $1{,}000$ steps, presumably due to the choice of larger time steps.
\end{remark}

\paragraph{Velocity formulation}

Table~\ref{tab:velocity_choices} compares the two velocity formulations from Section~\ref{subsec:pressure} on the NCSX case.
The residual traces agree to within $0.5\%$ after the first six steps.
The potential route costs $30\%$ less per step ($0.34$ against $0.49$ s), replacing the $k = 3$ Laplacian solve with a $k = 1$ one and the $\varepsilon$-shifted $k = 2$ solve with a $k = 1$ one.

\subsection{Further numerical results: Newton}
\label{subsec:further_newton}

This section presents further numerical results for Newton's descent method.

\begin{figure}
\centering
\begin{minipage}[t]{0.48\linewidth}
\centering
    \resizebox{\linewidth}{!}{\includegraphics{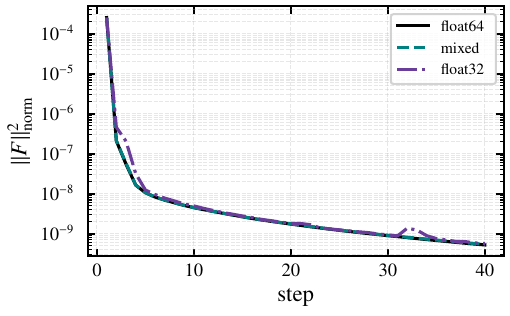}}
    \caption{Newton's method in single (inner solves at $10^{-5}$, $4.8$~min), mixed (single-precision solves, double-precision residual, refined to $10^{-8}$, $7.4$~min) and double precision (inner solves at $10^{-10}$, $9.0$~min).
    Wall times for the $40$ steps at $\mesh{16}{32}{32}$, $p = 2$, each step with at most $200$ MINRES iterations to a relative residual of $0.1$. All three reach the same residual.}
    \label{fig:newton_tol_sweep}
\end{minipage}\hfill
\begin{minipage}[t]{0.48\linewidth}
\centering
    \resizebox{\linewidth}{!}{\includegraphics{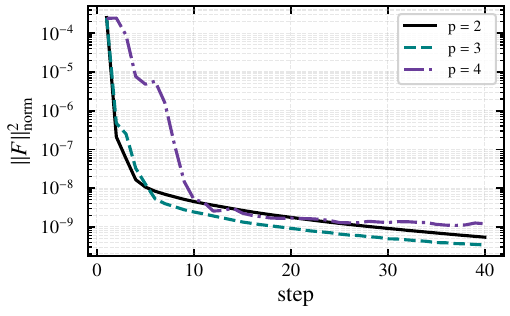}}
    \caption{Newton's method for $p = 2, 3, 4$, with $400$ MINRES iterations per step at $p = 4$ and $200$ otherwise.
    Increasing the order past $p = 2$ does not pay off.}
    \label{fig:newton_p_sweep}
\end{minipage}
\end{figure}

\paragraph{Precision}
Solver precision limits the accuracy of the descent direction and consequently the force residual that can be reached.
Figure~\ref{fig:newton_tol_sweep} shows that mixed and double precision agree. A larger tolerance of the inner solves, $\sqrt{\epsilon_{32}} \approx 3.5 \times 10^{-4}$, is too loose for the Newton direction: MINRES does not converge and the residual does not decrease.

Per step, mixed precision is $1.3 \times$ faster than float64, and single precision with the inner solves at $10^{-5}$ is $2.2 \times$ faster.

\paragraph{Order}
See Figure~\ref{fig:newton_p_sweep}.
Increasing the order past $p=2$ has no clear positive effect. After $40$ steps, $p = 3$ is at $3.4 \times 10^{-10}$ against $5.4 \times 10^{-10}$ for $p = 2$, at $1.3$ times the cost per step. At $p = 4$, the inexact Newton solve's MINRES iteration budget needs to be increased to $400$ and the run reaches $1.1 \times 10^{-9}$ at $3.1$ times the cost per step of $p = 2$. With $200$ iterations budget, the descent stalls above $10^{-5}$ as the additional fill-in in $\delta^2 \Energy$ complicates the Newton solve.

\paragraph{Hyperparameters}
Newton's method, in particular its inexact MINRES solve, comes with two hyperparameters that were determined by sweeps: the parallel-flow penalty $\kappa$ of Section~\ref{subsec:newton} and the budget of MINRES iterations per solve.

Table~\ref{tab:newton_penalty} and Table~\ref{tab:newton_inner} show the results. Any $\kappa$ between $1$ and $10$ gives a floor within a factor $1.6$ of the best and reaches $10^{-8}$ in $6$ to $10$ steps. The MINRES budget trades the cost per step against the number of steps: $50$ iterations need $25$ steps to $10^{-8}$ and floor at $4.8 \times 10^{-10}$, $200$ need $6$ steps and floor at $1.6 \times 10^{-10}$, and $400$ save one step at twice the cost.

\begin{table}[htb]
\centering
\begin{minipage}[t]{0.48\linewidth}
\centering
{\small\setlength{\tabcolsep}{3pt}
\centering
\begin{tabular}{l rrrr}
\toprule
$\kappa$ & \makecell[r]{$\normnorm{F}^2$ floor \\ $\times 10^{-10}$} & \makecell[r]{$< 10^{-8}$ \\ at step} & s/step & \makecell[r]{MINRES \\ it./step} \\
\midrule
$0$ & 200.02 & -- & 8.8 & 200 \\
$1$ & 2.30 & 6 & 8.8 & 200 \\
$3$ & 1.62 & 6 & 9.0 & 200 \\
$10$ & 2.55 & 10 & 8.9 & 200 \\
\bottomrule
\end{tabular}
}
\caption{Sweep to determine the best parallel-flow penalty $\kappa$. Without the penalty ($\kappa = 0$), $\normnorm{F}^2 < 10^{-8}$ is never reached as the descent direction is dominated by the field-aligned flows in the kernel of $\delta^2 \Energy$.}
\label{tab:newton_penalty}
\end{minipage}\hfill
\begin{minipage}[t]{0.48\linewidth}
\centering
{\small\setlength{\tabcolsep}{3pt}
\centering
\begin{tabular}{l rrrr}
\toprule
MINRES & \makecell[r]{$\normnorm{F}^2$ floor \\ $\times 10^{-10}$} & \makecell[r]{$< 10^{-8}$ \\ at step} & s/step & \makecell[r]{MINRES \\ it./step} \\
\midrule
$50$ & 4.80 & 25 & 2.6 & 50 \\
$100$ & 1.90 & 10 & 4.7 & 100 \\
$200$ & 1.62 & 6 & 9.0 & 200 \\
$400$ & 1.66 & 5 & 17.3 & 400 \\
\bottomrule
\end{tabular}
}
\caption{The inner solve on the same case at $\kappa = 3$. The choice of MINRES iterations is mostly a zero-sum trade-off between cost per step and the number of steps.}
\label{tab:newton_inner}
\end{minipage}
\end{table}

\begin{figure}[tb]
    \centering
    \includegraphics{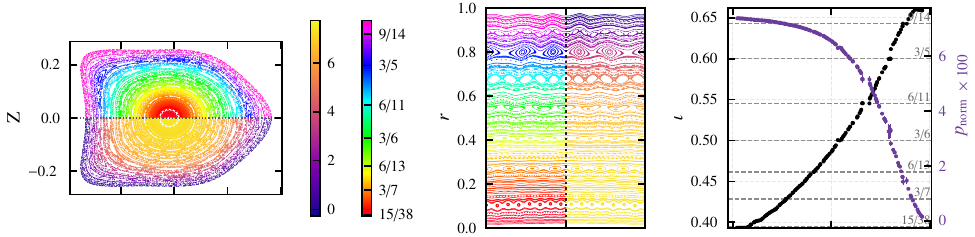} \\
    \includegraphics{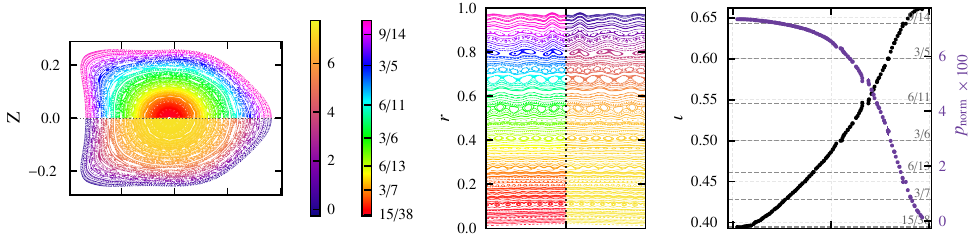} \\
    \includegraphics{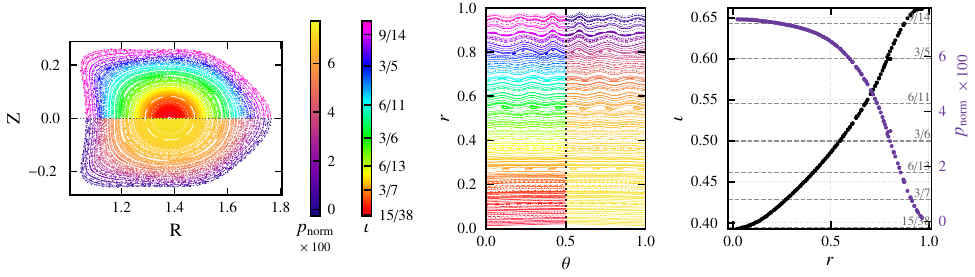}
    \caption{Poincaré sections at $\zeta = {0.5}$ for the Newton descent runs in Table~\ref{tab:newton_convergence}.
    Resolutions are $(n_r, n_\theta, n_\zeta) = \mesh{16}{32}{32}$ (top), $\mesh{24}{48}{48}$ (middle), and $\mesh{32}{64}{64}$ (bottom).
    The shape of grid-scale island features is, as expected, resolution-dependent. At large enough resolution, the $\tfrac35$ island chain does not form through numerical reconnection.}
    \label{fig:newton_sections}
\end{figure}

\paragraph{Resolution}
Figure~\ref{fig:newton_sections} shows the Poincar\'e sections of the relaxed states at three resolutions.
The islands opened by grid-scale reconnection shrink with resolution. At $n_r = 16$ the $\tfrac35$, $\tfrac{6}{11}$ and $\tfrac12$ chains are open, at $n_r = 24$ they are smaller, and at $n_r = 32$ the surfaces stay nearly nested.
Nonetheless, the final pressure profiles, rotational transform profiles, and force balances are in very strong agreement across resolution. 

\paragraph{Current sheets}

\begin{figure}[tb]
    \centering
    \includegraphics{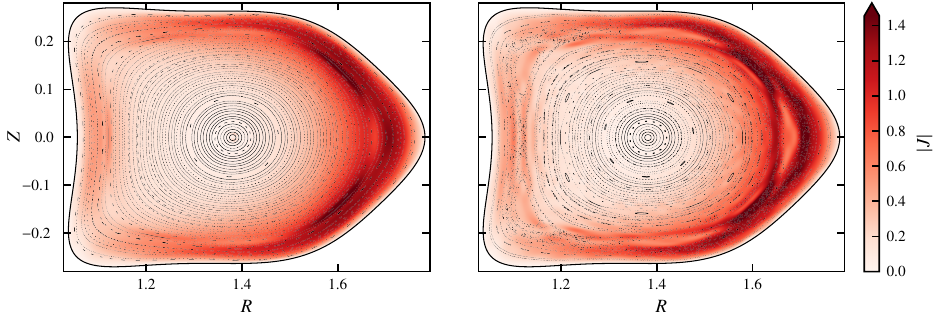}
    \caption{Current density $|J|$ at $\zeta = 0.5$ in the unseeded equilibrium of Figure~\ref{fig:islands_equilibrium}, $\mesh{48}{96}{96}$ (left), and in the seeded state of Figure~\ref{fig:islands_seeded32}, $\mesh{32}{64}{64}$ (right), on one colour scale. The black dots are the Poincar\'e crossings of every sixth traced field line.}
    \label{fig:current_sheets}
\end{figure}

Figure~\ref{fig:current_sheets} shows the current density $|J|$, $J = \widetilde{\curl} \, B$, at $\zeta = 0.5$ in the unseeded equilibrium of Figure~\ref{fig:islands_equilibrium} and in the seeded state of Figure~\ref{fig:islands_seeded32}.
In the nested state the current is a smooth crescent on the outboard side, the Pfirsch--Schlüter current of Section~\ref{subsec:rational_surfaces}.
In the seeded state it concentrates in thin bands along the separatrices of the $\tfrac35$ chain, and its maximum on this plane rises from $1.57$ to $1.99$.

\section{Hyperparameters}
\label{sec:hyperparameters}

\subsection{Code settings}
\label{subsec:code_settings}

We provide a complete list of all user-set MRX parameters and their default values in Table~\ref{tab:hyperparameters}.

\begin{table}[htb]
\caption{MRX relaxation parameter choices and default values, used by the reference run of Table~\ref{tab:runs}.}
% 2026-09-22: refactored (Tobias: "island seeding is automatic now"): grouped, notes shortened so the table fits the
% text width (it ran 280 pt over), the manual envelope-seed row replaced by the automatic energy-criterion seeding,
% restart row dropped, smoothing constant rounded to 0.07 (the runs: 0.0737; the code's SMOOTHING_C: 0.075).
% Defaults as in scripts/relax.py, mrx/relaxation.py, mrx/hessian.py (NEWTON_*) and mrx/precision.py (default_tol).
\centering
\footnotesize
\begin{tabular*}{\linewidth}{@{\extracolsep{\fill}}l l l@{}}
\toprule
parameter & default & notes \\
\midrule
\multicolumn{3}{@{}l}{\emph{Geometry and discretization}} \\
Equilibrium file (VMEC or GVEC) & -- & initial field and geometry \\
{Restart checkpoint} & {--} & {state of an earlier run as initial condition} \\
Symmetry \{stellarator, field period, none\} & stellarator & half a period, \ref{sec:stellarator_symmetry} \\
Splines $(n_r, n_\theta, n_\zeta)$ & $\mesh{32}{64}{64}$ & Fig.~\ref{fig:newton_h_sweep} \\
Degree $p$, quadrature points per cell & $2$, $p + 1$ & Fig.~\ref{fig:newton_p_sweep}, cost $\sim p^3$ \\
Knot vectors & uniform & Fig.~\ref{fig:mesh_refinement} \\
\addlinespace
\multicolumn{3}{@{}l}{\emph{Linear solvers}} \\
Precision \{float32, float64, mixed\} & float32 & Fig.~\ref{fig:newton_tol_sweep}, Sec.~\ref{subsec:seeding} \\
Tolerance & $10^{-5}$ & $10^{-10}$ in float64, $10^{-8}$ in mixed precision \\
\addlinespace
\multicolumn{3}{@{}l}{\emph{Relaxation}} \\
Method \{Newton, gradient descent\} & Newton & Section~\ref{subsec:newton} \\
Newton: MINRES iterations, tolerance & $(200, 0.1)$ & Tab.~\ref{tab:newton_inner}, $\delta t \leq 1$ \\
Newton: parallel-flow penalty $\kappa$ & $3$ & Tab.~\ref{tab:newton_penalty} \\
Gradient descent: smoothing $(\gamma, \varepsilon)$ & $(1, 0.07 \, \vavg{g_{rr}} \, h_r^2)$ & Fig.~\ref{fig:gamma_sweep} \\
Gradient descent: velocity \{Leray, potential\} & potential & Tab.~\ref{tab:velocity_choices} \\
CFL cap $C$ & $0.5$ & \eqref{eq:cfl} \\
Helicity correction $\sigma$ & off & Tab.~\ref{tab:helicity_preservation}, cost $+3$--$5\%$ per step \\
Maximum steps & $100$ / $2000$ & Newton / gradient descent \\
Steps per compiled chunk & $10$ / $200$ & diagnostics are done between chunks \\
Stopping criterion on $\normnorm{F}^2$ & $10^{-10}$ & mean over a chunk \\
\addlinespace
\multicolumn{3}{@{}l}{\emph{Topology changes}} \\
Island seeding (Section~\ref{subsec:seeding}) & off & \ref{subsec:seeding} \\
Resistivity $\eta \, \delta t$ & off & units of $\vavg{g_{rr}} \, h_r^2$, Section~\ref{sec:resistive} \\
Smoothing $\varepsilon$ of the reference field & $0.1$ & units of $\vavg{g_{rr}} \, h_r^2$, $0.03$ in Section~\ref{subsec:reconnection_demo} \\
\bottomrule
\end{tabular*}

\label{tab:hyperparameters}
\end{table}

\subsection{Numerical experiments}
\label{subsec:experiment_parameters}

Table~\ref{tab:runs} lists the parameters used for all numerical experiments reported in this work.

\begin{table}[htb]
\caption{The runs behind the paper's figures and tables.
The reference configuration is Newton's method on the NCSX li383 case at $\mesh{32}{64}{64}$, $p = 2$, half a field period with stellarator symmetry, mixed precision, solver tolerance $10^{-8}$, Newton-MR with the parallel-flow penalty $\kappa = 3$ and a budget of $200$ MINRES iterations, the step capped at $\delta t = 1$, and $150$ steps from the VMEC field to the floor in chunks of $30$.
The gradient descent runs use velocity smoothing with $\gamma = 1$ and $\varepsilon = 0.07 \, \vavg{g_{rr}} \, h_r^2$, the potential velocity and $5{,}000$ steps in chunks of $500$ without a floor stop, at $\mesh{16}{32}{32}$.
Each row names what the experiment sweeps and what else deviates from the reference.}
% 2026-09-22: checked row by row against the run records (relax.json of every run named in the paper); the rows of
% the earlier reconnection experiments went with their floats
\centering
\footnotesize
\setlength{\tabcolsep}{4pt}
\begin{tabular}{@{}>{\raggedright\arraybackslash}p{3.3cm} >{\raggedright\arraybackslash}p{2.8cm} >{\raggedright\arraybackslash}p{4.6cm} >{\raggedright\arraybackslash}p{4.7cm}@{}}
\toprule
experiment & shown in & swept & notes \\
\midrule
helicity preservation & Tab.~\ref{tab:helicity_preservation} & precision mixed / float64, $\sigma = 0$ / $\sigma \neq 0$ & gradient descent, $\gamma = 0$, $\mathrm{steps} = 1000$ \\
\addlinespace
Newton against gradient descent, wall time & Fig.~\ref{fig:newton} & Newton / gradient descent & $\mesh{16}{32}{32}$, Newton over the first $100$ steps of the $n_r = 16$ run \\
\addlinespace
Newton convergence in $n_r$ & Tab.~\ref{tab:newton_convergence}, Fig.~\ref{fig:newton_h_sweep}, Fig.~\ref{fig:newton_sections} & Newton, $n_r \in \{16, 24, 32, 48, 64\}$, and gradient descent at $n_r = 16$ & $\mathrm{steps}$ from $820$ down to $120$ (two to five hours each), no floor stop \\
\addlinespace
Newton: penalty, inner solve & Tab.~\ref{tab:newton_penalty}, Tab.~\ref{tab:newton_inner} & $\kappa \in \{0, 1, 3, 10\}$ and MINRES $\in \{50, 100, 200, 400\}$ & $\mesh{16}{32}{32}$, $\mathrm{steps} = 200$, no floor stop \\
\addlinespace
Newton: degree & Fig.~\ref{fig:newton_p_sweep} & $p \in \{2, 3, 4\}$, $400$ MINRES iterations at $p = 4$ & $\mesh{16}{32}{32}$, $\mathrm{steps} = 40$ \\
\addlinespace
Newton: solver tolerance, precision & Fig.~\ref{fig:newton_tol_sweep} & precision mixed / float64 / float32 (tol $10^{-5}$), $\mathrm{tol} \in \{10^{-6}, 10^{-8}, 10^{-10}\}$ in mixed & $\mesh{16}{32}{32}$, $\mathrm{steps} = 40$ \\
\addlinespace
nested equilibrium & Fig.~\ref{fig:islands_equilibrium} & -- & $\mesh{48}{96}{96}$, $\mathrm{steps} = 200$, no floor stop \\
\addlinespace
seed selection: relax, seed, relax & Tab.~\ref{tab:seed_selection}, Fig.~\ref{fig:islands_seeded32}, Fig.~\ref{fig:seed_trace} & every chain at $\aopt$ & the reference run, then $200$ steps after the seed, no floor stop \\
\addlinespace
reconnection between relaxed states & Tab.~\ref{tab:reconnection_demo}, Fig.~\ref{fig:reconnection_before}, Fig.~\ref{fig:reconnection_after} & seed none / $(6, 1)$ at $\aopt$ / $(5, 1)$ at $\aopt / 4$ & from the reference run and its seeded relaxations: $200$ steps with $\eta \, \delta t = 0.064 \, \vavg{g_{rr}} h_r^2$ against the nested state smoothed by $\varepsilon = 0.03 \, \vavg{g_{rr}} h_r^2$, then $100$ ideal steps, no floor stop \\
\addlinespace
velocity formulation & Tab.~\ref{tab:velocity_choices} & velocity potential / Leray & gradient descent \\
\addlinespace
velocity smoothing & Fig.~\ref{fig:gamma_sweep} & $\gamma \in \{0, 1\}$, $\mathrm{steps} \in \{5000, 7000\}$ & gradient descent \\
{verification against an analytic equilibrium} & {Tab.~\ref{tab:landreman_convergence}} & {$n \in \{8, 12, 16, 24, 32\}$, $p \in \{2, 3\}$} & {the equilibrium of~\citep{landreman_analytic_2026}, $\mesh{n}{2n}{2n}$, float64, no relaxation} \\
\addlinespace
quasi-axisymmetric shape optimization & Fig.~\ref{fig:shape_optimization}, Fig.~\ref{fig:shape_errors}, Fig.~\ref{fig:vacuum_qa_poincare} & displacement $0$ and $10$~mm RMS, draws $0$--$4$ & vacuum field of the QA device, $\mesh{24}{48}{24}$, $p = 3$, float64, augmented Lagrangian (\ref{subsec:shape_constraints}), stop rules of~\ref{subsec:shape_constraints} \\
\bottomrule
\end{tabular}

\label{tab:runs}
\end{table}

\FloatBarrier
\bibliographystyle{elsarticle-num}
\bibliography{bibliography}

\end{document}